\documentclass[%
twocolumn,
superscriptaddress,
nofootinbib,
 amsmath,amssymb,
 aps, prd
]{revtex4-2}

\usepackage{graphicx}
\usepackage{tensor}
\usepackage{siunitx}
\usepackage{xcolor} 
\usepackage[normalem]{ulem} 
\usepackage{lipsum} 

\usepackage[pdfborder={0 0 0}, plainpages=false, hyperfigures=true]{hyperref}

\newcommand{\beq}{\begin{equation}}
\newcommand{\eeq}{\end{equation}}
\newcommand{\beqn}{\begin{eqnarray}}
\newcommand{\eeqn}{\end{eqnarray}}

\newcommand{\lo}{\mathrel{\raise.3ex\hbox{$<$}\mkern-14mu
    \lower0.6ex\hbox{$\sim$}}}
\newcommand{\go}{\mathrel{\raise.3ex\hbox{$>$}\mkern-14mu
    \lower0.6ex\hbox{$\sim$}}}

\newcommand{\WSU}{\affiliation{Department of Physics \& Astronomy,
	Washington State University, Pullman, Washington 99164, USA}}
\newcommand{\UNH}{\affiliation{Department of Physics \& Astronomy, University of New Hampshire, 9 Library Way, Durham NH 03824, USA}}
\newcommand{\TAPIR}{\affiliation{TAPIR, Walter Burke Institute for Theoretical Physics, MC 350-17, California Institute of Technology, Pasadena, California 91125, USA}}
\newcommand{\Cornell}{\affiliation{Cornell Center for Astrophysics and Planetary Science, Cornell University, Ithaca, New York, 14853, USA}}
\newcommand{\MPI}{\affiliation{Max Planck Institute for Gravitational Physics (Albert Einstein Institute), Am M{\H u}hlenberg 1, 14476 Potsdam, Germany}}

\newcommand{\UCBP}{\affiliation{Department of Physics, University of California, Berkeley, Berkeley, CA 94720, USA}}

\begin{document}
\title{Robustness of neutron star merger simulations to changes in neutrino transport and neutrino-matter interactions}
\author{Francois Foucart}\UNH
\author{Patrick Chi-Kit Cheong}\UCBP\UNH
\author{Matthew D. Duez}\WSU
\author{Lawrence E. Kidder}\Cornell
\author{Harald P. Pfeiffer}\MPI
\author{Mark A. Scheel}\TAPIR

\begin{abstract}
Binary neutron star mergers play an important role in nuclear astrophysics: their gravitational wave and electromagnetic signals carry information about the equation of state of cold matter above nuclear saturation density, and they may be one of the main sources of r-process elements in the Universe. Neutrino-matter interactions during and after merger impact the properties of these electromagnetic signals, and the relative abundances of the produced r-process elements. Existing merger simulations are however limited in their ability to realistically model neutrino transport and neutrino-matter interactions. Here, we perform a comparison of the impact of the use of state-of-the art two-moment or Monte-Carlo transport schemes on the outcome of merger simulations, for a single binary neutron star system with a short-lived neutron star remnant ($(5-10)\,{\rm ms}$). We also investigate the use of different reaction rates in the simulations. While the best transport schemes generally agree well on the qualitative impact of neutrinos on the system, 
differences in the behavior of the high-density regions can significantly impact the collapse time and the properties of the hot tidal arms in this metastable merger remnant. The chosen interaction rates, transport algorithm, as well as recent improvements by Radice {\it et al} to the two-moment algorithms can all contribute to changes at the $(10-30)\%$ level in the global properties of the merger remnant and outflows. The limitations of previous moment schemes fixed by Radice  {\it et al} also appear sufficient to explain the large difference that we observed in the production of heavy-lepton neutrinos in a previous comparison of Monte-Carlo and moment schemes in the context of a low mass binary neutron star system.
\end{abstract}

\maketitle

\section{Introduction}

The collisions of two neutron stars have proven themselves over the last decade to be particularly interesting events for the study of nuclear physics. Gravitational waves, gamma-ray bursts, and kilonovae all carry information about the properties of cold neutron rich matter above nuclear saturation density, thus providing us with a rare opportunity to study the high-density state of quantum chromodynamics. This was most clearly demonstrated by the multimessenger detection of GW170817~\cite{TheLIGOScientific:2017qsa,GBM:2017lvd,Monitor:2017mdv,2017Sci...358.1559K,2017ApJ...848L..19C,Cowperthwaite:2017dyu,2017Sci...358.1583K,2017ApJ...848L..32M,2017ApJ...848L..18N,2017Natur.551...75S,2017ApJ...848L..16S,2017ApJ...848L..27T,2017Sci...358.1565E,Mooley:2018dlz}, though joint detections of gamma-ray bursts and kilonovae~\cite{Tanvir:2013,Berger:2013,jin:15,Yang:2015pha,Rastinejad:2022zbg,Troja:2022yya,Yang:2022qmy} could also provide useful information if our ability to model these signals was not so limited by uncertainties in numerical simulations and nuclear physics inputs. Neutron star mergers are also probably the systems with the strongest observational evidence for the production of r-process elements in the Universe~\cite{2017Natur.551...80K}; yet how much these mergers practically contribute to nucleosynthesis remains very uncertain, as is the abundance patterns of the elements that they produce. These two open questions have in fact clearly overlapping uncertainties, as the radioactive decay of the ashes of the r-process is what powers kilonovae. If we could gain a proper understanding of the connection between kilonova signals and the properties of the matter ejected by mergers, and from there to the properties of the merging compact objects, we would improve our ability to understand both the equation of state of dense matter and the role of mergers in astrophysical nucleosynthesis.

Besides observational uncertainties, our ability to understand kilonova signals and r-process nucleosynthesis from mergers is limited by a number of theoretical unknowns in the properties of neutron-rich nuclei (reaction rates, opacities,...)~\cite{Barnes:2020nfi,Zhu:2020eyk}, as well as uncertainties in the simulation of neutron star mergers and their matter outflows. For merger simulations, the most important issues are likely the poorly resolved impact of magnetic fields on the generation of matter outflows~\cite{Kiuchi2014}, and the impact of neutrino transport and neutrino-matter interactions on the composition of these outflows~\cite{Wanajo2014,Foucart:2020qjb}. Here, we focus on the latter issue, which directly impacts the outcome of nucleosynthesis~\cite{Lippuner2015} and the color and duration of kilonovae~\cite{2013ApJ...775...18B}.

Numerical algorithms for general relativistic neutrino transport in merger simulations have made very significant progresses over the last decade, moving from existing order-of-magnitude accurate leakage schemes~\cite{1997A&A...319..122R,Rosswog:2003rv,Deaton2013,Neilsen:2014hha} to hybridization of the leakage scheme with an approximate two-moment~\cite{Wanajo2014,Sekiguchi:2015} and one-moment~\cite{Radice:2016} transport algorithm, standalone two-moment transport~\cite{FoucartM1:2015,Sun:2022vri}, two-moment transport complemented with the evolution of the neutrino number density~\cite{Foucart:2016rxm}, significant improvements to the methods used to take implicit timesteps in high-density regions~\cite{Radice:2021jtw}, low-order Monte-Carlo transport~\cite{Foucart:2020qjb}, higher-order time-stepping with Monte-Carlo~\cite{Kawaguchi:2022tae}, and the potential use of Monte-Carlo transport as a closure for two-moment schemes~\cite{Foucart:2017mbt,Izquierdo:2023fub}. Even the best of these schemes either remain approximation to the equations of radiation transport (two-moment schemes), or can only be used at very low accuracy (Monte-Carlo). Additionally, there has been only limited attention paid to the role of the chosen interaction rates in simulations using the most modern transport scheme.

A number of comparisons of neutrino transport schemes have already been performed through direct simulations of binary neutron star mergers. In~\cite{FoucartM1:2016}, we used the SpEC code to compare neutrino leakage with an early version of our two-moment transport scheme in a low-mass binary neutron star system, finding sigificant differences in the composition of the remnant and outflows between the two schemes (average $Y_e$ rising from $0.1$ to $0.2$ in that system) -- a result in agreement with the interpretation of the neutron-poor outflows previously seen in the first two-moment simulations of Wanajo {\it et al}~\cite{Wanajo2014}. The same binary neutron star configuration was also evolved with leakage in~\cite{Mosta:2020hlh} and two-moment schemes in~\cite{Curtis:2023zfo}, for a system collapsing to a black hole on relatively short time scales. The collapse time itself was significantly impacted by the choice of transport scheme, with the two-moment simulation collapsing after $12\,{\rm ms}$ and the leakage simulation after $22\,{\rm ms}$. In~\cite{Foucart:2016rxm} we showed for the same system that the ad-hoc choices made for the energy spectrum of neutrinos when computing neutrino-matter interactions similarly impacted outflow composition, though mainly in the polar regions. Radice {\it et al}~\cite{Radice:2021jtw} used the THC code to perform comparisons of moment simulations including their one-moment scheme~\cite{Radice:2016}, two-moment scheme with Eddington closure, and two-moment scheme with Minerbo closure for a low-mass (non-collapsing) and a high-mass (collapsing) systems. The one-moment scheme is found to differ from the two-moment scheme by a factor of two in neutrino luminosities (still closer to the two-moment scheme than a leakage scheme likely would be), and lower electron fraction of the outflows, while the different two-moment schemes are in significantly better agreement. In~\cite{Foucart:2020qjb}, we compared Monte-Carlo transport with the SpEC gray two-moment scheme in a low-mass system, finding $O(10\%)$ differences in the neutrino luminosities and outflow composition (and larger errors for the heavy-lepton luminosity). We note of course that these results focused on direct comparisons of neutrino transport scheme are just a fraction of the community effort going to the study of neutron star mergers (see e.g.~\cite{BaiottiReview2016,Radice:2020ddv,Kiuchi:2024lpx} and references therein).

Here, we continue this effort to address the capabilities of various neutrino transport scheme by studying a binary neutron star merger with a collapse timescale of $\sim 10\,{\rm ms}$, a case already studied with a hybrid two-moment+leakage scheme by Sekiguchi {\it et al}~\cite{Sekiguchi:2016}. We used the same physical system in~\cite{Foucart:2024kci} to study the impact of residual eccentricity and equation of state implementation on merger results. We perform a direct comparison of the improvements made to two-moment schemes by Radice {\it et al}~\cite{Radice:2021jtw} with results obtained using the preexisting SpEC two-moment algorithm~\cite{Foucart:2016rxm}, as well as a comparison of Monte-Carlo and two-moment transport in a collapsing binary neutron star system. We additionally consider the impact of the inclusion of different reactions in the simulations by considering simulations using NuLib rates in which we ignore pair processes for electron type neutrinos~\cite{OConnor:2015}, and simulations using the energy-integrated rates of~\cite{Rosswog:2003rv} including an approximate treatment of pair processes.

In Sec.~\ref{sec:methods}, we describe our initial conditions and numerical methods, with particular emphasis on the implementation within SpEC of the improvements to the two-moment scheme proposed by Radice {\it et al}~\cite{Radice:2021jtw}. Sec.~\ref{sec:results} discuss the outcome of the simulations in terms of the merger remnant, early matter ejection, and neutrino distribution. We focus on the early evolution, up to $\sim 3\,{\rm ms}$ post-merger. This allows us to study the dynamical ejecta and formation of the compact remnant. We also follow two of the simulations through collapse to a black hole.

\section{Methods}
\label{sec:methods}

\subsection{Initial Conditions}

In all simulations presented here, we consider the merger of neutron stars with gravitational masses $1.4M_\odot$ and $1.3M_\odot$. Both stars are initially non-rotating, and start their evolution about 6 orbits before merger. We start from the exact same initial conditions as in the low eccentricity simulations of~\cite{Foucart:2024kci} ($e\sim 0.003-0.004$), using either the Compose~\cite{Typel:2013rza,Oertel:2016bki} or StellarCollapse~\cite{OConnor2010} version of the SFHo equation of state. These initial conditions were generated with the Spells initial data solver~\cite{Pfeiffer2003,FoucartEtAl:2008}. In~\cite{Foucart:2024kci}, we showed that in the absence of neutrinos and magnetic fields, simulations using the two versions of this equation of state cannot be differentiated within estimated numerical errors. We note that the use of this equation of state results in neutron stars with circumferertial radius $R=11.9\,{\rm km}$ for the two masses simulated. This case uses the same binary parameters system as simulation SFHo-130-140 of Sekiguchi {\it et al}~\cite{Sekiguchi:2016}.

\subsection{GR-Hydro methods}

The evolution of the metric and fluid variables follow the exact same methods as in our neutrinoless simulations of the same system~\cite{Foucart:2024kci}. These methods are described in more details in~\cite{Duez:2008rb,Foucart:2013a}. To summarize, we evolve Einstein's equations in the generalized harmonic formalism~\cite{Lindblom2006} on a pseudospectral grid using p-adaptive mesh refinement~\cite{Szilagyi:2009qz}, and the fluid equations in the flux-conservative Valencia formalism~\cite{Banyuls1997} using high-order shock capturing methods (WENO5 reconstruction~\cite{Liu1994200,Jiang1996202,Borges} and HLL approximate Riemann solver~\cite{HLL}) on a finite volume grid with fixed mesh refinement. The grid resolution on the finest grid, which covers both neutron stars, is $\Delta x \sim (160-200)\,{\rm m}$ during inspiral\footnote{The grid spacing changes as our grid contracts to follow the motion of the neutron stars; we interpolate the solution on a new grid with $\Delta x\sim 200\,{\rm m}$ every time the grid contracts by a factor of $0.8$.}, and $\Delta x \sim 160\,{\rm m}$ during and after merger. During and after merger, each refinement level has $256^3$ cells, with the grid spacing increasing by a factor of $2$ between refinement level. The finest level thus covers a cube of side $40\,{\rm km}$. We use a total of 4 refinement levels in these simulations. In~\cite{Foucart:2024kci}, we also presented a higher resolution simulation, which provides an estimate of numerical errors due to the coupled evolution of the metric and fluid.

\subsection{Neutrino transport}

The main objective of this manuscript is to study the impact of neutrino transport and neutrino-matter interaction rates on the merger. We consider the following variations:

{\it Base transport algorithm}: We consider either a two-moment scheme with Minerbo closure~\cite{1981MNRAS.194..439T,shibata:11}, or the Monte-Carlo transport scheme recently implemented in SpEC~\cite{Foucart:2021mcb}. The two-moment scheme is generally expected to behave better in high-density regions (with caveats discussed below for some implementations of this scheme), but is limited by the lack of spectral information in the decoupling region and the use of approximate analytical closure for the pressure tensor in the optically thin region. The Monte-Carlo scheme, on the other hand, is expected to converge to the correct solution of Boltzmann's equations, but very slowly and with significant shot noice at the number of packets affordable in our simulations. In addition, the use of implicit Monte-Carlo in high-opacity regions should further slow down convergence of the scheme in those regions~\cite{Foucart:2021mcb}.

{\it Detailed methods for two-moment schemes}: Radice {\it et al}~\cite{Radice:2021jtw} recently developed improved methods for the use of two-moment schemes in merger simulations. These include changes to the treatment of numerical fluxes in high-density regions, as well as an improved treatment of the implicit solve required to evolve the moment equations in time. The preexisting SpEC methods, on the other hand, used an approximate implicit solve, linearizing the problem around $E=F^i=0$ (with $E$, $F^i$ the neutrino energy and flux density in the laboratory frame) and ignoring terms quadratic and higher in the fluid velocity. In this manuscript, we use simulations implementing both the modified fluxes in high-density regions, and the improved treatment of the implicit solve from~\cite{Radice:2021jtw}. We will refer to these methods as ``SpEC'' and ``Radice'', although all simulations described in this manuscript are performed with the SpEC code. Our implementation of the methods of~\cite{Radice:2021jtw} in SpEC, which differ in minor ways from the original, are described in more details in Sec.~\ref{sec:M1details}. We note that an intermediate model accounting for the fluid velocity at all orders was implemented in the BAM code by Schianchi {\it et al}~\cite{Schianchi:2023uky}; that method is not tested in the simulations presented here.

{\it Neutrino-matter interactions}: The simulations using Monte-Carlo methods use a NuLib~\cite{OConnor:2015} table with 16 energy groups, logarithmically spaced up to $528\,{\rm MeV}$. The table includes $\nu_e$ absorption on neutrons, $\bar\nu_e$ absorption on protons, as well as production of $\nu_\mu \bar\nu_\mu$ and $\nu_\tau\bar\nu_\tau$ pairs from $e^+e^-$ annihilation and nucleon-nucleon Bremstrahlung. The pair production channels are not included for electron type neutrinos. All inverse reactions are also included, in such a way that the emissivity $\eta$ and absorption opacity $\kappa_a$ satisfy Kirchoff's law $\eta/\kappa_a = B_\nu$, with $B_\nu$ the black-body distribution function of neutrinos in equilibrium with the fluid, integrated over the relevant energy bin. We also include in the table elastic scattering on protons, neutrons, $_2^4$He, and heavy nuclei. For the two-moment method, we first consider a gray version of the NuLib table, i.e. by taking a weighted average of the opacities that assumes an equilibrium distribution of neutrinos, then correcting for the estimated average energy of the neutrinos as described below. We also consider approximate gray emission and absorption rates calculated on the fly, using the rates of~\cite{Rosswog:2003rv}. In this latter version of the interaction rates, we use the same reactions as for the NuLib table, but now adding the contribution from pair production to the electron type neutrinos and additionally considering plasmon decay $\gamma\gamma\leftrightarrow \nu\bar\nu$. The two sets of reaction rates also make different approximations for the spectrum of neutrinos and blocking factors, even for reactions included in both sets of interaction rates.

\subsection{Simulations performed}

The specific combinations of methods used in this manuscript include:
\begin{enumerate}
\item Monte-Carlo transport with an energy-dependent NuLib table and a maximum of $10^8$ packets per species (MC-HR)
\item Monte-Carlo transport with an energy-dependent NuLib table and a maximum of $4\times 10^7$ packets per species (MC-LR; MC-LR and MC-HR start from the same pre-merger evolution, as before merger we do not need many neutrino packets to capture the very small impact of neutrinos on the evolution)
\item Two-moment transport using Radice's fluxes and implicit solve, and on-the-fly calculation of interaction rates (M1-Radice).
\item Two-moment transport with SpEC's fluxes and implicit solve, as well as on-the-fly calculation of interaction rates (M1-SpEC).
\item Two-moment transport using SpEC's fluxes and implicit solve, and the gray NuLib table (M1-NuLib)
\item No neutrinos at all. This is simulation ``SC-MR'' from~\cite{Foucart:2024kci}, renamed ``Hydro'' here, and thus not a new simulation in this manuscript.
\end{enumerate}
These simulations are also listed on Table~\ref{tab:sim}.

\begin{table}
\begin{tabular}{c|c|c|c|c}
ID & Transport & Fluxes & Implicit Solve & Rates\\
\hline
MC-HR & MC (HR) & NA & NA & NuLib\\
MC-LR & MC (LR) & NA & NA & NuLib\\
M1-Radice & M1 & Radice & Radice & On-the-fly \\
M1-SpEC & M1 & SpEC & SpEC & On-the-fly \\
M1-NuLib & M1 & SpEC & SpEC & GrayNuLib\\
Hydro & None & NA & NA & NA
\end{tabular}
\caption{List of simulations performed for this manuscript. We list the transport scheme used (M1 or MC), the method used to calculate interaction rates, and, for M1 simulations, the method used to calculate numerical fluxes and to perform the implicit solve. The ``Hydro'' simulation is performed without neutrino transport (simulation ``SC-MR'' from~\cite{Foucart:2024kci})}
\label{tab:sim}
\end{table}

All simulations are performed to at least $3\,{\rm ms}$ post-merger. The MC-HR simulation as well as the M1-NuLib simulation are followed up to the collapse of the remnant to a black hole.

\subsection{Implementation details for the two-moment scheme}
\label{sec:M1details}

Most of the simulations presented in this manuscript use some variations of the two-moment scheme in which we evolve the energy density, momentum density, and number density of neutrinos in the laboratory frame. This formalism was first developed in~\cite{1981MNRAS.194..439T,shibata:11}, and its SpEC implementation is described in~\cite{FoucartM1:2015,Foucart:2016rxm}. Improvements to the two-moment methods, and especially to the treatment of the neutrino spectrum and implicit time-stepping, were presented in~\cite{Radice:2021jtw}. We follow the methods of~\cite{Foucart:2016rxm} for the `SpEC' simulations and of~\cite{Radice:2021jtw} for the `Radice' simulations; in this section, we will present a brief overview of the moment methods, then focus on implementation details of the `Radice' algorithm within the SpEC code.

The main equations evolved in the two-moment scheme are
\beqn
\partial_t \tilde E + \partial_i \left( \alpha \tilde F^i - \beta^i \tilde E\right)  &=& \alpha \tilde P^{ij} K_{ij} 
-\tilde F^j \partial_j \alpha \notag\\
&& - \alpha \tilde S^\mu n_\mu\\
\partial_t \tilde F_j + \partial_i \left( \alpha \tilde P^i_j - \beta^i \tilde F_j \right) &=& -\tilde E \partial_j \alpha
+ \tilde F_k \partial_j \beta^k  \notag\\
&& + \frac{\alpha}{2} \tilde P^{ik} \partial_j \gamma_{ik}  
+ \alpha \tilde S^\mu \gamma_{j\mu}\\
\partial_t \tilde N + \partial_j \left(\alpha \sqrt{\gamma} \mathcal{F}^j -\beta^j \tilde N\right)
&=& \alpha \sqrt{\gamma} C_{(0)}
\eeqn
with $E$ the neutrino energy density, $F^i$ the neutrino momentum density, $N$ the neutrino number density, and tilde quantities denoting densitized variables (i.e. $\tilde E = \sqrt{\gamma}E$, with $\gamma$ the determinant of the spatial 3-metric $\gamma_{ij}$).
The metric is decomposed according to
\beq
ds^2 = -\alpha^2 dt^2 + \gamma_{ij} \left(\beta^i dt + dx^i\right)\left(\beta^j dt + dx^j\right)
\eeq
with $\alpha$ the lapse and $\beta^i$ the shift vector. The term $S^\alpha$ includes coupling between the neutrinos and the fluid. Its exact form will depend on the interactions included in the simulations. In our current simulations, we account for an isotropic emissivity in the fluid frame $\eta$, an absorption opacity $\kappa_A$, and an isotropic elastic scattering opacity $\kappa_S$. Then
\beq
S^\alpha = \eta u^\mu - \kappa_A J u^\mu - (\kappa_A + \kappa_S) H^\mu
\eeq
with $u^\mu$ the 4-velocity of the fluid, $J$ the neutrino energy density in the fluid frame, and $H^\mu$ the neutrino momentum density in the fluid frame. Finally, $C_{(0)}$ is given by
\beq
C_{(0)} = \eta_N - \kappa_N \frac{J}{\langle\nu\rangle}
\eeq
with $\eta_N, \kappa_N$ the number emissivity and absorption opacity, and $\nu$ the average neutrino energy. We use
\beq
\langle \nu \rangle = \frac{W (E-F_iv^i)}{N}
\eeq
with $v^i$ the spatial 3-velocity of the fluid. The main approximations of a gray two-moment scheme comes in the analytical form used to estimate the pressure tensor $P^{ij}$ and the number flux $\mathcal{F}^j$, as well as in the assumptions made about the neutrino spectrum that go into the calculations of the energy-averaged $\eta, \eta_N, \kappa_A, \kappa_S, \kappa_N$. We refer the reader to~\cite{Foucart:2016rxm} for more details about our standard choices for these calculations and a discussion of the impact of these approximations on the outcome of the radiation transport problem in neutron star mergers.

{\it Numerical fluxes}: The first important change when going from the standard SpEC algorithm to the methods of~\cite{Radice:2021jtw} is the way the flux terms on the left-hand-side of the evolution equations are computed, i.e. all the divergence terms of the form $\partial_i f^i$. In low-density regions, these fluxes can be treated with the same high-order shock-capturing methods as the similar terms coming into the evolution of the fluid variables. In high optical-depth regions, however, the numerical diffusion that is naturally introduced by these methods can become larger than the physical diffusion rates of neutrinos through dense matter. Then, using these methods can lead to a large overestimate of the neutrino diffusion rate. In previous SpEC simulations, we used a complex procedure described in~\cite{Audit2002,FoucartM1:2015} to drive the flux in the evolution of $\tilde E$ to the desired solution of the diffusion equation when $\kappa_A \Delta x \gg 1$. Radice {\it et al}~\cite{Radice:2021jtw} however pointed out that a simpler and more reliable methods is to use a weighted average of a `low-order' flux taken from shock capturing methods (Lax-Friedrich flux), and a `high-order' flux that does not include diffusive shock-capturing methods (second order finite differencing). The detailed calculations of the fluxes in SpEC exactly matches that of~\cite{Radice:2021jtw}.

{\it Time stepping method}: The second modification to our algorithm is in the method used to capture the neutrino-matter coupling (i.e. the source terms $S^\alpha$). In the two moment equations, most terms in the evolution of $(\tilde E, \tilde F^i)$ are treated explicitly (in SpEC, using a second-order Runge-Kutta algorithm). The source term $C_{(0)}$ is treated implicitly, but the evolution of $\tilde N$ can be trivially solved as a post-processing steps after we obtain the values of $(\tilde E, \tilde F^i)$ at the end of a time step. The terms involving $\tilde S^\alpha$ are more complex to handle. In high-opacity regions, they need to be treated implicitly, due to the very short timescale for neutrino-matter equilibration. We do this neutrino species by neutrino species, given a guess for the fluid properties at the end of a time step (see below for the method used to guess at the fluid temperature and electron fraction; the other properties of the fluid are assumed constant during a time step in this calculation). Practically, this requires solving an equation of the type
\beq
U^{n+1} = U^n + S^n_{\rm exp} dt + S^{n+1}_{\rm imp} dt
\eeq
with $U=(\tilde E,\tilde F^i)$ the evolved variables, $S_{\rm exp}$ the explicit source terms and fluxes, and $S_{\rm imp}$ the implicit source terms. The superscripts $n$ and $n+1$ refer to values at the beginning and end of a time step (or half step for the intermediate step of the Runge-Kutta algorithm). In previous simulations, we used a linearization of $S^{n+1}$ around $E=F^i=0$ to solve this equation, neglecting terms quadratic in the fluid velocity. Radice {\it et al}~\cite{Radice:2021jtw} instead derived the full jacobian matrix $\partial(S_{\rm imp})/\partial{U}$ for generic $U$. In simulation M1-Radice, we solve the implicit problem using the dog-legged method and the jacobian matrix of~\cite{Radice:2021jtw}. We note that this does not exactly match the method of~\cite{Radice:2021jtw}. This is in part because we use a different time stepping algorithm, and in part because we make a slightly different approximation in regions of high optical depth. In~\cite{Radice:2021jtw}, the authors note that in these regions the solver can fail, and that this issue can be resolved by calculating the jacobian under the assumption of an `optically thick' closure for the pressure tensor. In our simulation, we do not wait for a failure and automatically assume the optically thick closure if the equilibration timescale $\tau_{\rm eq} = 1/\sqrt{\kappa_A (\kappa_A+\kappa_S)} \leq 0.1 \alpha  dt$, with $dt$ the time step. We also treat all source terms explicitly if $\tau_{\rm eq} \geq 10 \alpha dt$. Finally, we use as initial guess for the solution
\beq
U^{n+1}_{\rm guess} = U^{*,n} (1-f) + U_{\rm eq} f 
\eeq
with $U_{\rm eq}$ representing the values of the evolved variables for neutrinos in equilibrium with the fluid, discussed below, $U^{*,n}$ the value of the evolved variables {\it after application of the explicit source terms}, and
\beq
f = \exp{\left(-\tau_{\rm eq}/(\alpha dt)\right)}.
\eeq

{\it Guess for temperature and electron fraction in high-opacity regions}: in dense, hot regions, the temperature $T$ and electron fraction $Y_e$ of the fluid may be stiffly coupled to the neutrinos. Calculating interaction rates using the fluid variables at the beginning of a time step is then occasionally unstable. In simulation M1-Radice, we updated our methods to guess at the fluid properties at the end of a step following methods similar to Radice {\it al}~\cite{Radice:2021jtw}. Specifically, we calculate the total (fluid plus neutrinos) specific internal energy $\epsilon_{\rm tot}$ and total (fluid and neutrinos) lepton number fraction $Y_{e,{\rm tot}}$ before performing an implicit solve (but after adding the explicit components of the right hand side to the neutrino moments). We then solve for the values of $T$, $Y_e$ such that
\beqn
\epsilon_{\rm tot} &=& \epsilon_{\rm fluid}(T,Y_e) + \sum_i \frac{\eta_{\nu_i}(T,Y_e)}{\rho\kappa_{\nu_i}(T,Y_e)}\\
Y_{e,{\rm tot}} &=& Y_e + \frac{\eta^N_{\nu_e}(T,Y_e)}{\kappa^N_{\nu_e}(T,Y_e)}\frac{m_b}{\rho} - \frac{\eta^N_{\bar\nu_e}(T,Y_e)}{\kappa^N_{\bar\nu_e}(T,Y_e)}\frac{m_b}{\rho}
\eeqn
with $m_b$ the mass of a baryon, $\rho$ the baryon number density, and the sum over $i$ being over all neutrino species. Practically, this comes down to requiring that neutrinos reach equilibrium with the fluid, and that we satisfy conservation of energy and lepton number. Let us now call $(T_0,Y_{e,0})$ the temperature and electron fraction before the implicit step, and $(T_{\rm eq}, Y_{e,{\rm eq}})$ the solutions of the system of equations above. When calculating neutrino-matter interactions, we then use as our guess for $T$ and $Y_e$
\beq
T = f T_0 + (1-f) T_{\rm eq};\,\, Y_e = f Y_{e,0} + (1-f) Y_{e,{\rm eq}}
\eeq
and
\beq
f = \max{\left(0,\min{\left[1,2(\alpha dt-\max_i{\left(\frac{1}{\sqrt{\kappa_{\nu_i,a} (\kappa_{\nu_i,a}+\kappa_{\nu_i,s})}}\right)}\right]}\right)}
\eeq
with $\alpha$ the lapse and $dt$ the time step. Practically, $f\approx 1$ when the equilibration timescale between neutrinos and the fluid is long compared to the time step, and $f\approx 0$ if that timescale is short.

{\it Correction of interaction rates in optically thin regions}: In optically thin regions, it is quite common for the average energy of neutrinos to be sigificantly higher than the average energy that neutrinos would have if they were in equilibrium with the fluid. Given the strong dependence of interaction rates on neutrino energy, this causes uncorrected gray schemes to underestimate interaction rates in these regions. In SpEC, we calculate an estimated neutrino temperature $T_{\nu}$ as~\cite{Foucart:2016rxm}
\beq
T_\nu = \frac{F_2(\eta_\nu)}{F_3(\eta_\nu)}W \frac{E-F_i v^i}{N}
\eeq
with $W$ the Lorentz factor, $N$ the neutrino number density (in the laboratory frame),
\beq
F_k(\eta) = \int_0^\infty dx \frac{x^k}{1+exp(x-\eta)}
\eeq
the Fermi integrals, $\eta=\mu/T$, $T$ the fluid temperature and $\mu$ the chemical potential. We then correct opacities by a factor of $(T_\nu/T)^2$, following the scaling of opacities with neutrino temperature for $T_\nu \approx T$. With the more advanced implicit solver of Radice, this is occationally unstable and we apply a few corrections. First, we do not correct interaction rates when $T_\nu < T$. Second, we take advantage of the fact that neutrinos quickly equilibrate with the fluid when $\kappa_a dt \gtrsim 1$ to ignore corrections in high-opacity regions. Practically, we take
\beq
a = \frac{T_\nu^2}{T^2} (1-f) + f;\,\, f = \frac{\alpha dt \kappa_a}{\alpha dt \kappa_a +1} 
\eeq
with $a$ the multiplicative factor applied to opacities. We note that $a\rightarrow 1$ when $\kappa_a \alpha dt \gg 1$. Finally, we follow Radice {\it et al}~\cite{Radice:2021jtw} and apply the correction to both emissivities and opacities, rather than solely to opacities, in order to still drive the neutrino distribution function to the correct equilibrium energy density. While in this manuscript the changes are only applied to simulation M1-Radice, we believe that they would likely be beneficial to simulations using our old algorithm as well. We also note that a more advanced algorithm to solve for $T_\nu$ without explicitly relying on the approximate scaling $\kappa_a \propto T_\nu^2$ has since been developed in Andresen {\it et al}~\cite{Andresen:2024mtt}, but is not currently implemented in SpEC.

We note that these are the only modifications made to the SpEC code; in particular, we use the same methods to estimate the shape of the neutrino energy spectrum and number flux density $\mathcal{F}^j$ as in~\cite{Foucart:2016rxm}, even though these methods differ from the default choices made in THC.

\subsection{Diffusion test}

\begin{figure}
\includegraphics[width=0.95\columnwidth]{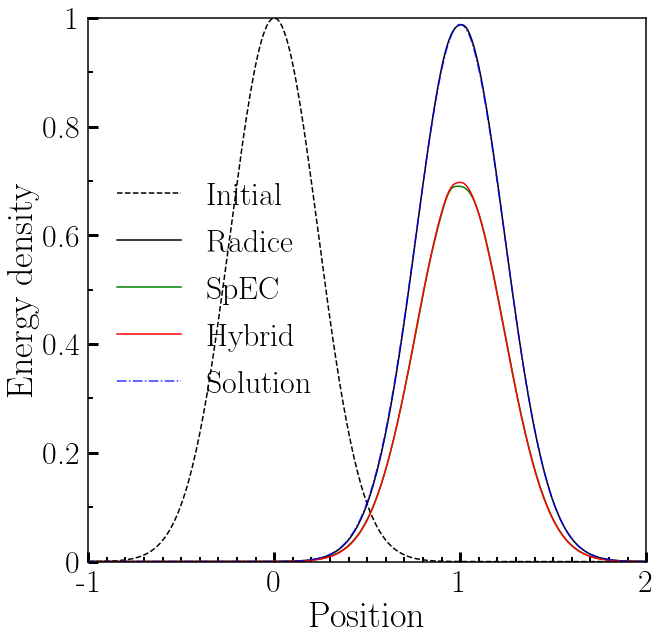}
 \caption{Diffusion of a pulse of radiation through a medium of high scattering opacity $\kappa_s=10^3$. We show the initial pulse, as well as results at $t=2$ for the analytical solution of the diffusion equation, the solution obtained with the `Radice' methods (which lies exactly underneath the analytical results), the solution obtained with the `SpEC' methods, and a hybrid algorithm using the `Radice' fluxes and the `SpEC' implicit solve. The last two provide nearly identical (but erroneous) results.}
\label{fig:diff}
\end{figure}

To illustrate the impact of the changes made to the ability of the SpEC code to handle regions with high scattering opacity, we reproduce the diffusion test of Radice {\it et al}~\cite{Radice:2021jtw}. In this test, a pulse of radiation $E=e^{-9x^2}$ is evolved in a fluid moving at velocity $v^x=0.5$ and with scattering opacity $\kappa_s = 10^3$. We evolve the pulse over a time $\Delta t=2$, in slab geometry, with grid spacing $\Delta x = 0.0098$ and Courant factor of $0.25$. Each grid cell thus has an optical depth $\tau = 9.8$. The results of the test are shown in Fig.~\ref{fig:diff}. We see that the M1-Radice code correctly reproduces the expected solution. The two simulations using the linearization of the sources around $E=F^i=0$ and only including $O(v/c)$ terms in the Jacobian do not, regardless of the method used to calculate fluxes in high-opacity regions. As opposed to Radice {\it et al}~\cite{Radice:2021jtw}, we find that the energy of the pulse decreases when using the approximate implicit solve, though the results are otherwise equally problematic. In Schianchi {\it et al}~\cite{Schianchi:2023uky}, the same test was performed using all velocity terms in the linearization of the implicit problems, in which case they find much better agreement with the analytical solution. It is thus likely that the approximations made in the linearization of the implicit problem in the original SpEC code are responsible for its issues with this diffusion test.

\section{Simulation results}
\label{sec:results}

\subsection{Merger dynamics}

Before contact, the various simulations presented here behave nearly identically, and follow largely the same evolution as the simulations without neutrino transport described in~\cite{Foucart:2024kci}. Neutrino emission mostly comes from the surface of the star, which is heated by numerical dissipation. One millisecond before merger, we observe a luminosity of $\sim 2\times 10^{51}\,{\rm erg/s}$ for each species of neutrinos, with only $\sim 20\%$ variations between simulations and neutrino species. As a result, the neutron stars are very slightly cooler and more neutron rich in simulations with neutrino transport than in simulations without transport: the mass weighted average temperature is $\sim 1\%$ lower in most of our simulations than in~\cite{Foucart:2024kci}, and the mass weighted electron fraction $\sim 0.2\%$ lower. The exceptions are simulation M1-Radice, for which the temperature is nearly identical to that of the simulation without transport, and the MC simulation, for which both temperature and electron fraction are nearly identical to the fluid-only evolution (despite, in both cases, similar neutrino luminosities). As these effects are small, resolution dependent, and do not impact the dynamics of the binary, we do not investigate these differences further.

\begin{figure}
\includegraphics[width=0.95\columnwidth]{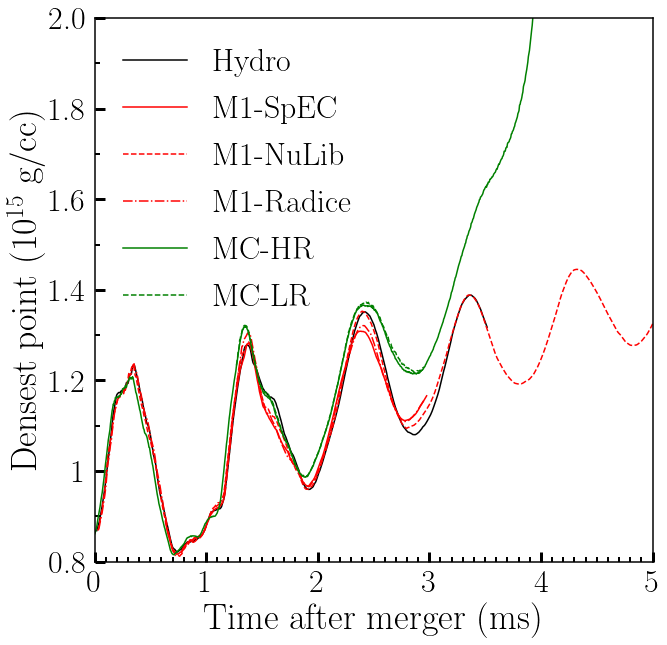}
 \caption{Maximum value of the baryon density on the grid as a function of time, for all 6 simulations.}
\label{fig:rhomax}
\end{figure}

There is no clear, local definition of the merger time for binary neutron star mergers. In this manuscript, as in~\cite{Foucart:2024kci}, we define as {\it merger} the time at which the maximum baryon density on the grid increases to $3\%$ above its initial value. This happens soon after contact, as the two neutron star cores collide. We follow all simulations until at least $3.0\,{\rm ms}$ post-merger, and additionally continue MC-HR and M1-NuLib until collapse of the remnant to a black hole.

\begin{table}
\begin{tabular}{c|c|c|c|c|c}
{\bf Sim} & $M_{\rm disk}$ & $\langle T_{\rm disk} \rangle$ & $\langle Y_{e,\rm disk} \rangle$ & $\langle T_{\rm rem} \rangle$ & $\langle Y_{e,\rm rem} \rangle$\\
\hline
 & $M_\odot$ & MeV & & MeV & \\\hline
MC-HR & 0.054 & 10.8 & 0.126 & 28.2 & 0.066\\
MC-LR & 0.056 & 11.0 & 0.131 & 29.0 & 0.068\\
M1-Radice & 0.055 & 10.9 & 0.133 & 23.9 & 0.062\\
M1-SpEC & 0.050 & 10.1 & 0.118 & 25.6 & 0.053\\
M1-NuLib & 0.045 & 10.2 & 0.145 & 24.8 & 0.064\\ 
Hydro & 0.048 & 11.9 & 0.046 & 26.3 & 0.057
\end{tabular}
\caption{Remnant properties $2.5\,{\rm ms}$ post-merger. We show the mass below $10^{13}$ g/cc ({\it disk mass}), as well as the average temperature and electron fraction of that low-density matter, and the average temperature and electron fraction of the entire remnant.}
\label{tab:rem}
\end{table}

The most visible difference between the various simulations is the evolution of the maximum baryon density and eventual collapse of the remnant to a black hole, shown on Fig.~\ref{fig:rhomax}. The Monte-Carlo simulations exhibit higher maximum baryon density, with the differences remaining small up to $\sim 2.5\,{\rm ms}$ post-merger. As the remnant is close to collapse, this is however enough to lead to a divergence of the evolution once the MC simulation reaches high enough densities to trigger collapse. As a result, we observe noticeable difference in the collapse time. The MC-HR simulation collapses after $4\,{\rm ms}$. The two-moment and pure hydrodynamics simulations are longer lived. To estimate the difference in collapse time, we followed the M1-NuLib simulation up to black hole formation, and find that collapse happens $8.5\,{\rm ms}$ post-merger for that simulation. That simulation was chosen as the most likely to collapse quickly among the two-moment simulations (it has slightly higher maximum density than the others, as shown on Fig.~\ref{fig:rhomax}); this is thus likely to be a lower bound on the difference in collapse time between Monte-Carlo and the two-moment methods. The maximum density of the MC-LR simulation is close enough to that of MC-HR that the number of packets in the MC simulation does not appear to be a driving factor for these differences. We note however that the collapse time in such a system is very sensitive to small variations in the evolution (and would also be impacted e.g. by magnetic fields, which are not included here, and by numerical resolution). These stark differences in collapse time are due to very small changes in the evolution of the system in the few milliseconds following merger, and likely particularly important for the binary evolved here, as it sits right at the threshold of collapse at merger. This result is quite similar to the different collapse times observed between leakage and two-moment schemes by Curtis {\it et al}~\cite{Curtis:2023zfo}. We note that the system evolved in this work was alsoe simulated by Sekiguchi {\it et al}~\cite{Sekiguchi:2016} using their mixed two-moment/leakage transport scheme. They find a collapse time of $\sim 10\,{\rm ms}$, fairly close to our two-moment results\footnote{The exact collapse time is not provided for this configuration in Sekiguchi {\it et al}~\cite{Sekiguchi:2016}, it is $\sim 10,{\rm ms}$ for a range of simulations with the same total mass and SFHo equation of state, but different mass ratios.}

Whether the MC or M1 scheme is most accurate here is a difficult question to answer. On the one hand, the MC scheme properly takes into account the full energy spectrum of neutrinos in dense region, which could lead it to better capture transport effects in the remnant (specifically, to better capture the behavior of low-energy neutrinos). On the other hand, for monochromatic radiation, the M1 schemes (and particularly M1-Radice) are more accurate in their treatment of high-opacity regions, where the MC scheme has to rely on a low-accuracy implementation of implicit Monte-Carlo methods and noisy estimates of the radiation pressure. What is clear is that the choice of transport scheme -- leakage, two-moment, or Monte-Carlo -- can have a significant impact on the collapse time of such metastable remnants.

\begin{figure*}
\includegraphics[width=0.31\textwidth]{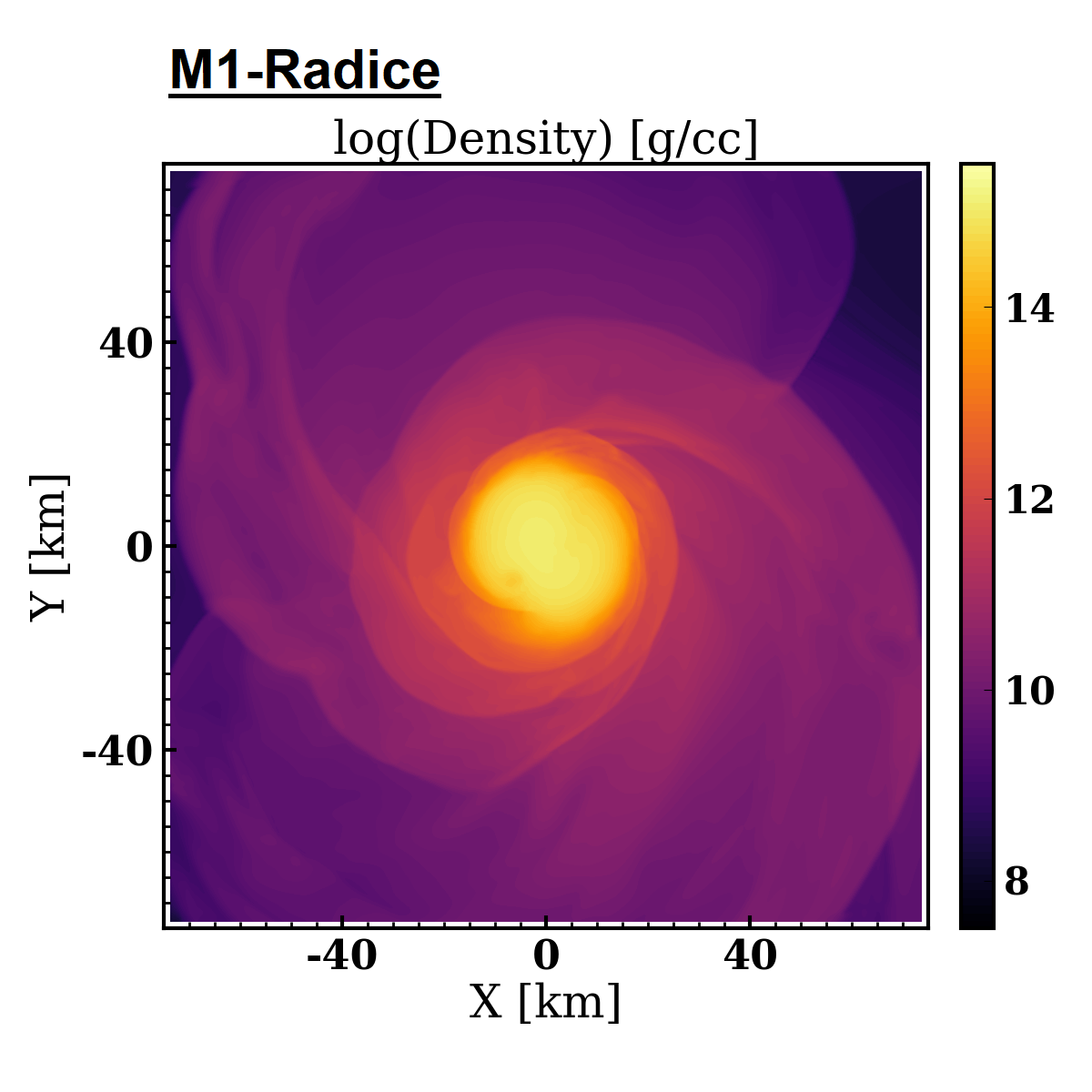}
\includegraphics[width=0.31\textwidth]{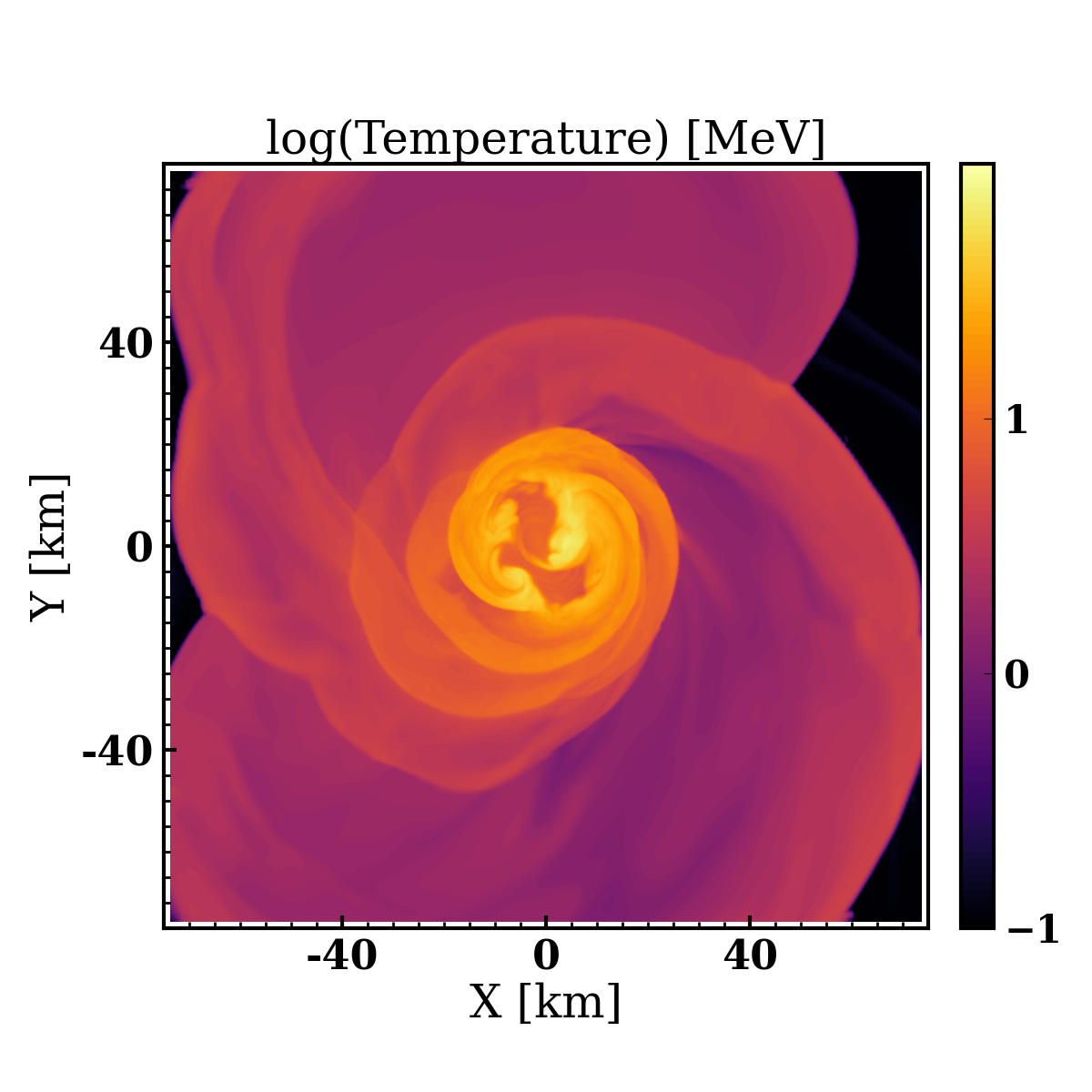}
\includegraphics[width=0.31\textwidth]{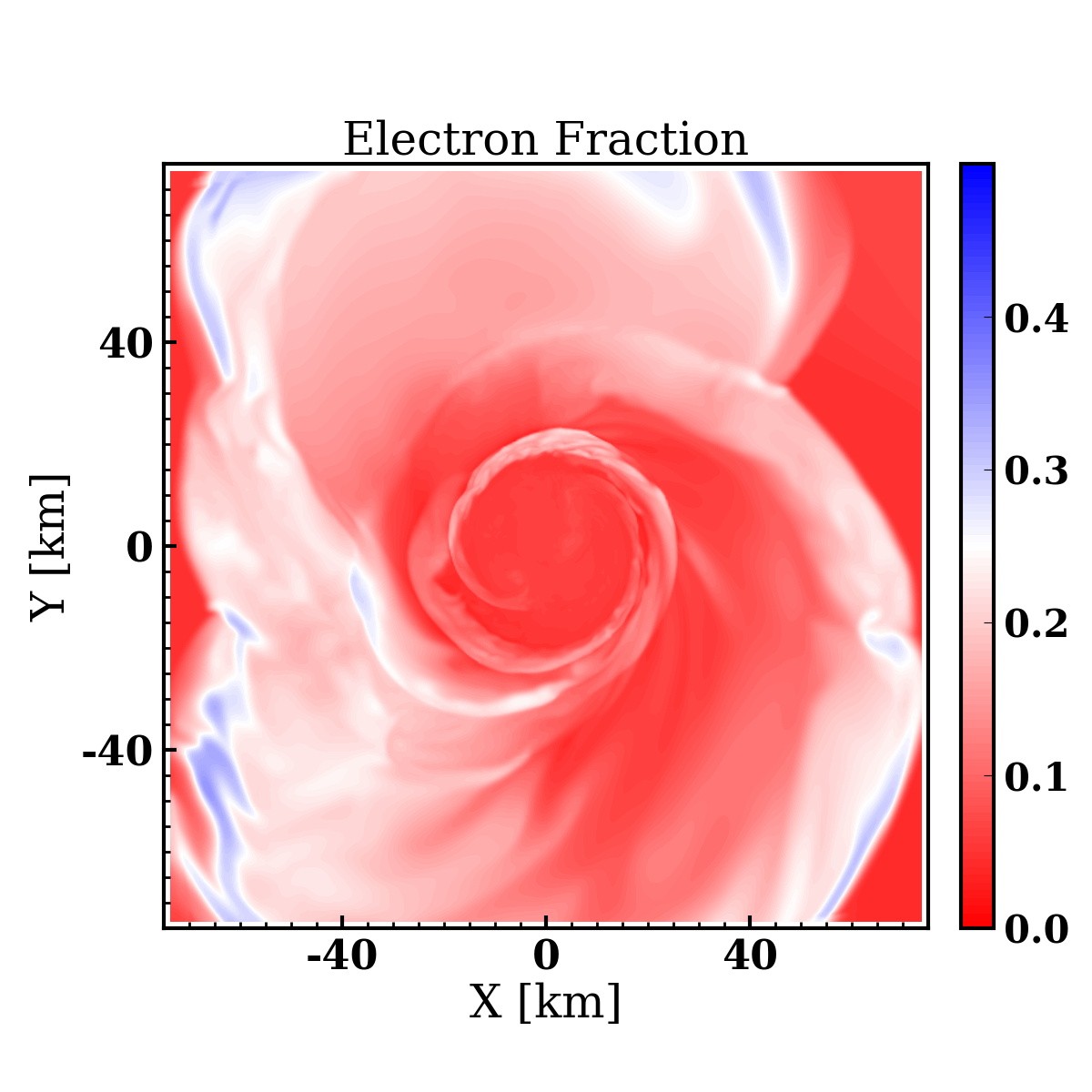}\\
\includegraphics[width=0.31\textwidth]{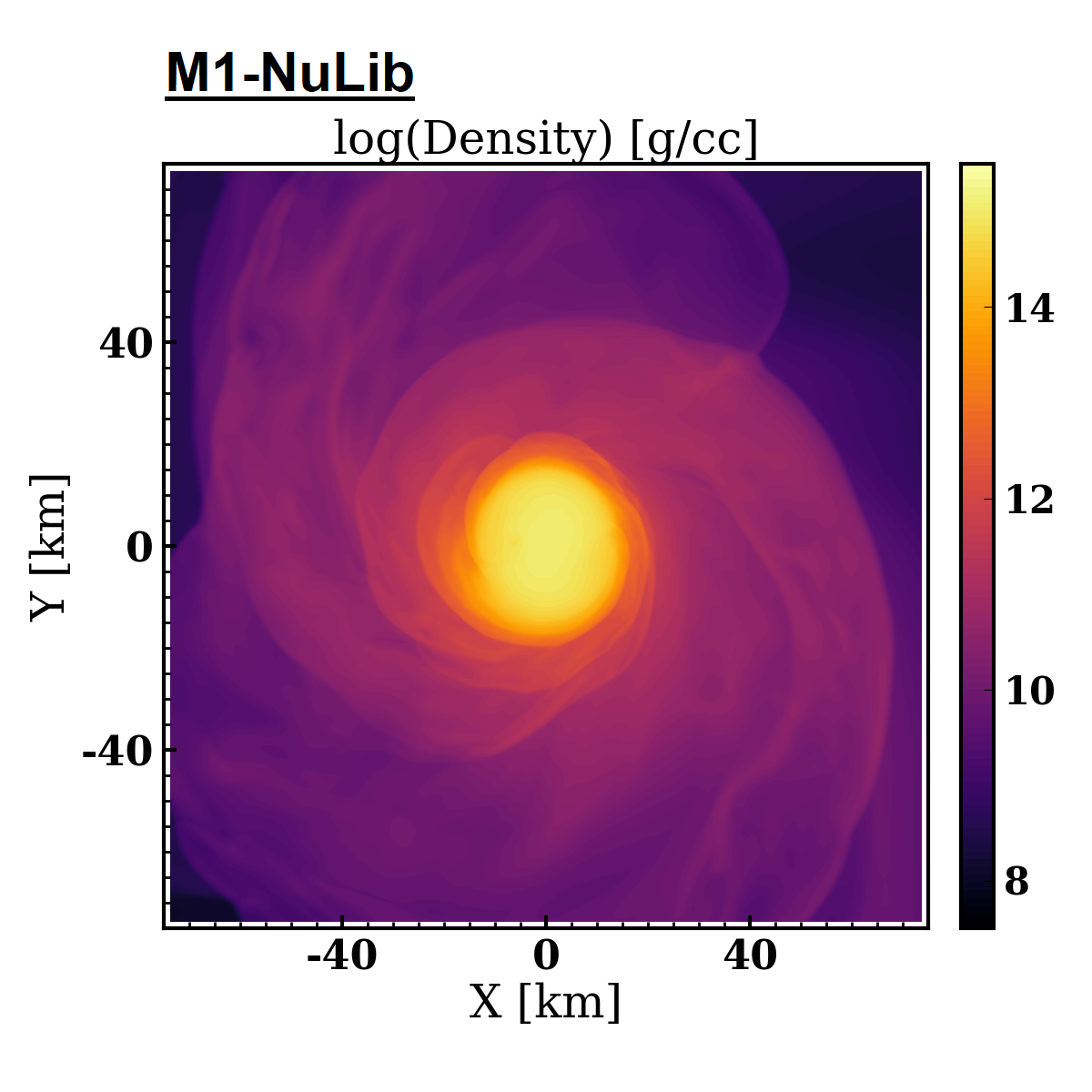}
\includegraphics[width=0.31\textwidth]{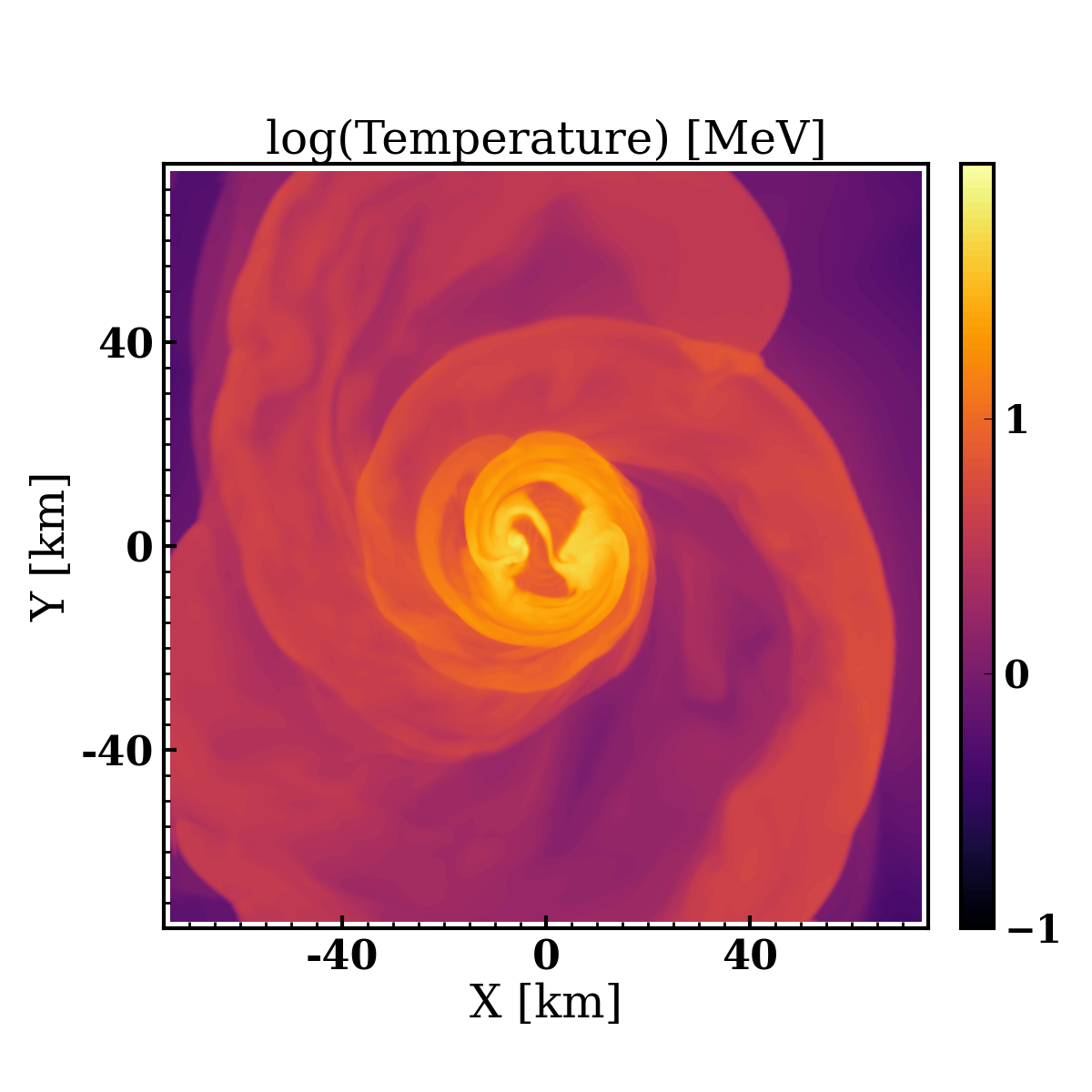}
\includegraphics[width=0.31\textwidth]{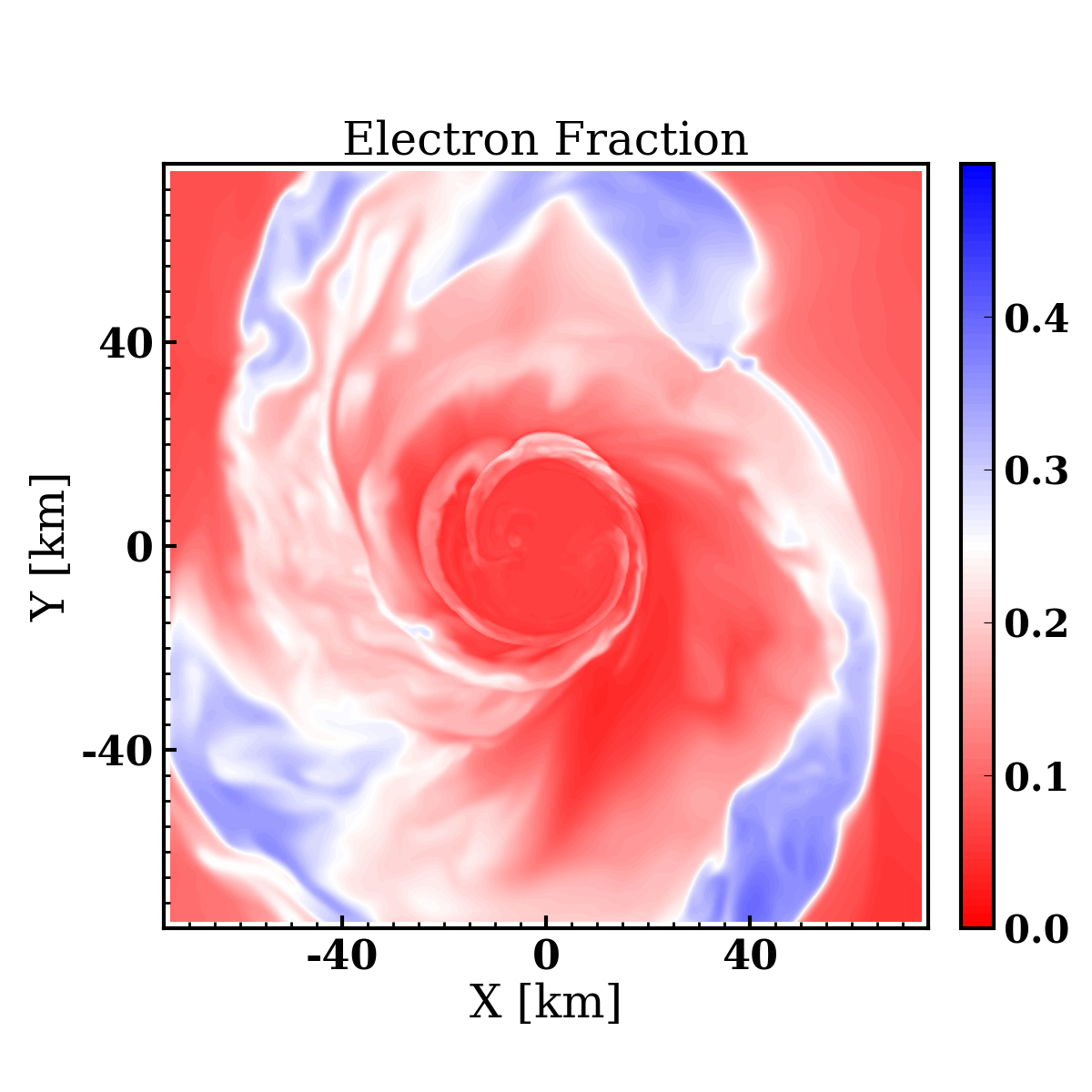}\\
\includegraphics[width=0.31\textwidth]{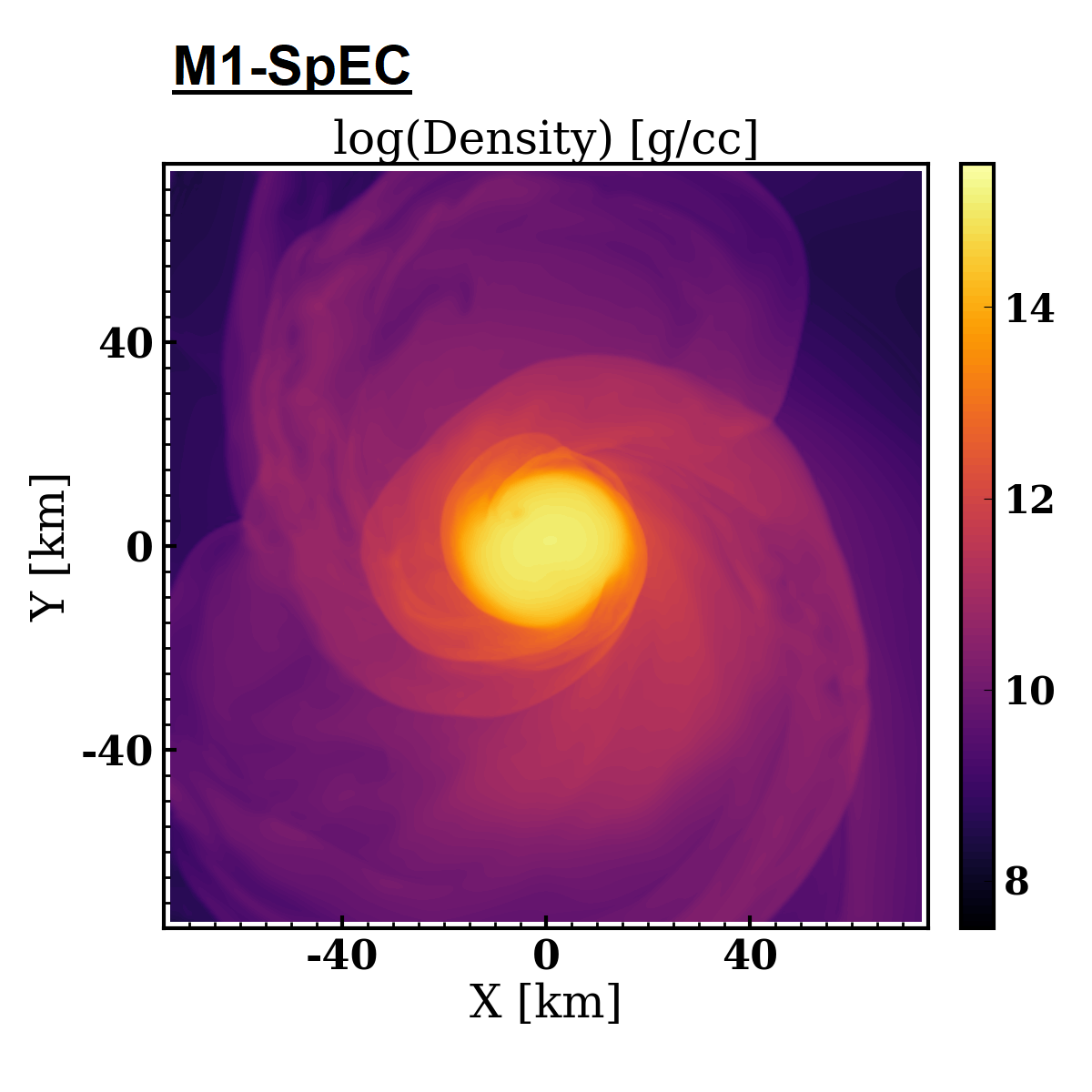}
\includegraphics[width=0.31\textwidth]{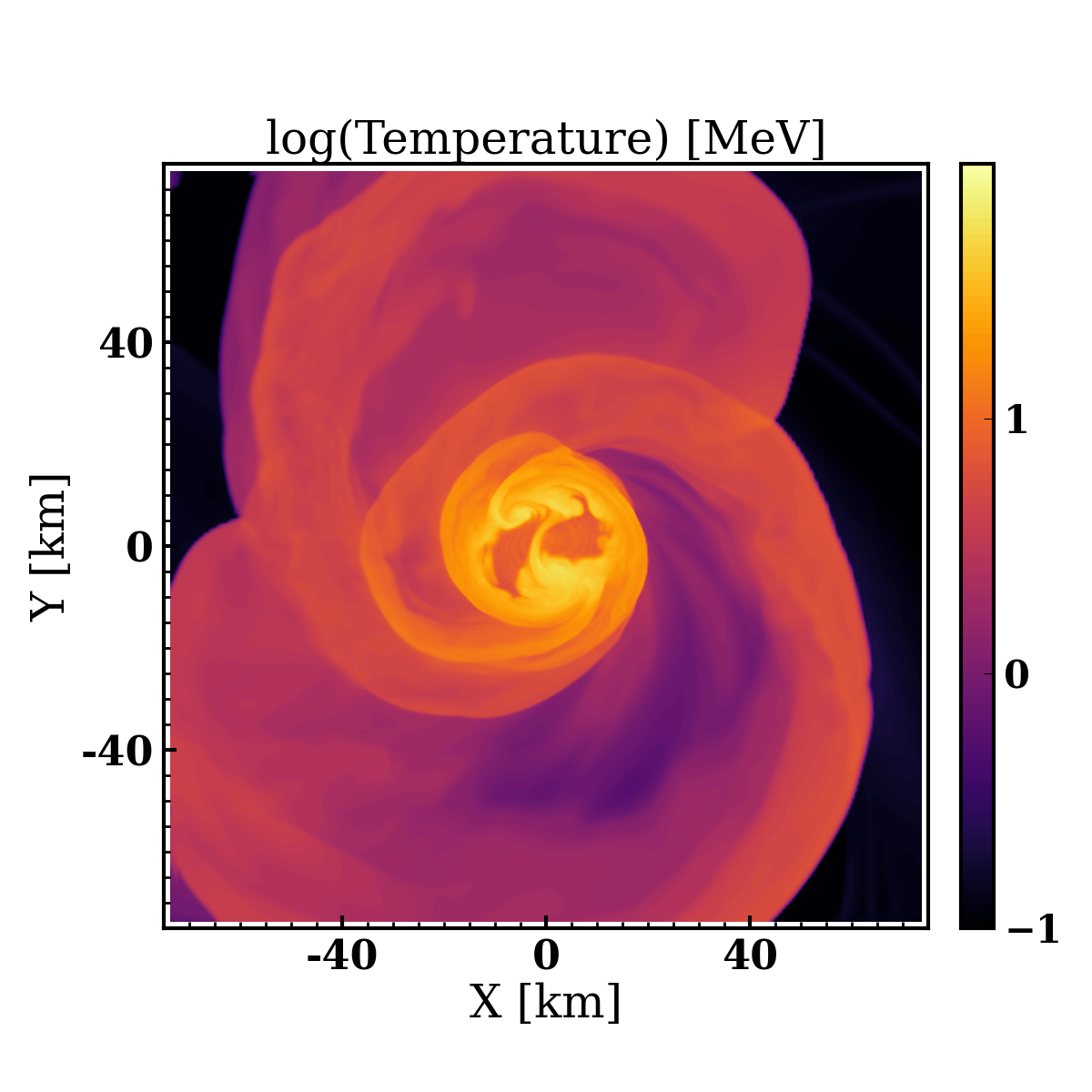}
\includegraphics[width=0.31\textwidth]{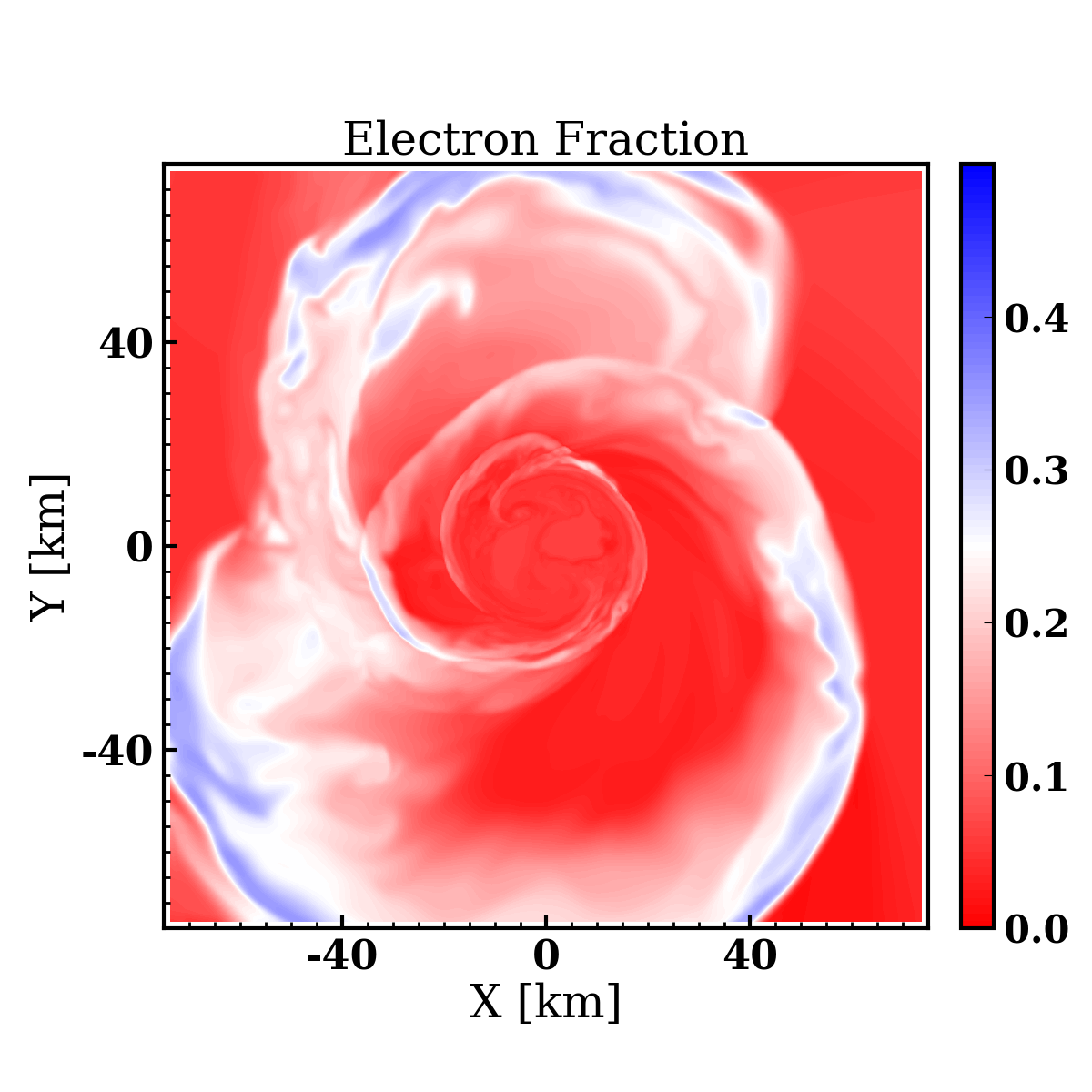}\\
\includegraphics[width=0.31\textwidth]{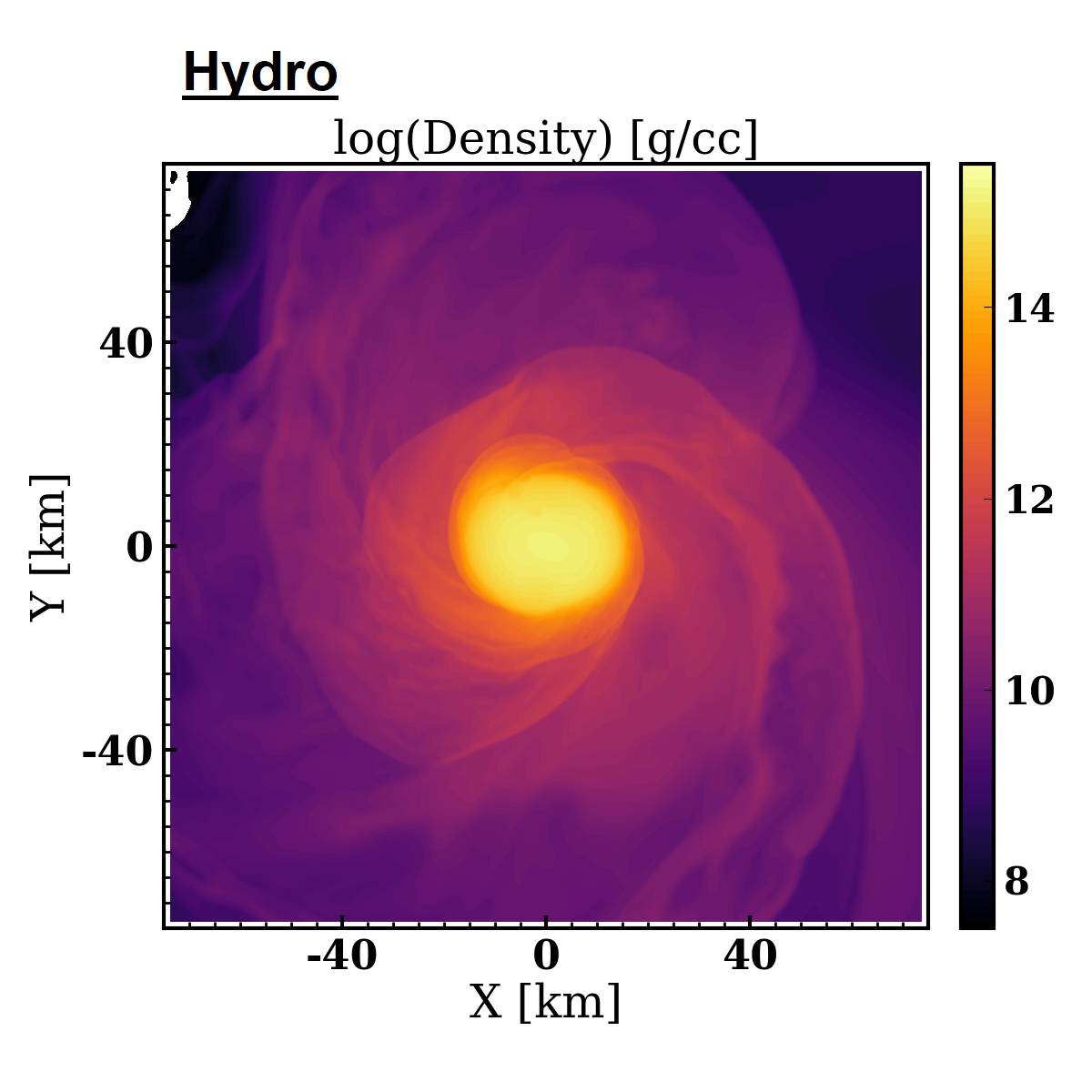}
\includegraphics[width=0.31\textwidth]{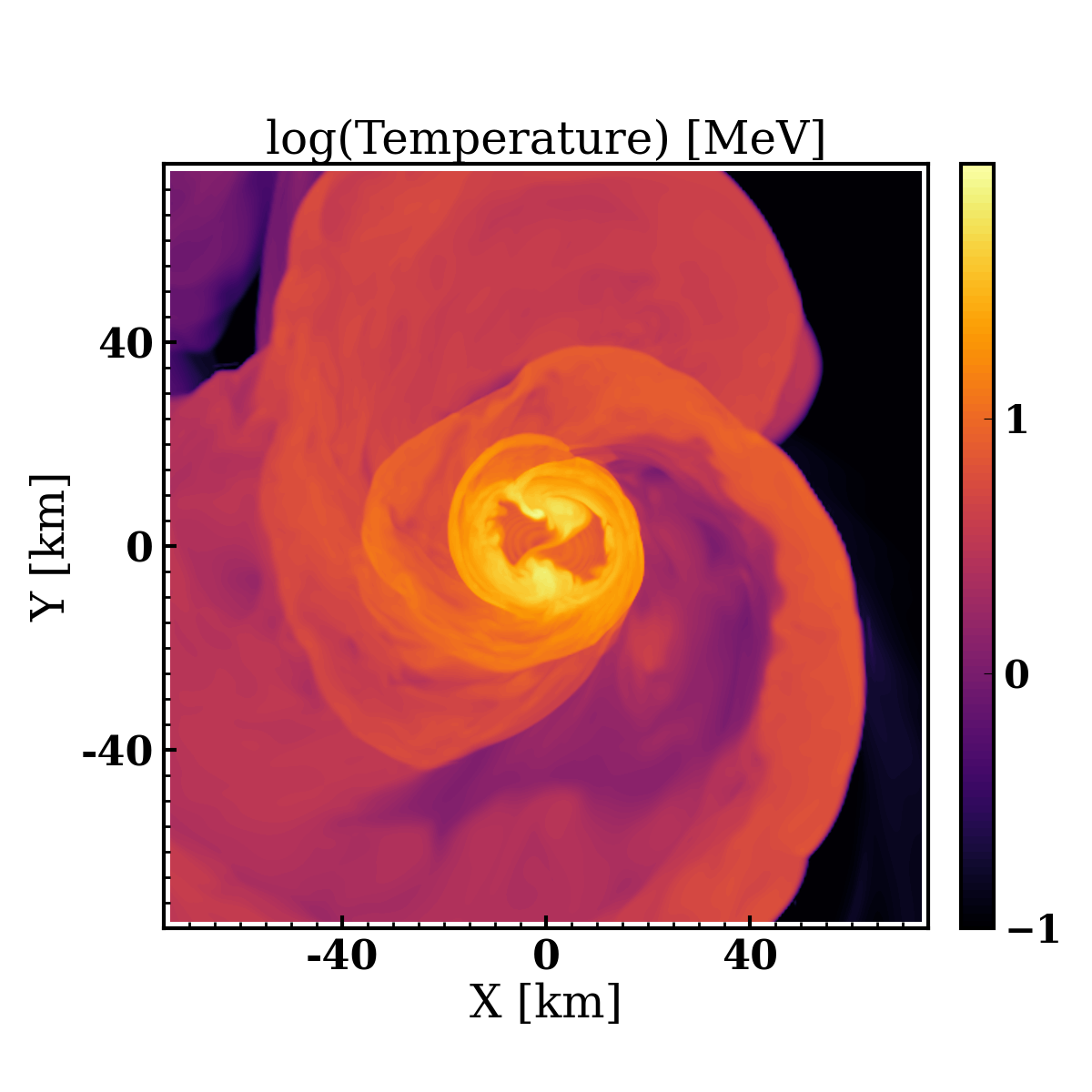}
\includegraphics[width=0.31\textwidth]{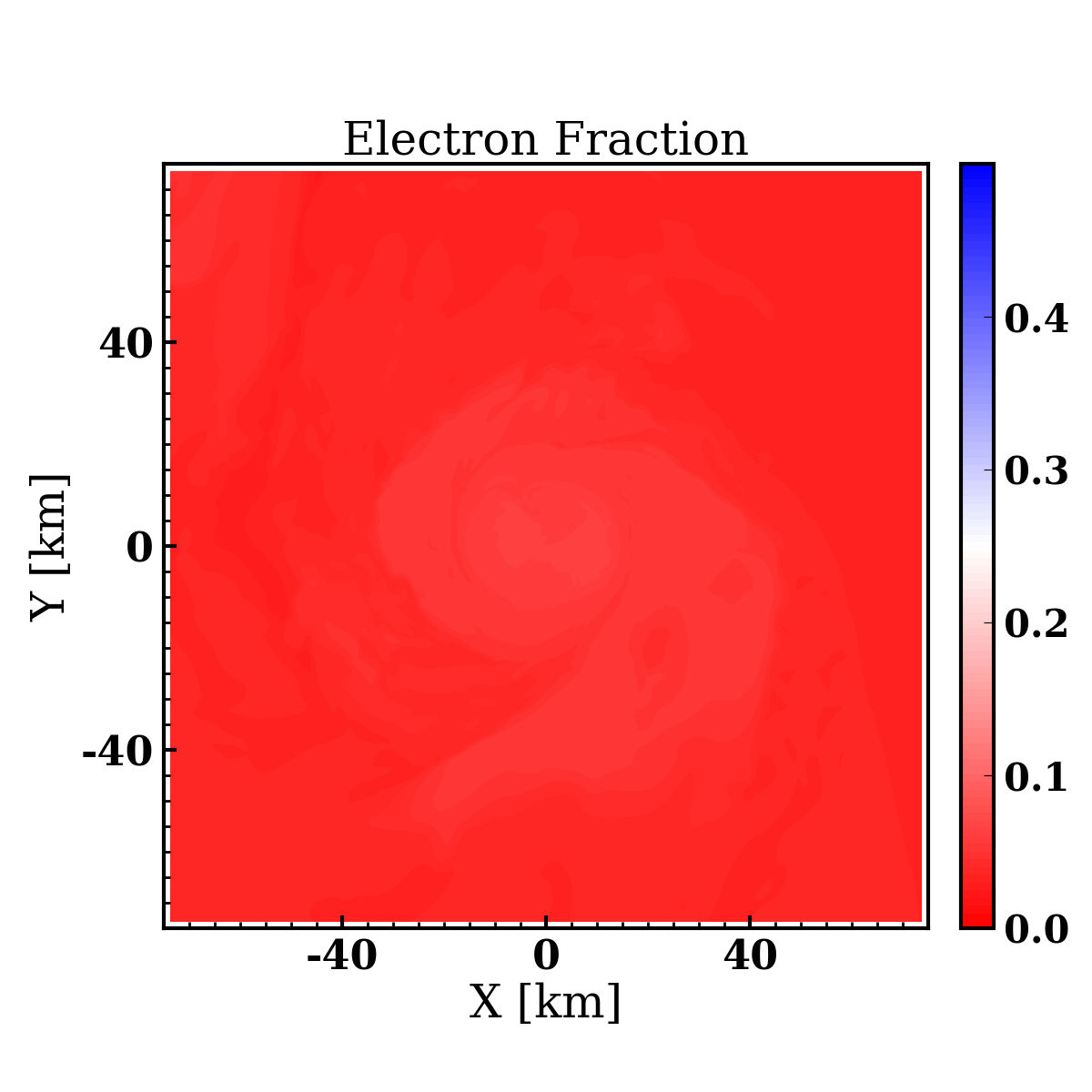}
 \caption{Horizontal slice through the merger remant $2.5\,{\rm ms}$ post-merger. We show the baryon density (left), temperature (middle) and electron fraction (right) for simulations M1-Radice (top), M1-NuLib (2nd row), M1-SpEC (3rd row) and Hydro (bottom).}
\label{fig:vis_hor}
\end{figure*}

\begin{figure*}
\includegraphics[width=0.31\textwidth]{Rho_Radice_Hor_2p5}
\includegraphics[width=0.31\textwidth]{Temp_Radice_Hor_2p5}
\includegraphics[width=0.31\textwidth]{Ye_Radice_Hor_2p5}\\
\includegraphics[width=0.31\textwidth]{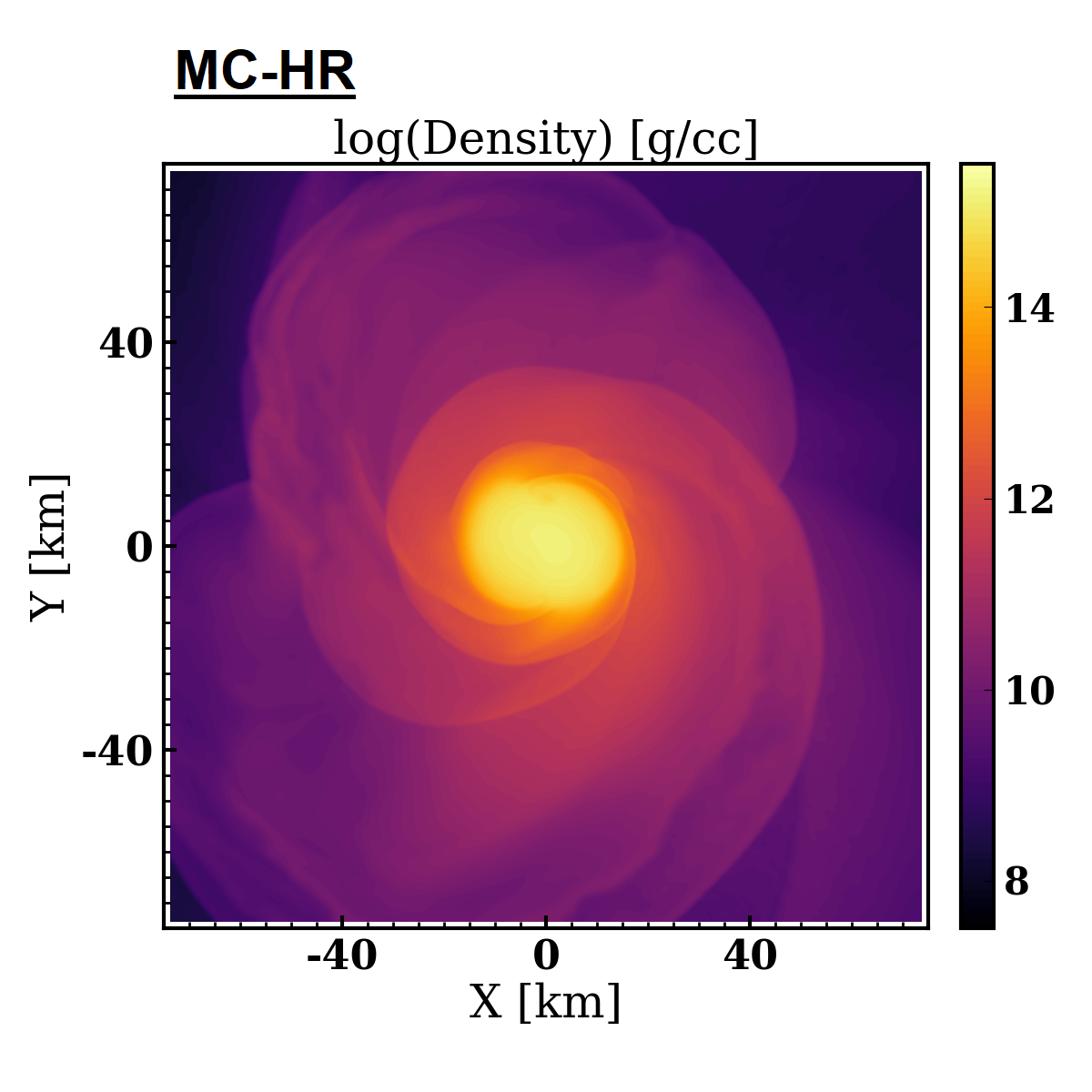}
\includegraphics[width=0.31\textwidth]{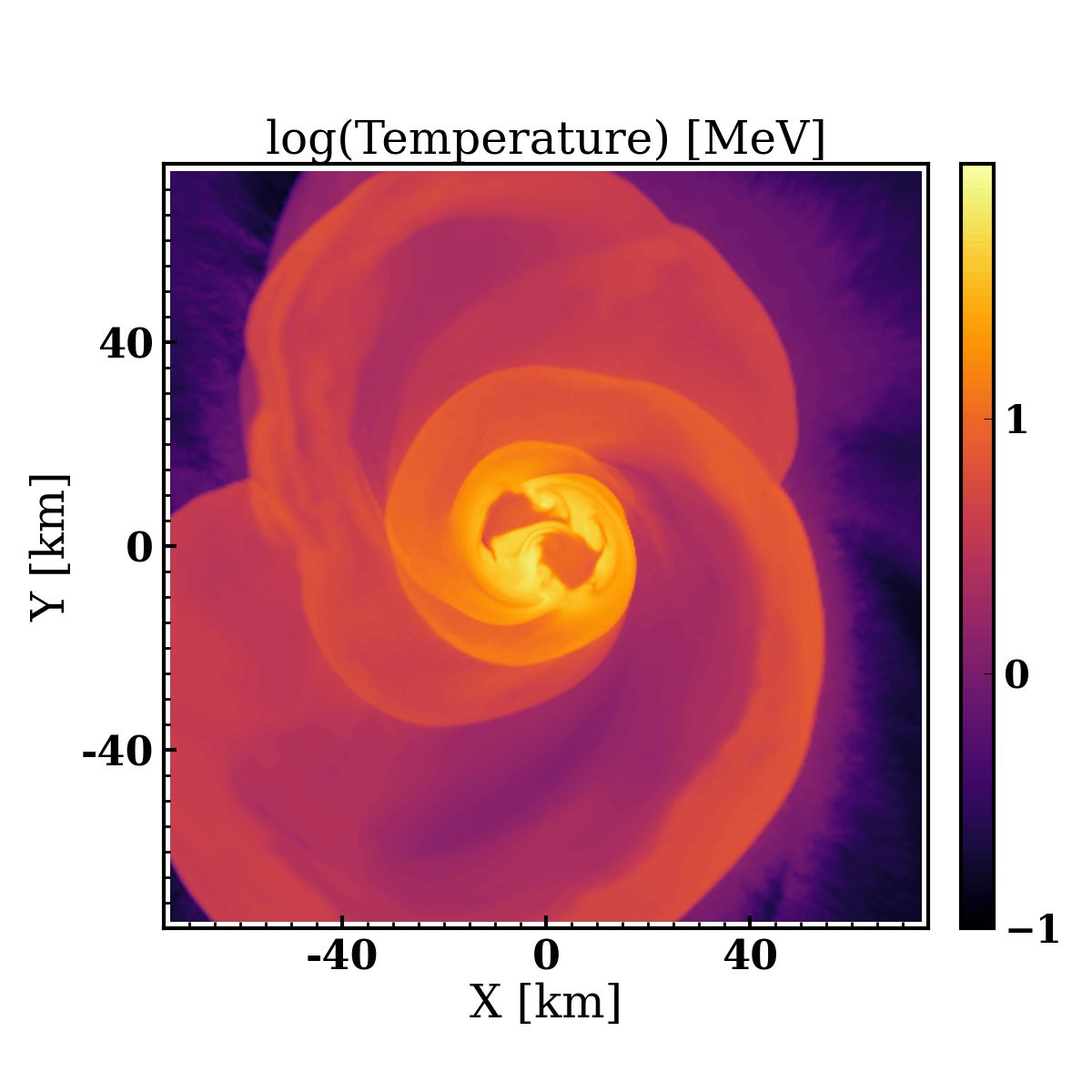}
\includegraphics[width=0.31\textwidth]{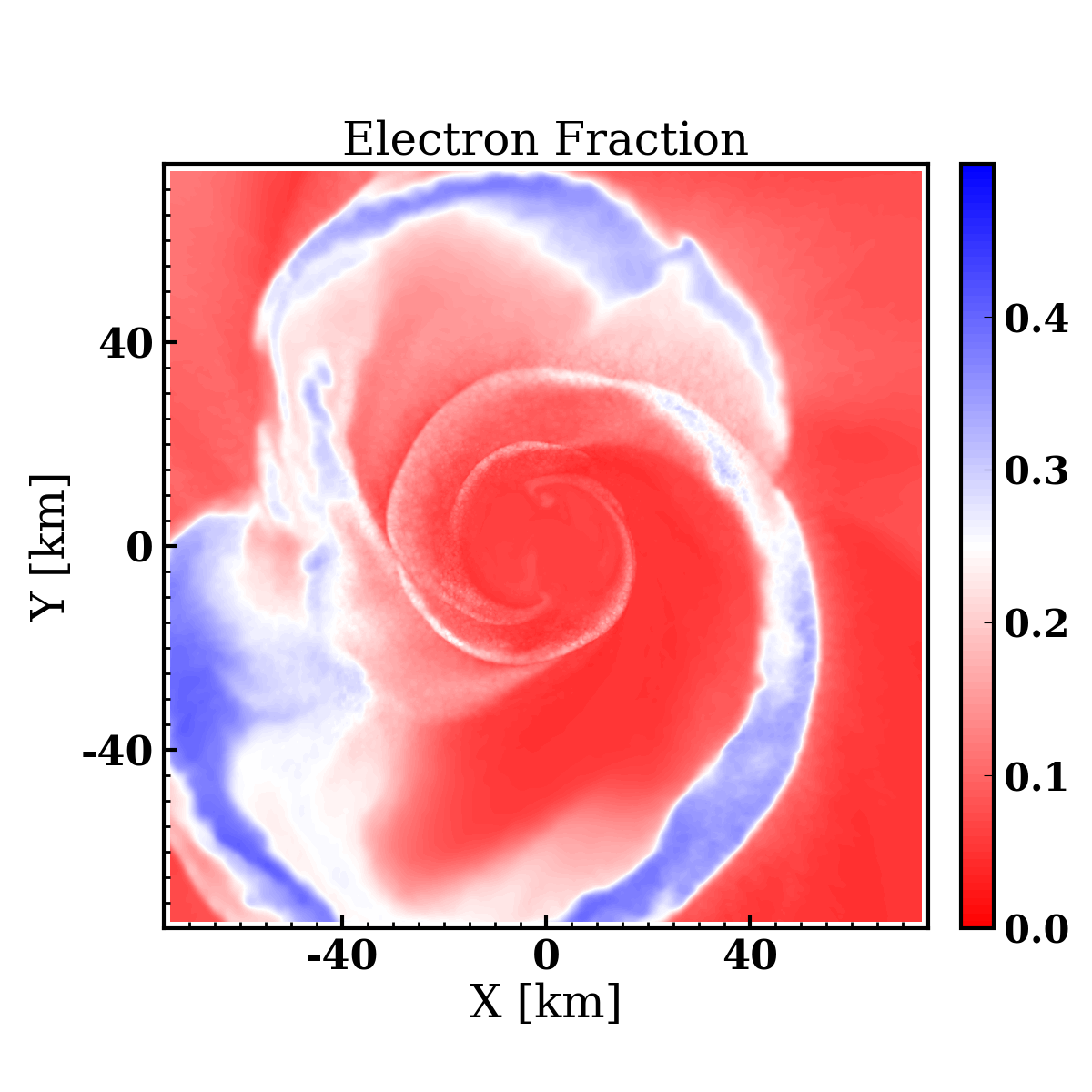}\\
\includegraphics[width=0.31\textwidth]{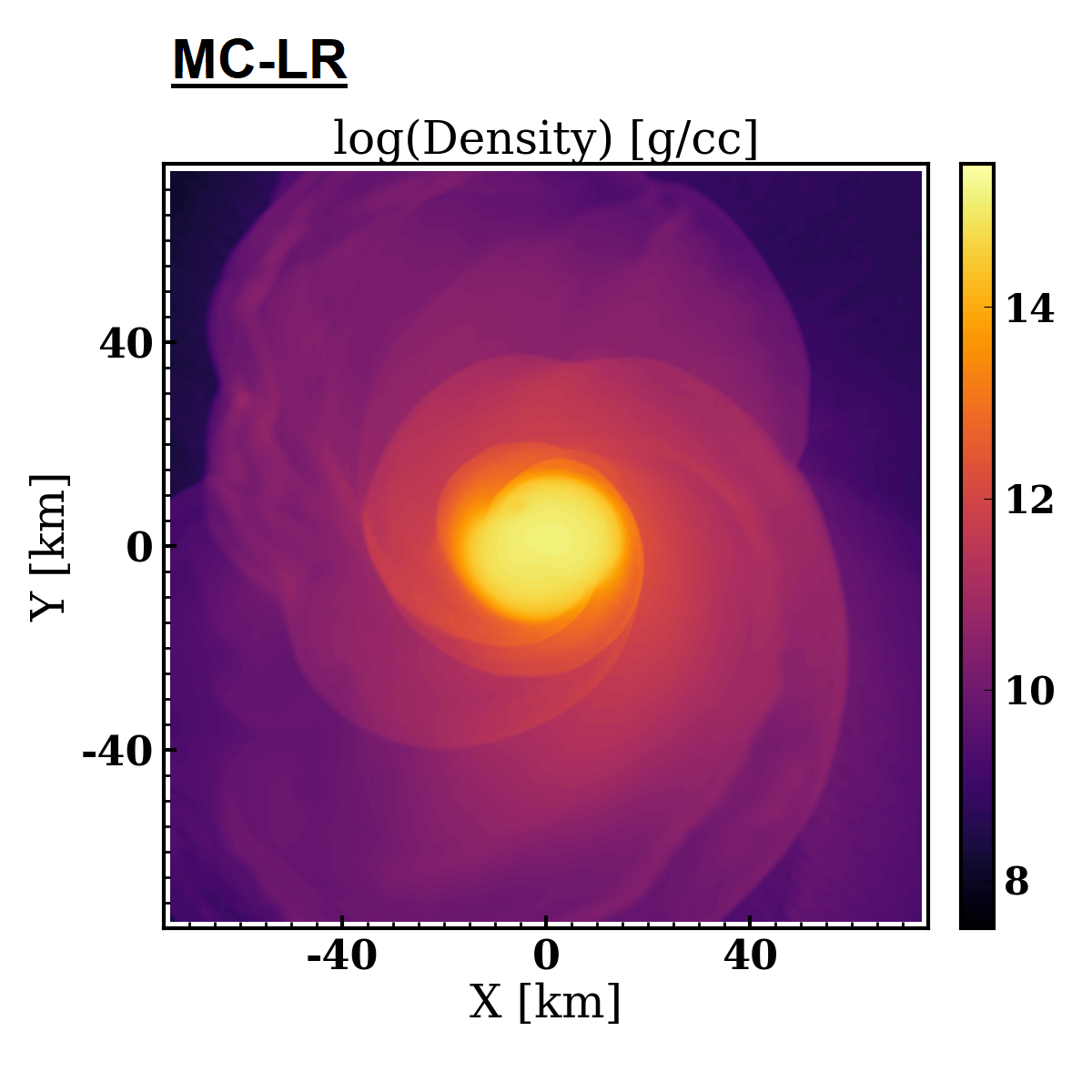}
\includegraphics[width=0.31\textwidth]{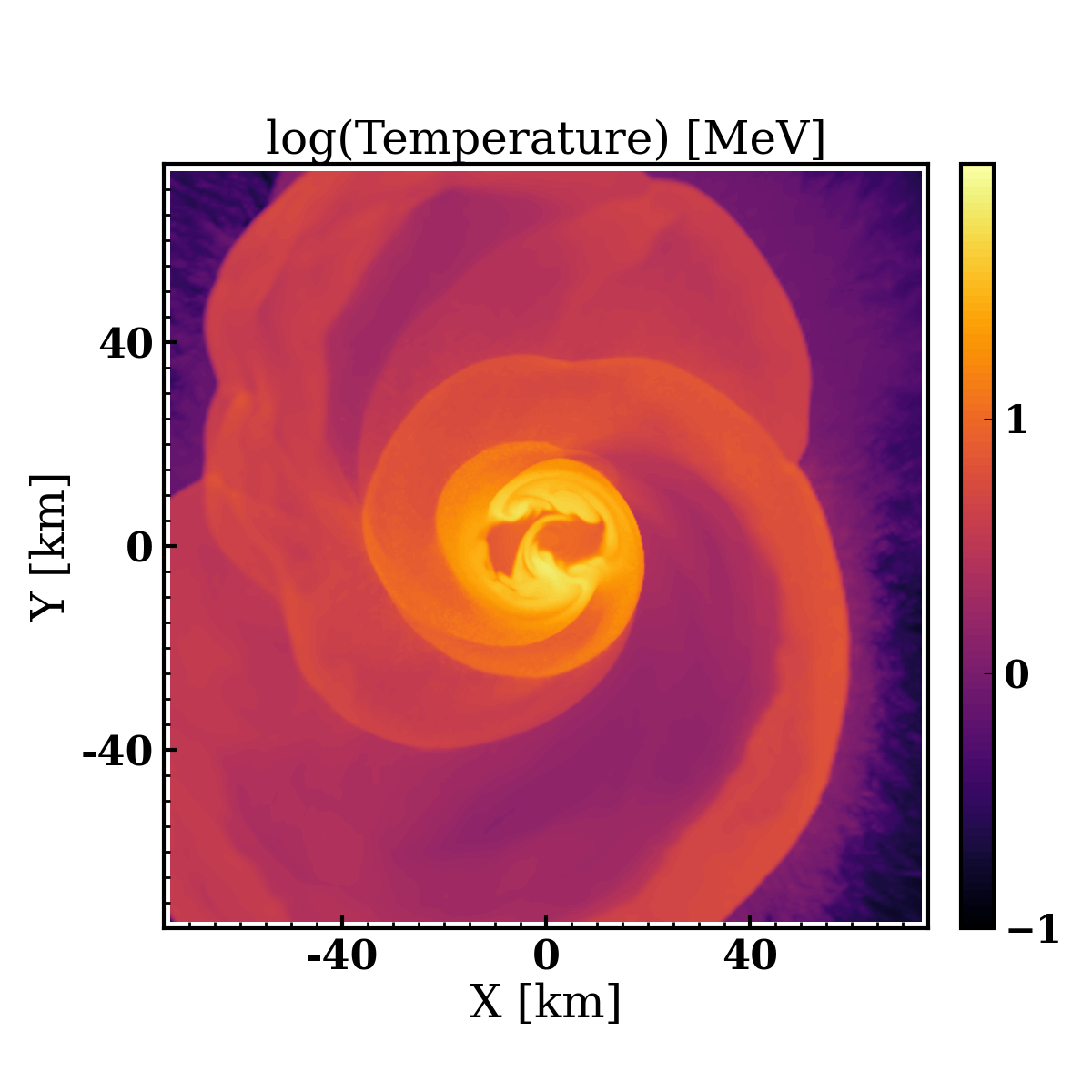}
\includegraphics[width=0.31\textwidth]{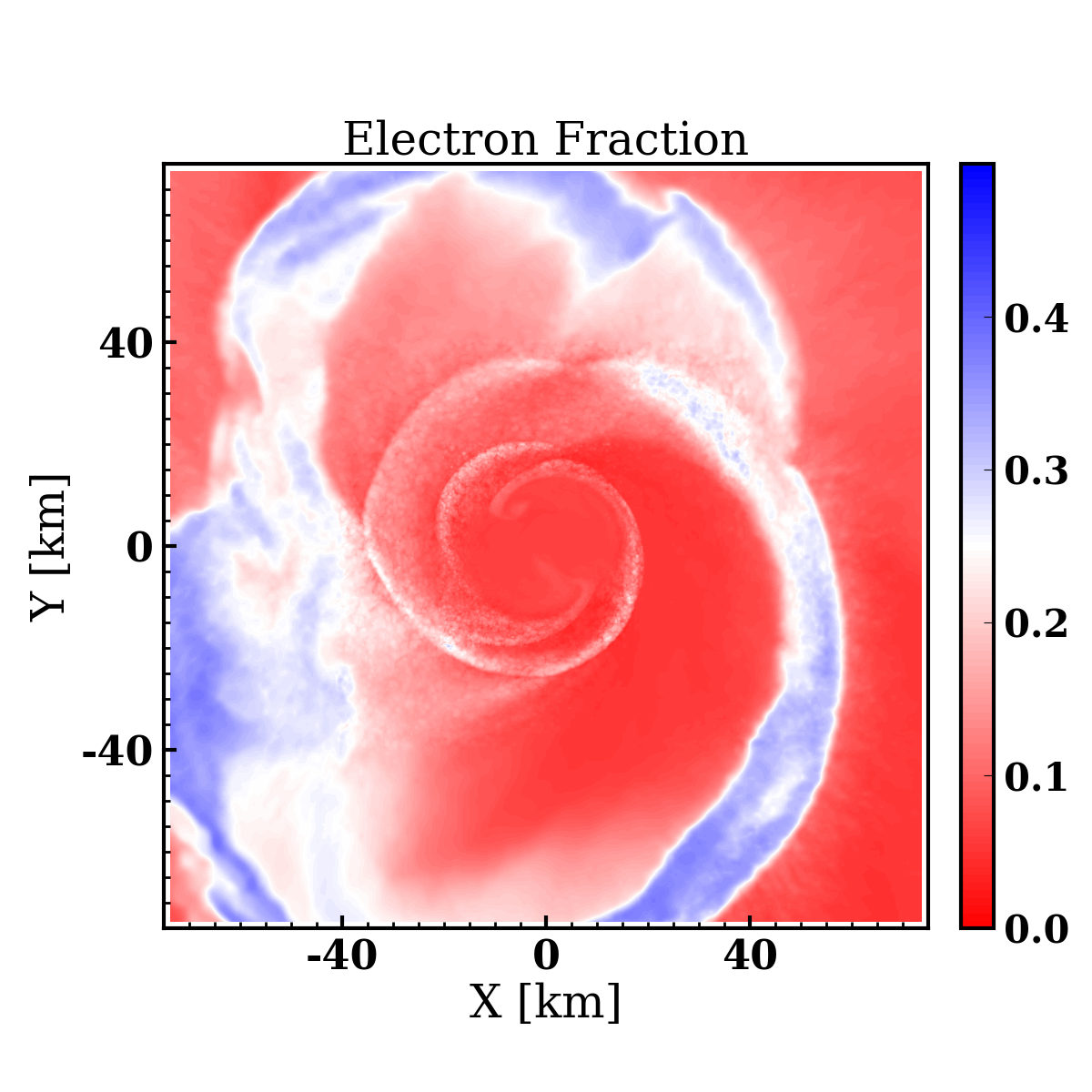}
 \caption{Same as Fig.~\ref{fig:vis_hor} but for simulations M1-Radice (top), MC-HR (middle) and MC-LR (bottom).}
\label{fig:vis_hor_mc}
\end{figure*}

\begin{figure*}
\includegraphics[width=0.31\textwidth]{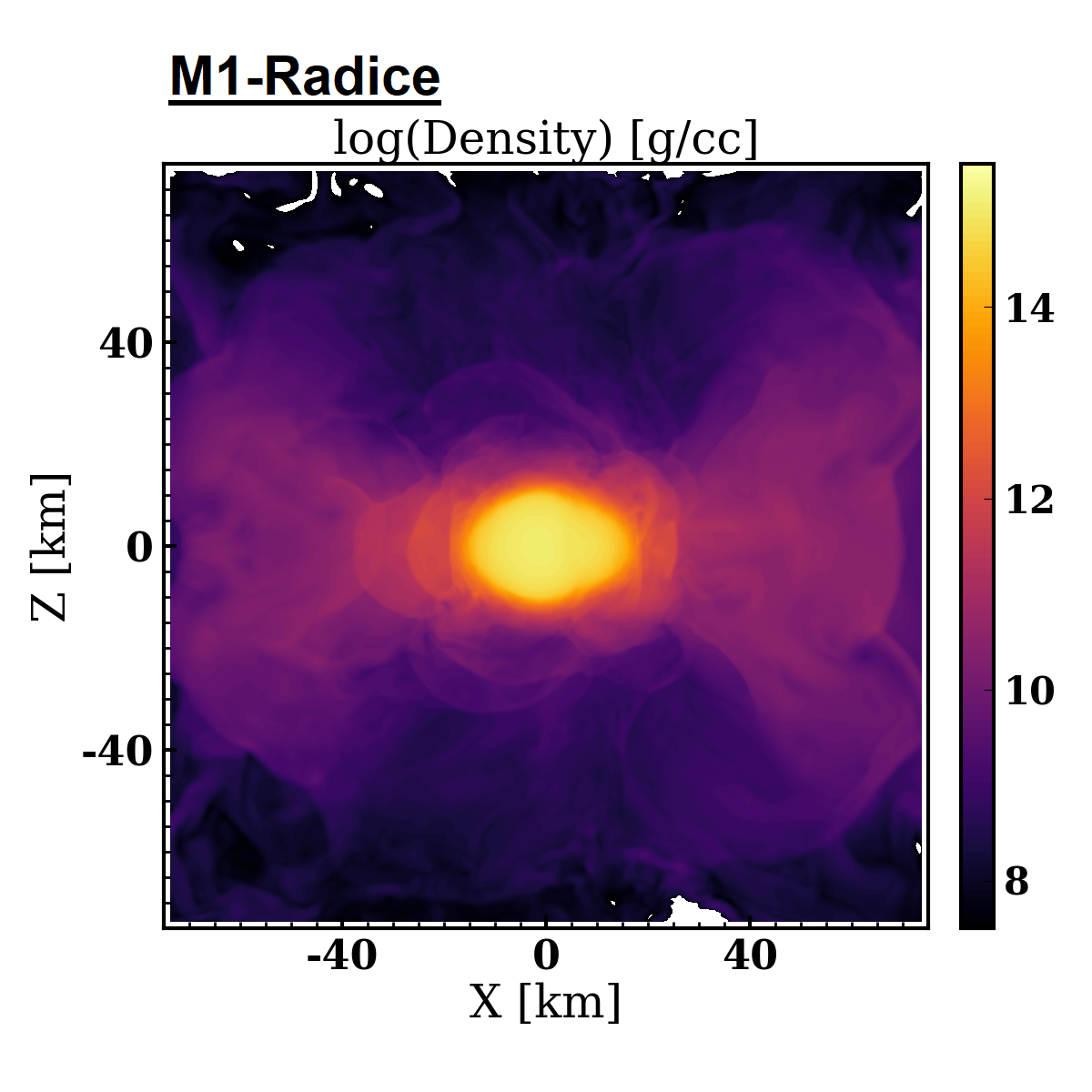}
\includegraphics[width=0.31\textwidth]{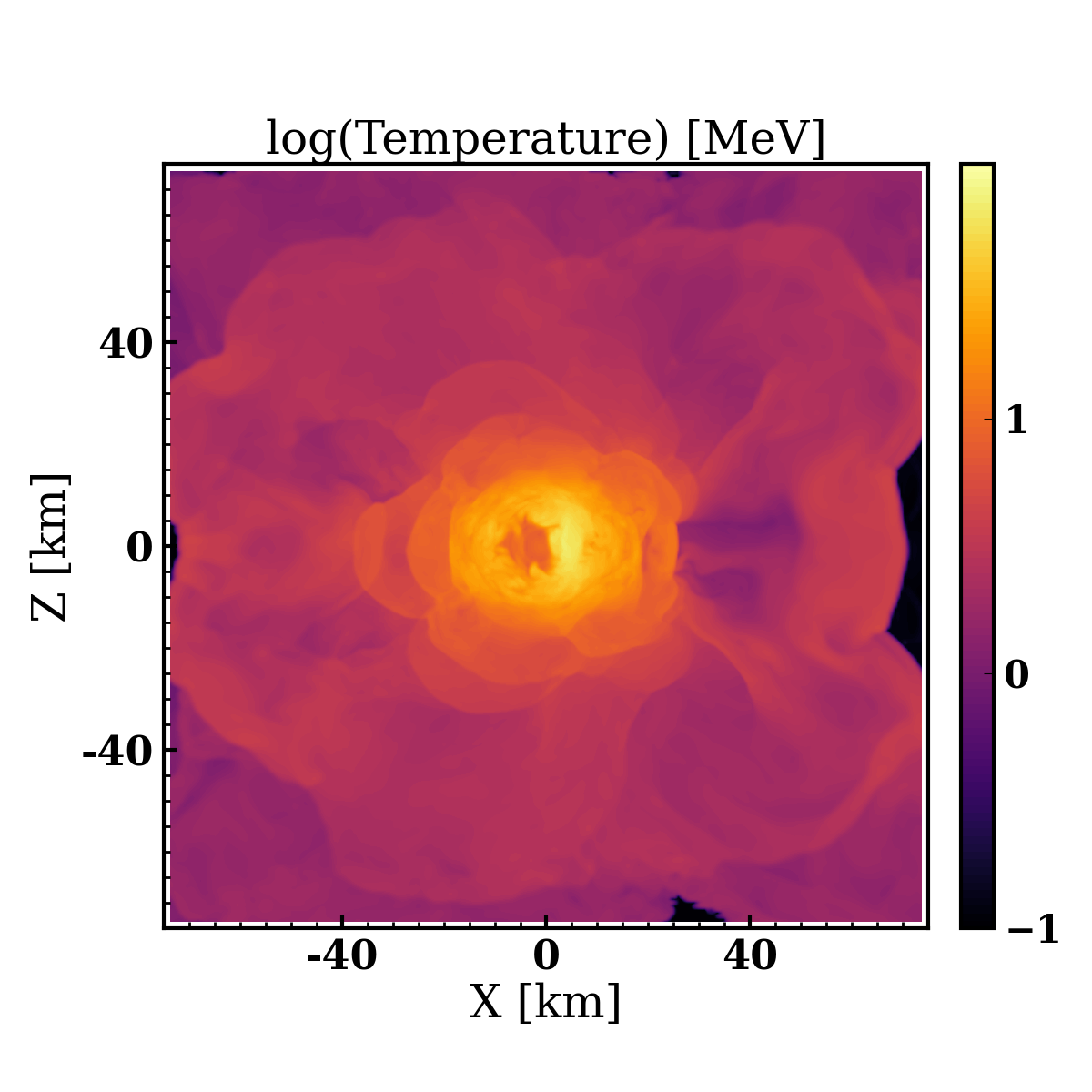}
\includegraphics[width=0.31\textwidth]{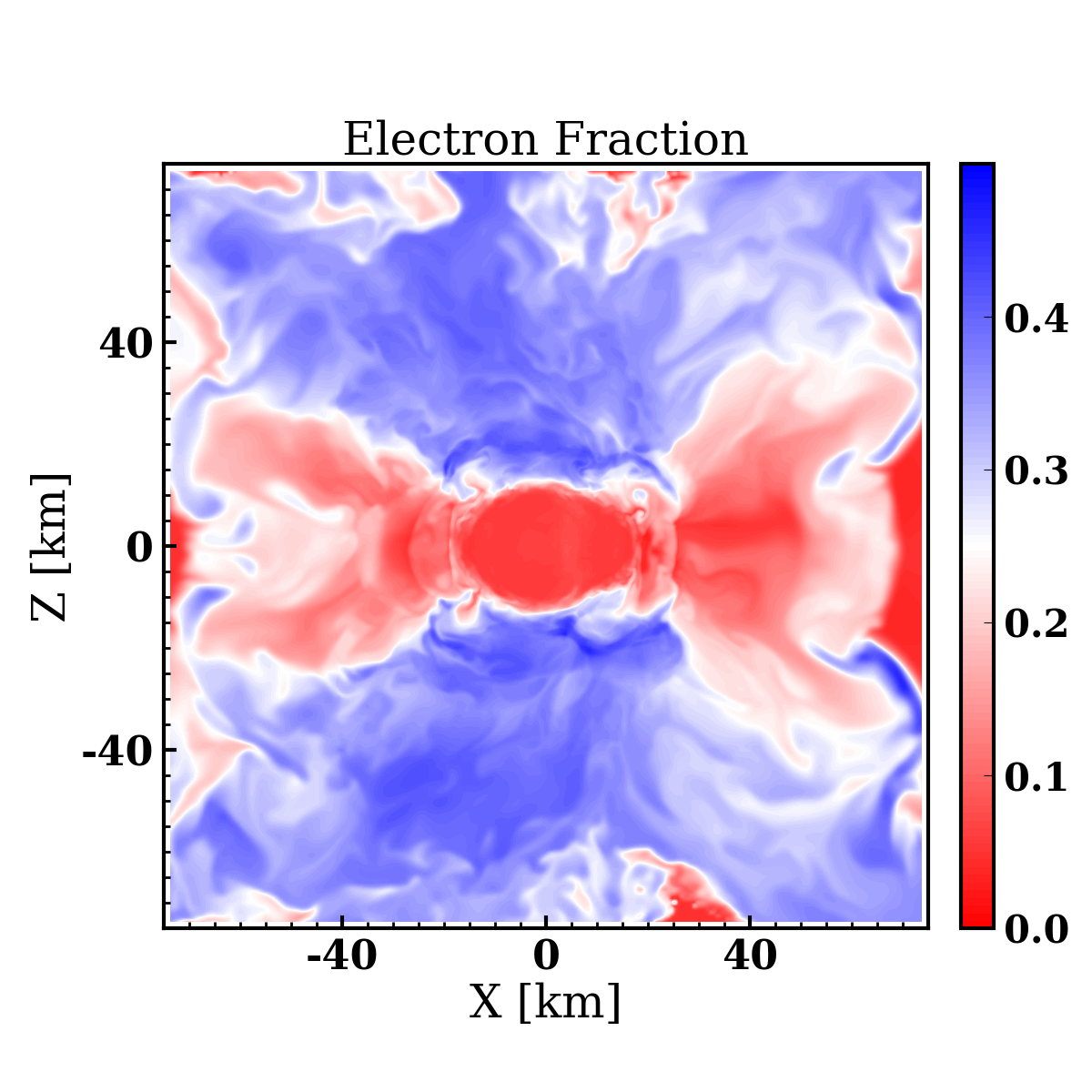}\\
\includegraphics[width=0.31\textwidth]{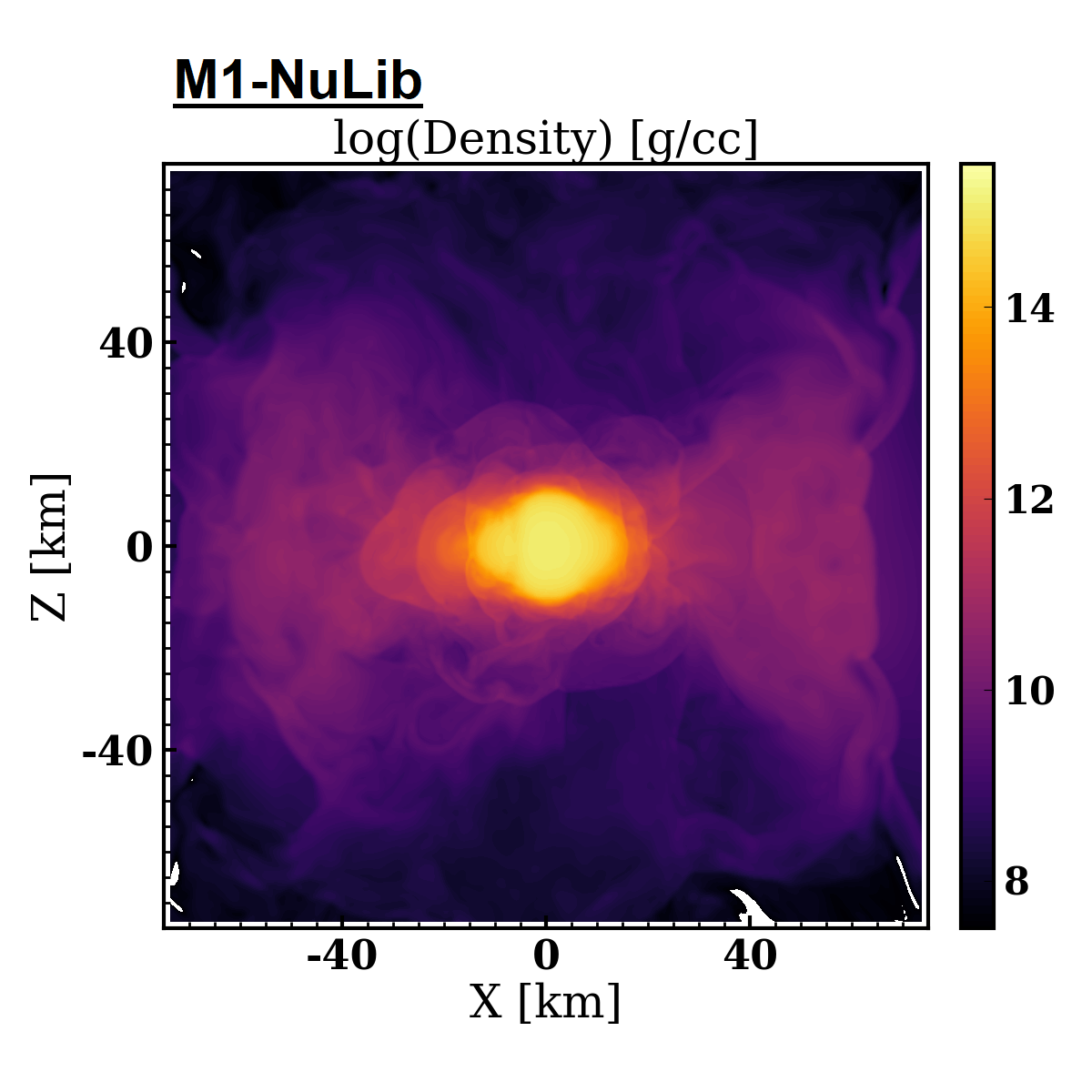}
\includegraphics[width=0.31\textwidth]{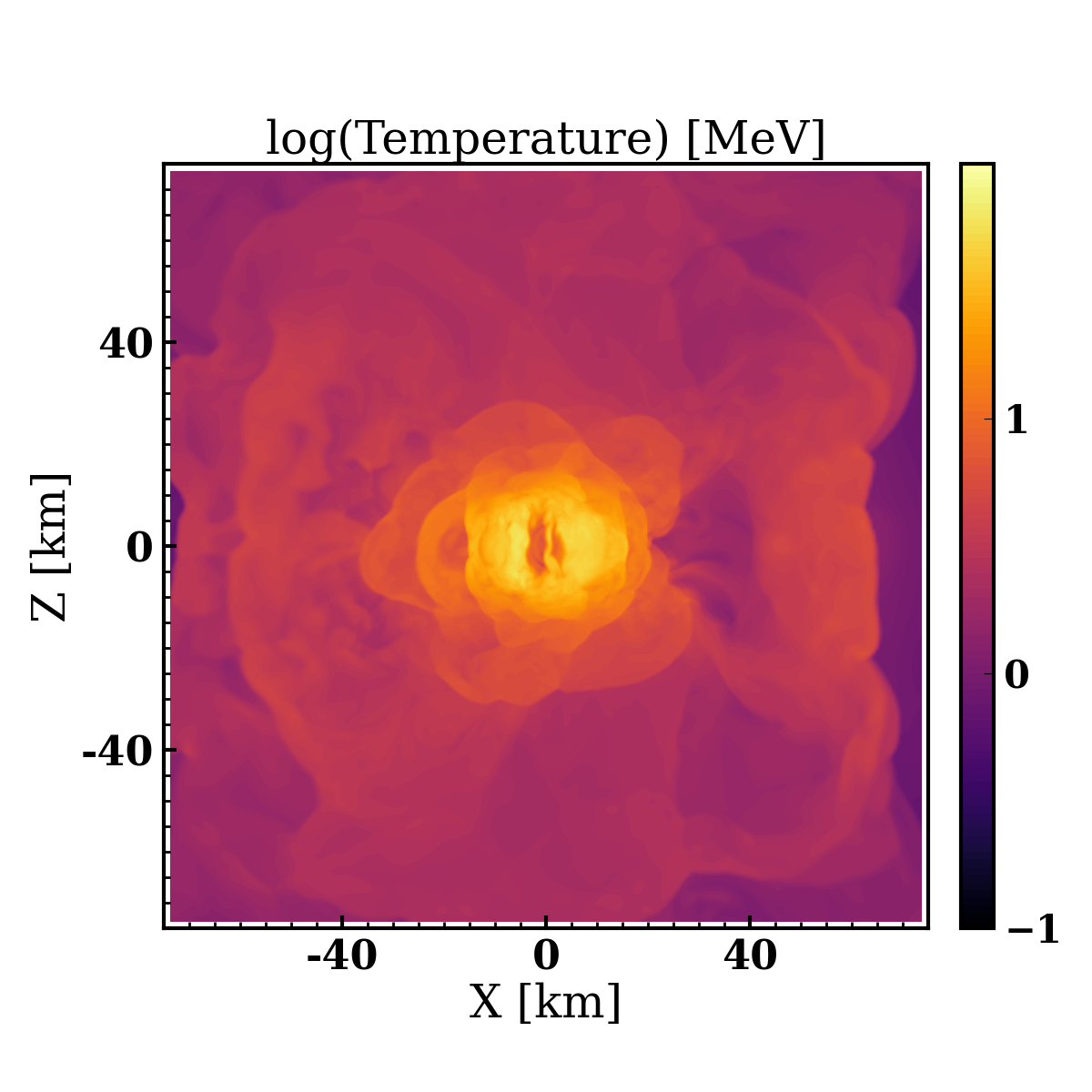}
\includegraphics[width=0.31\textwidth]{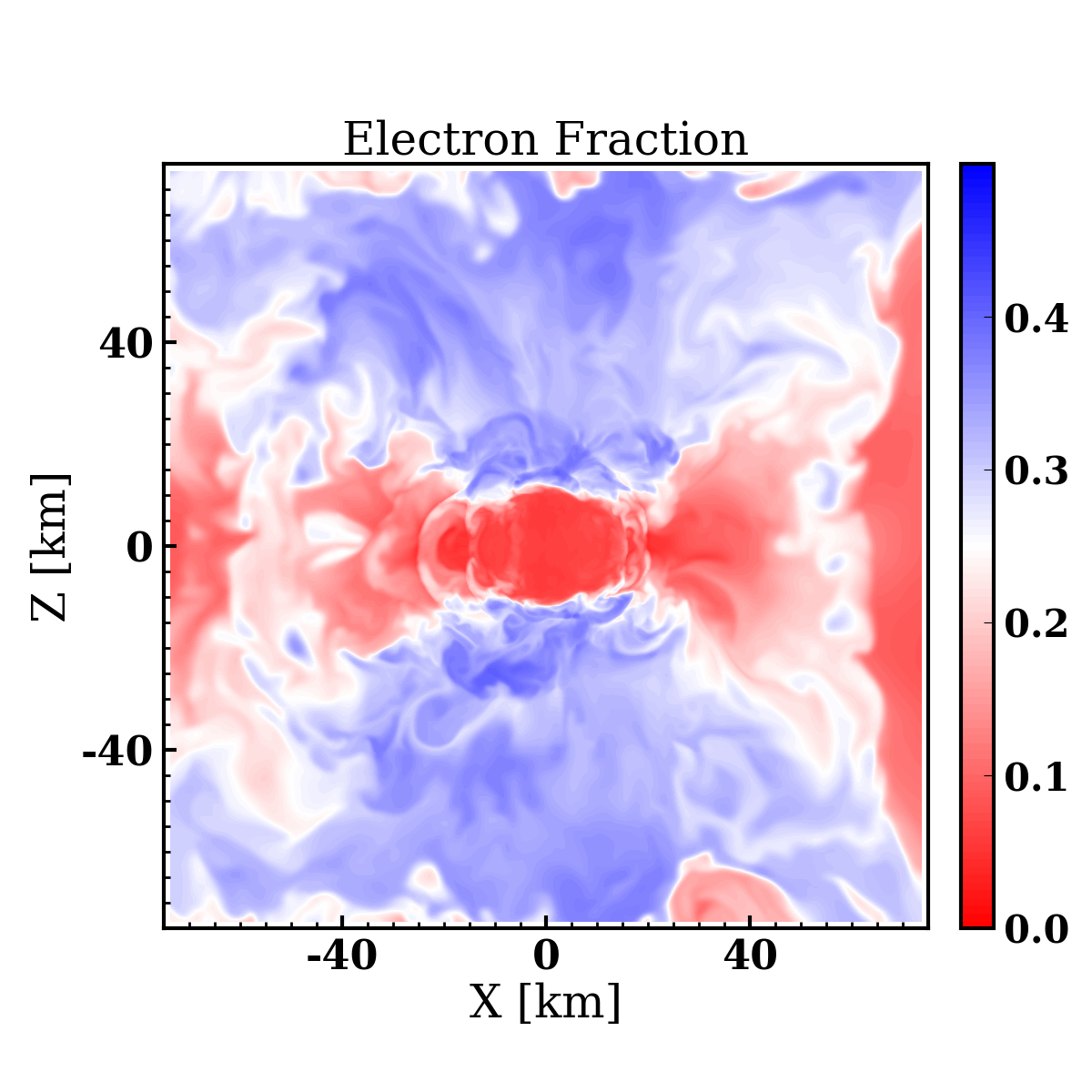}\\
\includegraphics[width=0.31\textwidth]{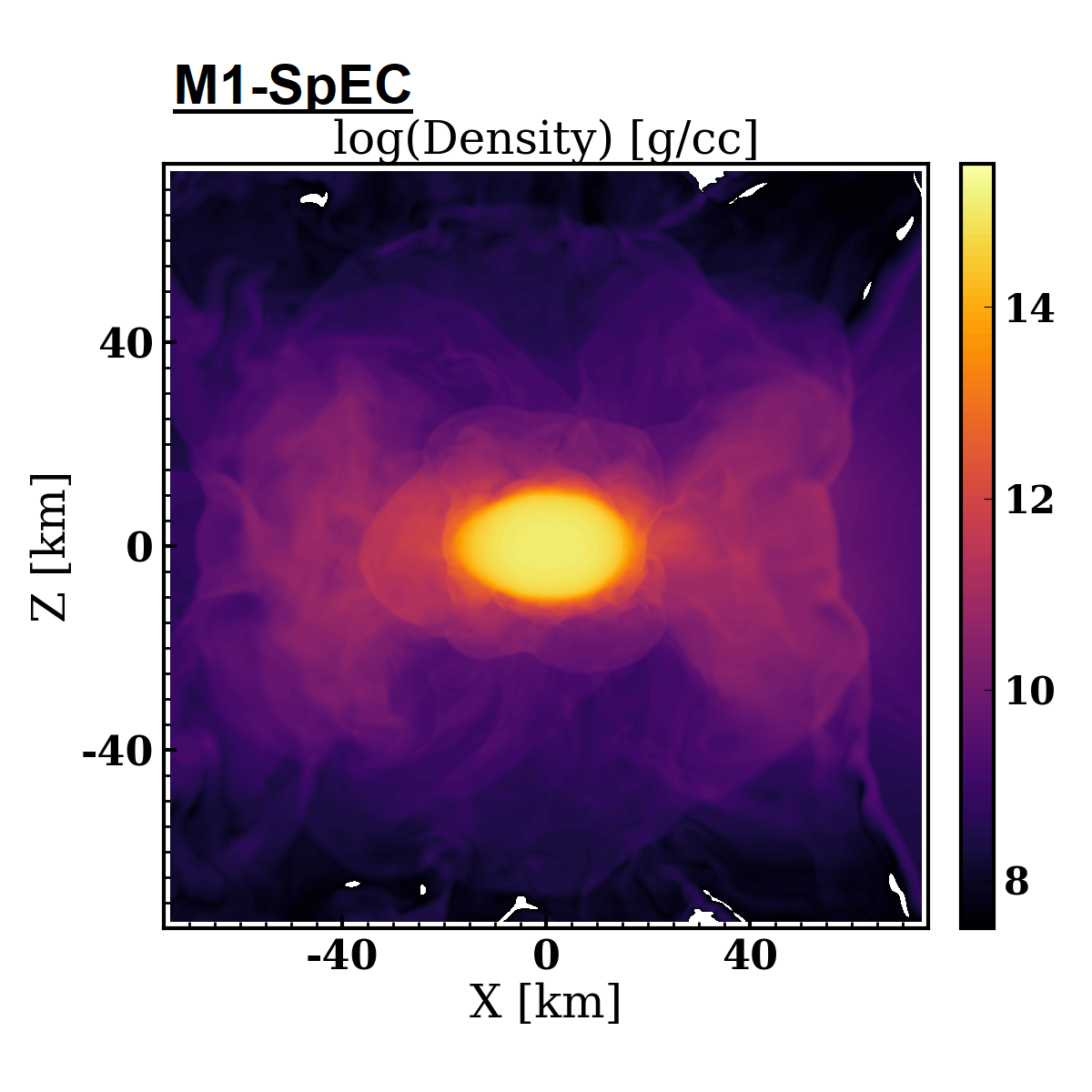}
\includegraphics[width=0.31\textwidth]{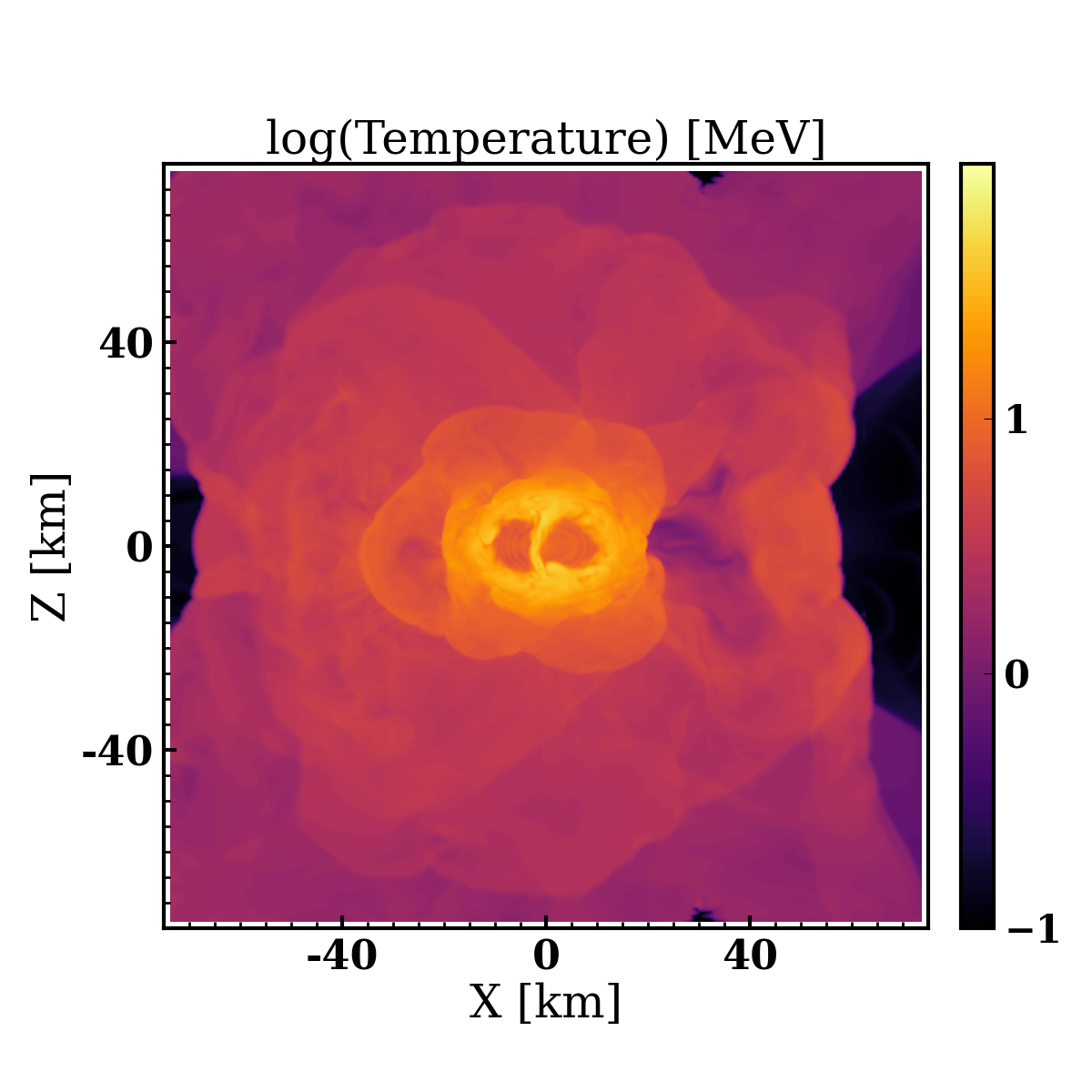}
\includegraphics[width=0.31\textwidth]{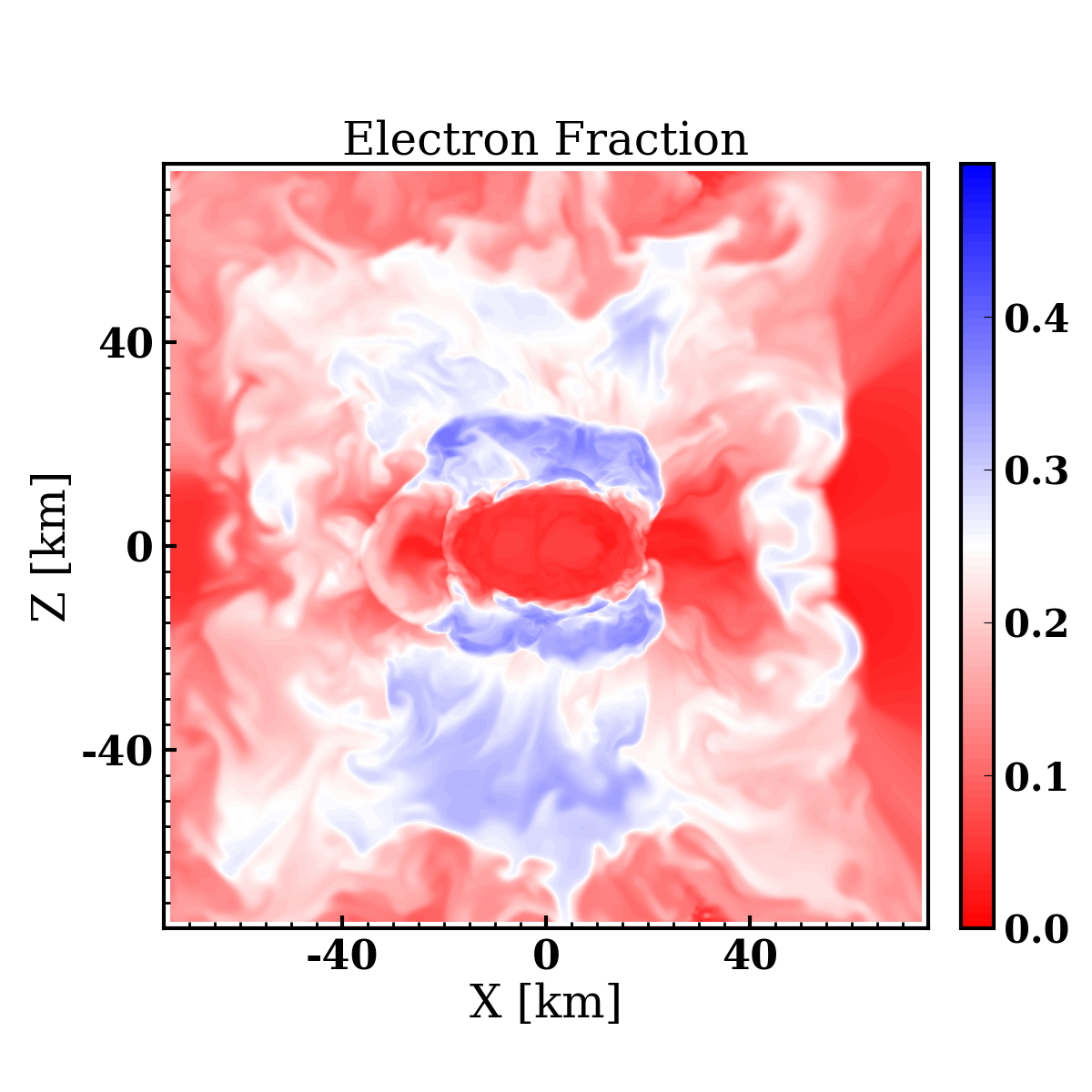}\\
\includegraphics[width=0.31\textwidth]{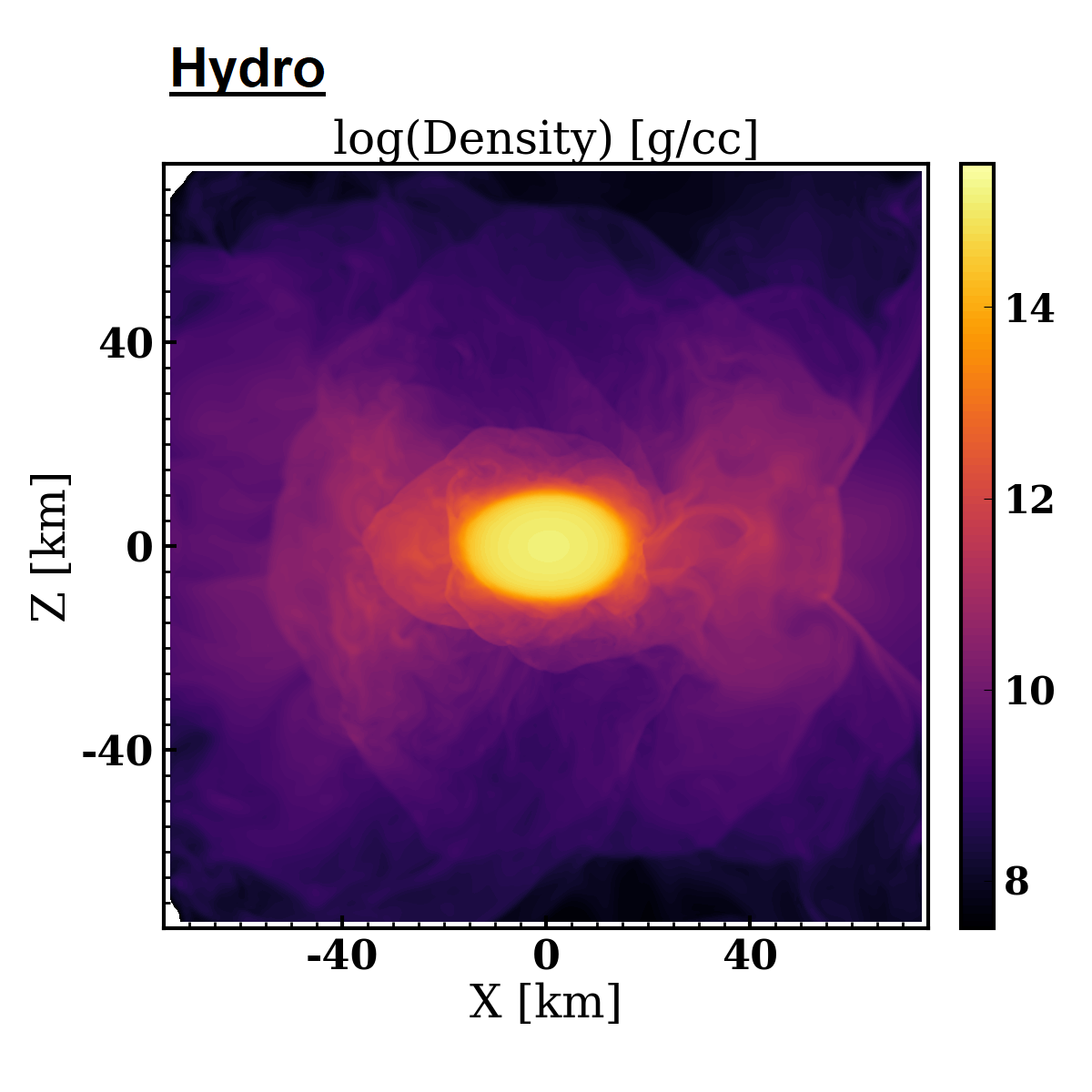}
\includegraphics[width=0.31\textwidth]{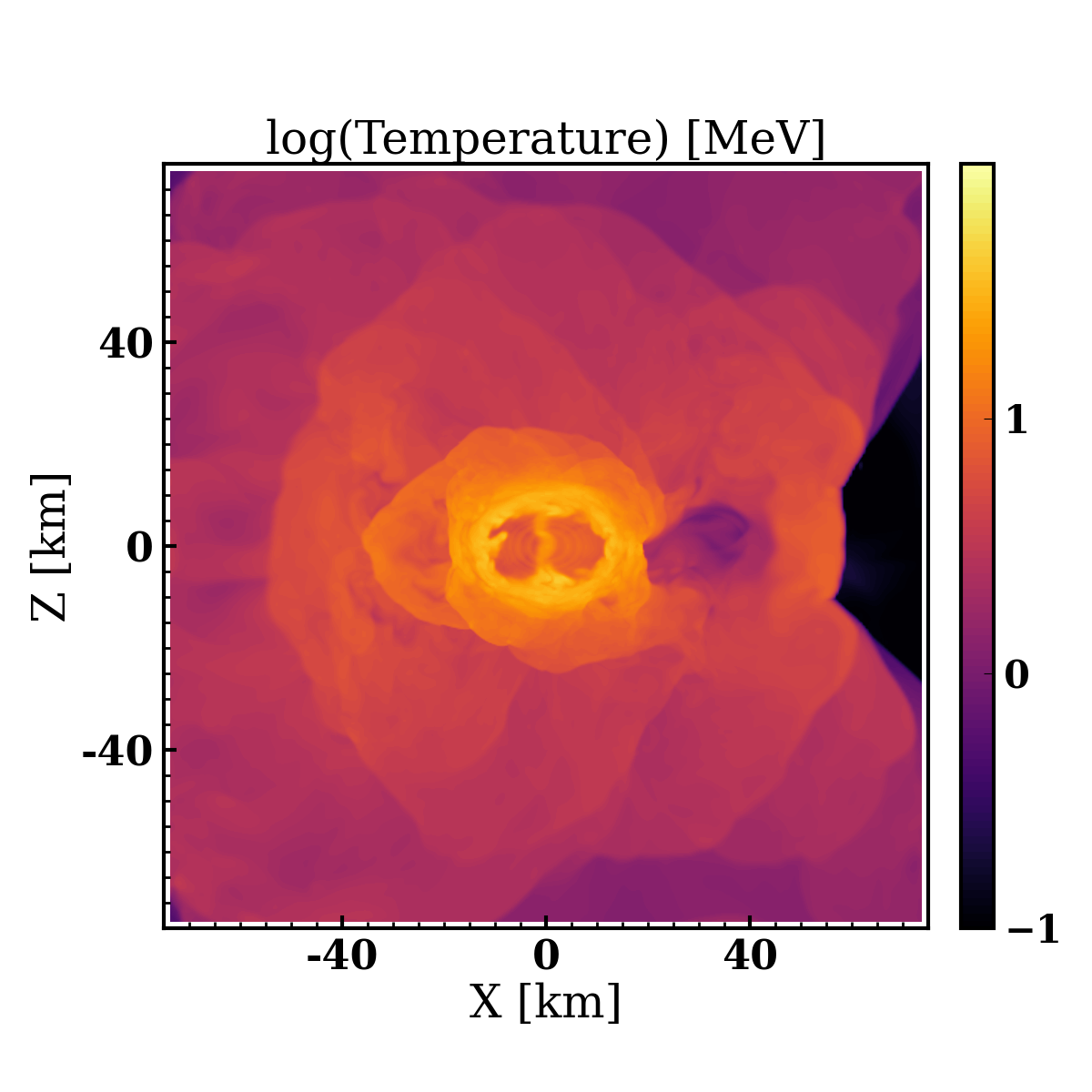}
\includegraphics[width=0.31\textwidth]{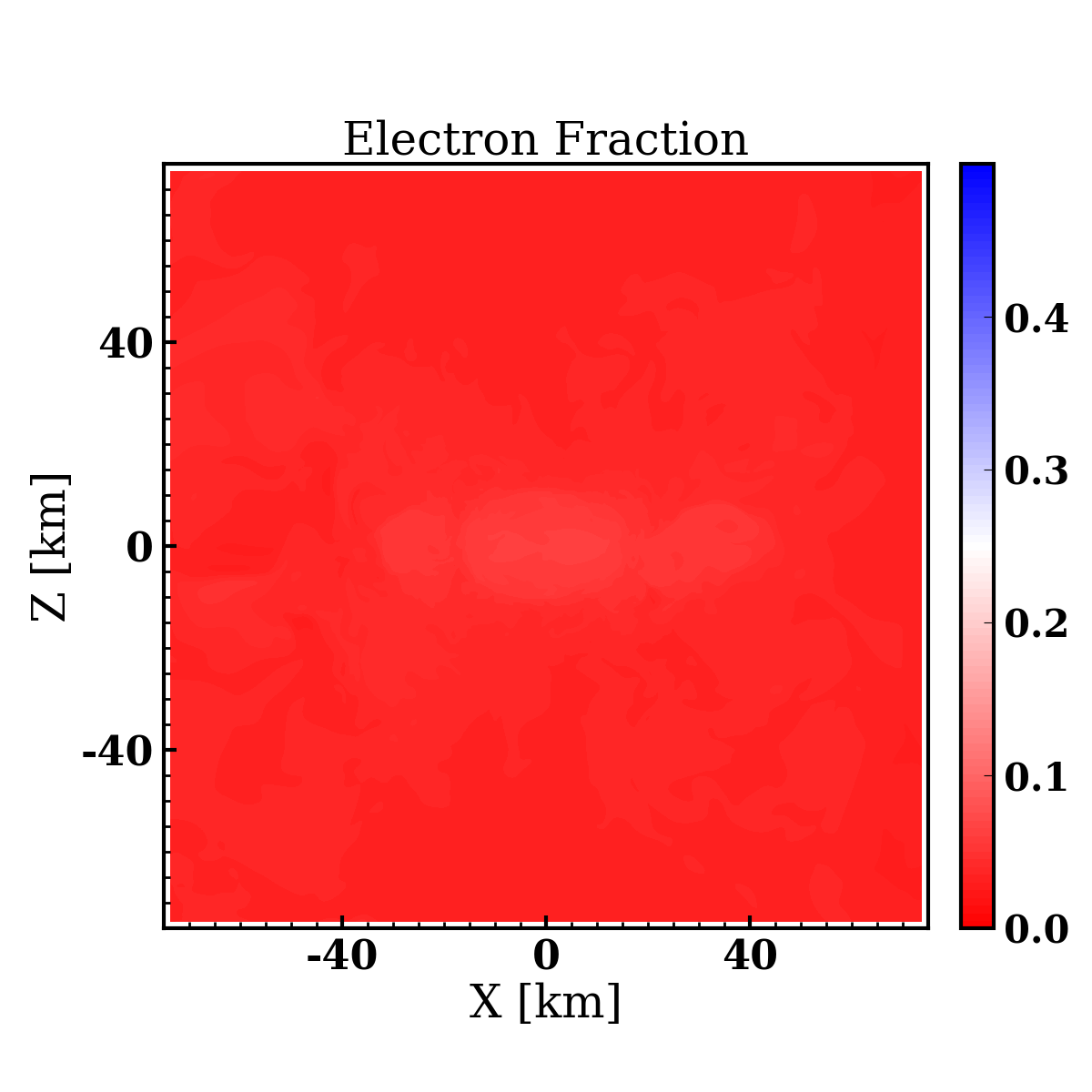}
 \caption{Vertical slice through the merger remant $2.5\,{\rm ms}$ post-merger. We show the baryon density (left), temperature (middle) and electron fraction (right)  for simulations M1-Radice (top), M1-NuLib (2nd row), M1-SpEC (3rd row) and Hydro (bottom).}
\label{fig:vis_ver}
\end{figure*}

\begin{figure*}
\includegraphics[width=0.31\textwidth]{Rho_Radice_Ver_2p5}
\includegraphics[width=0.31\textwidth]{Temp_Radice_Ver_2p5}
\includegraphics[width=0.31\textwidth]{Ye_Radice_Ver_2p5}\\
\includegraphics[width=0.31\textwidth]{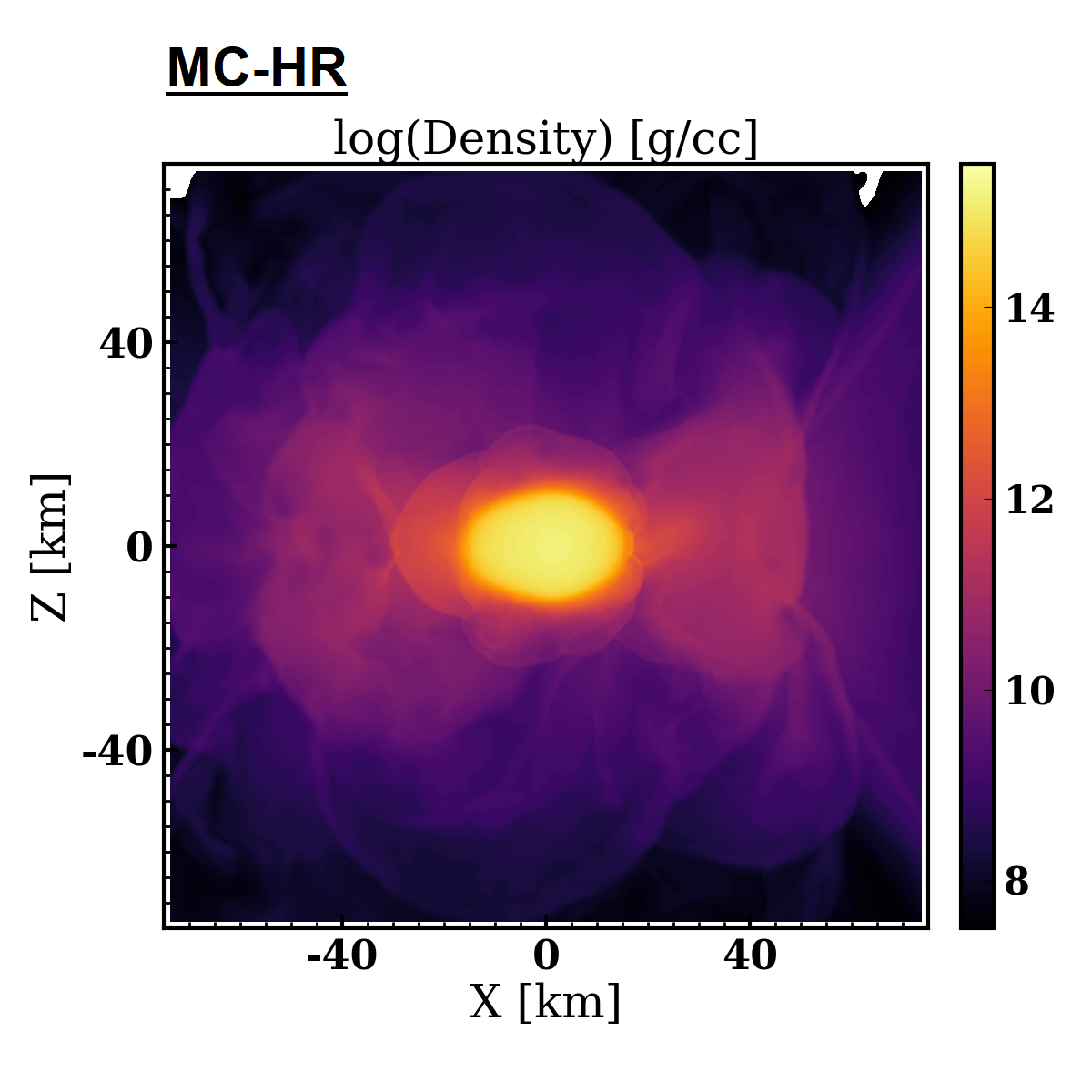}
\includegraphics[width=0.31\textwidth]{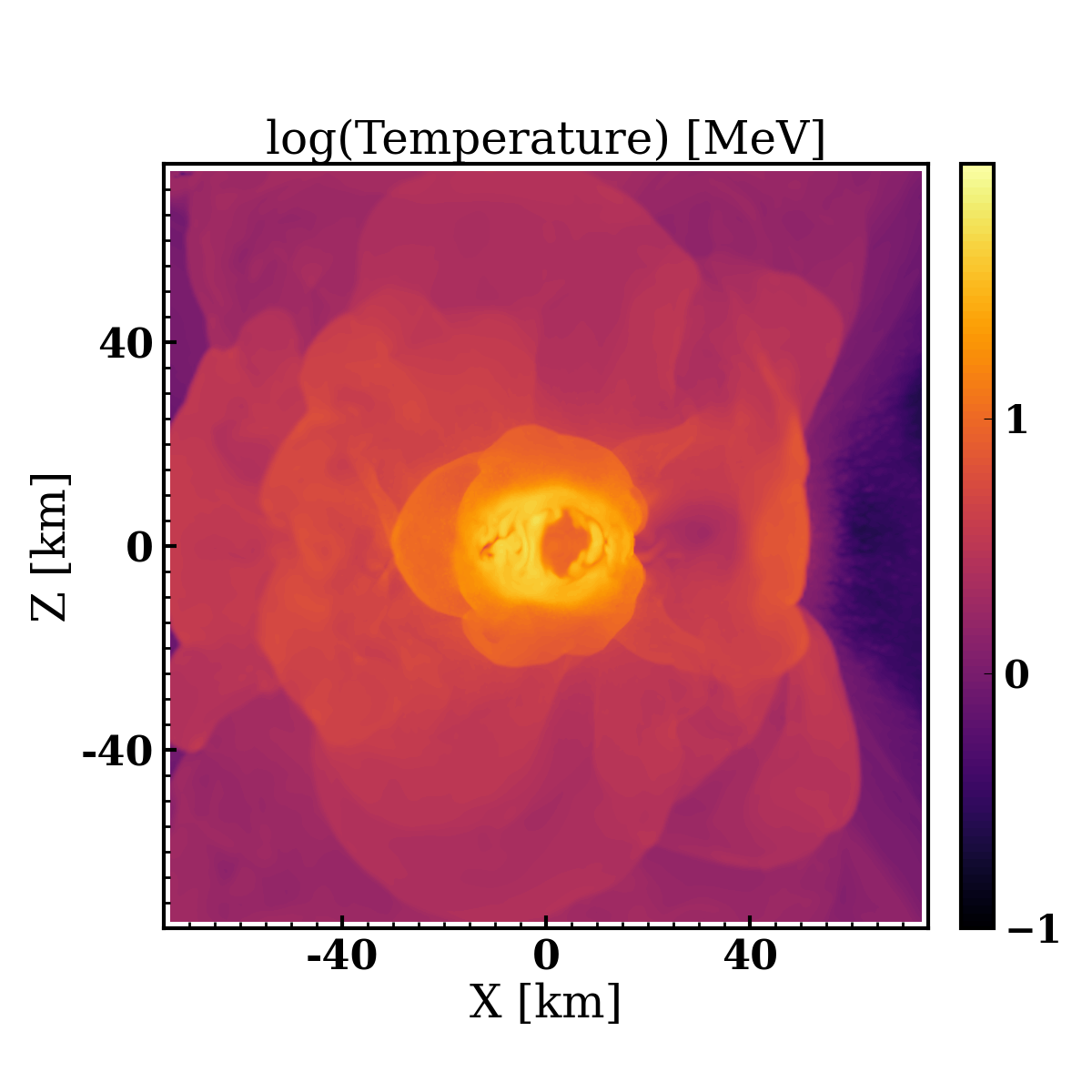}
\includegraphics[width=0.31\textwidth]{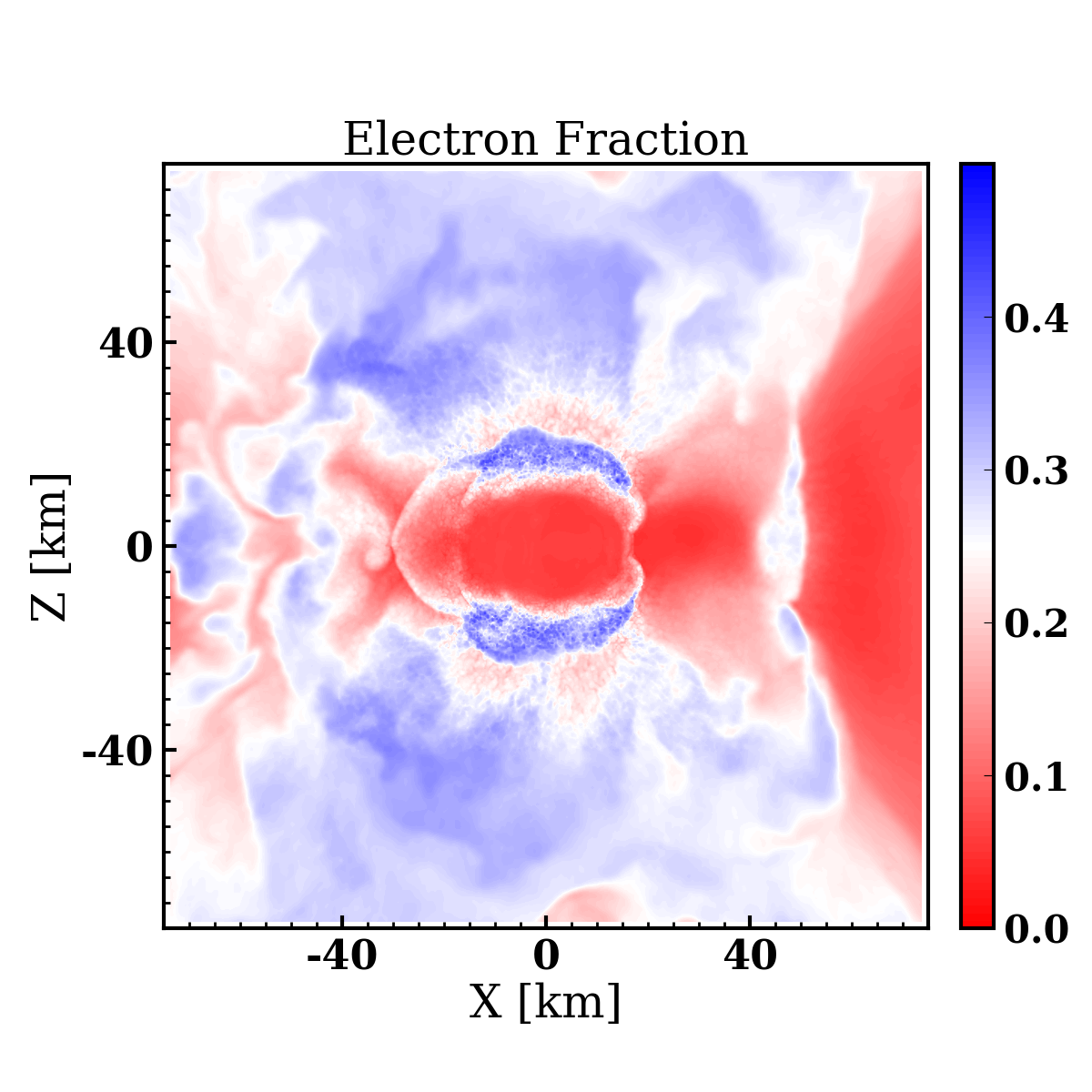}\\
\includegraphics[width=0.31\textwidth]{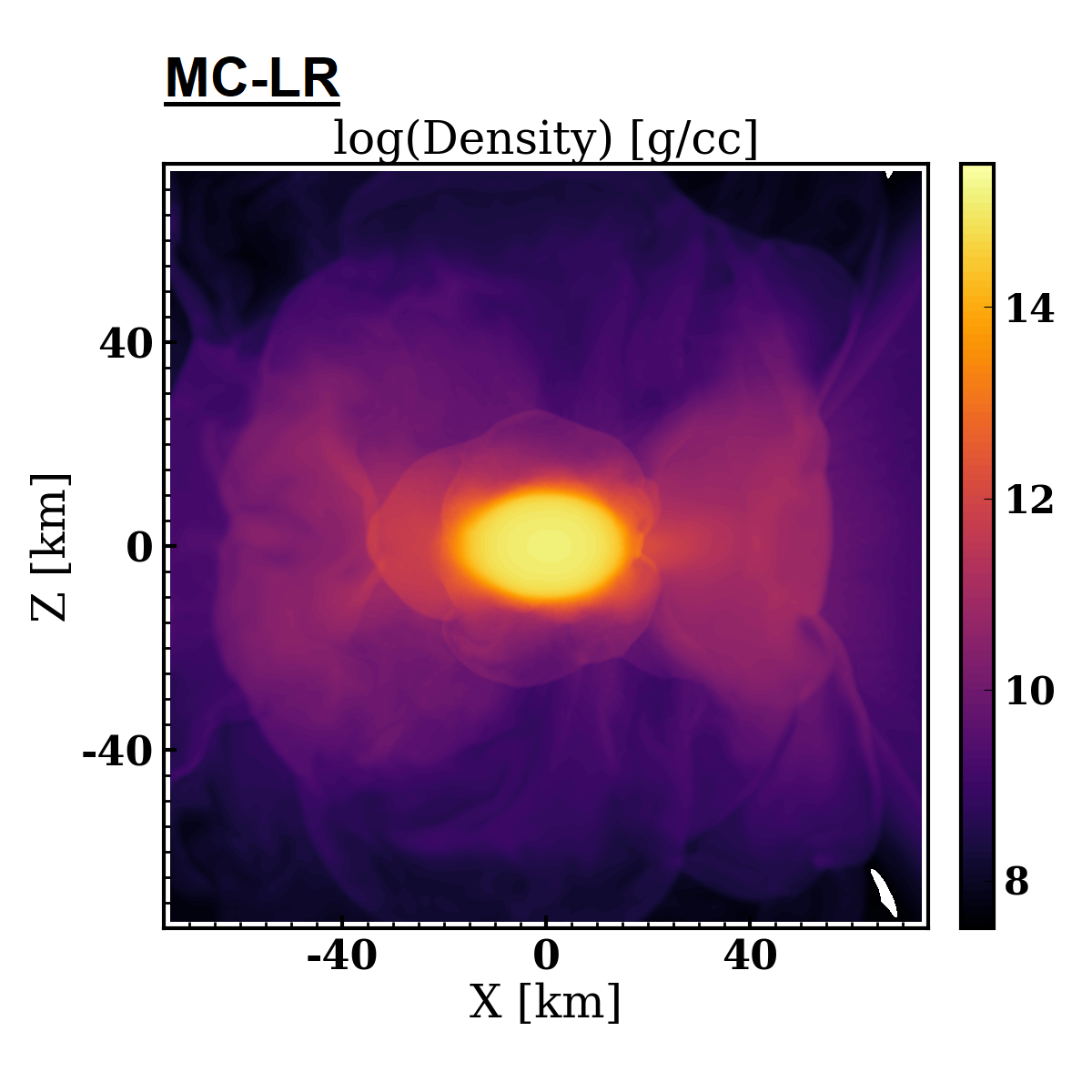}
\includegraphics[width=0.31\textwidth]{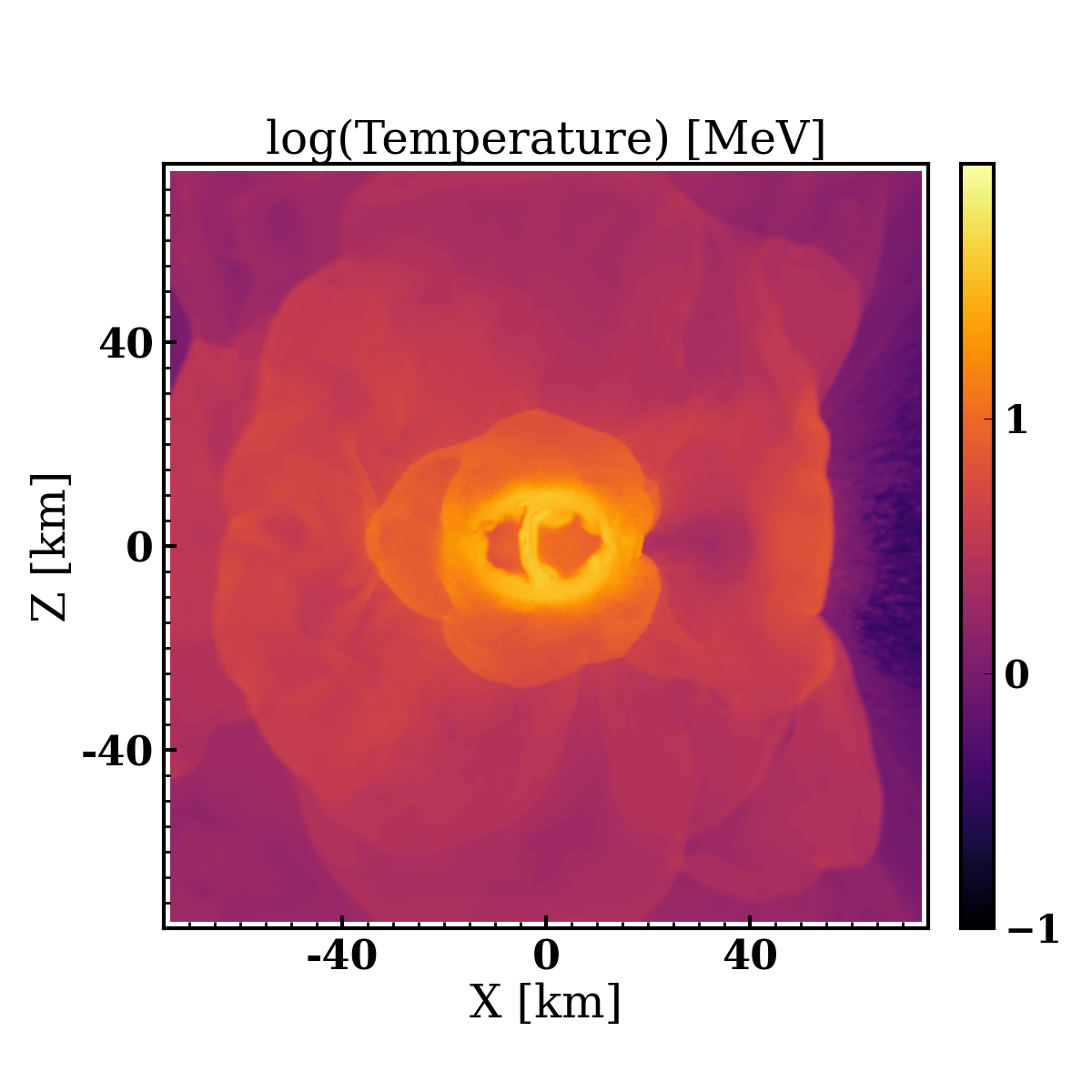}
\includegraphics[width=0.31\textwidth]{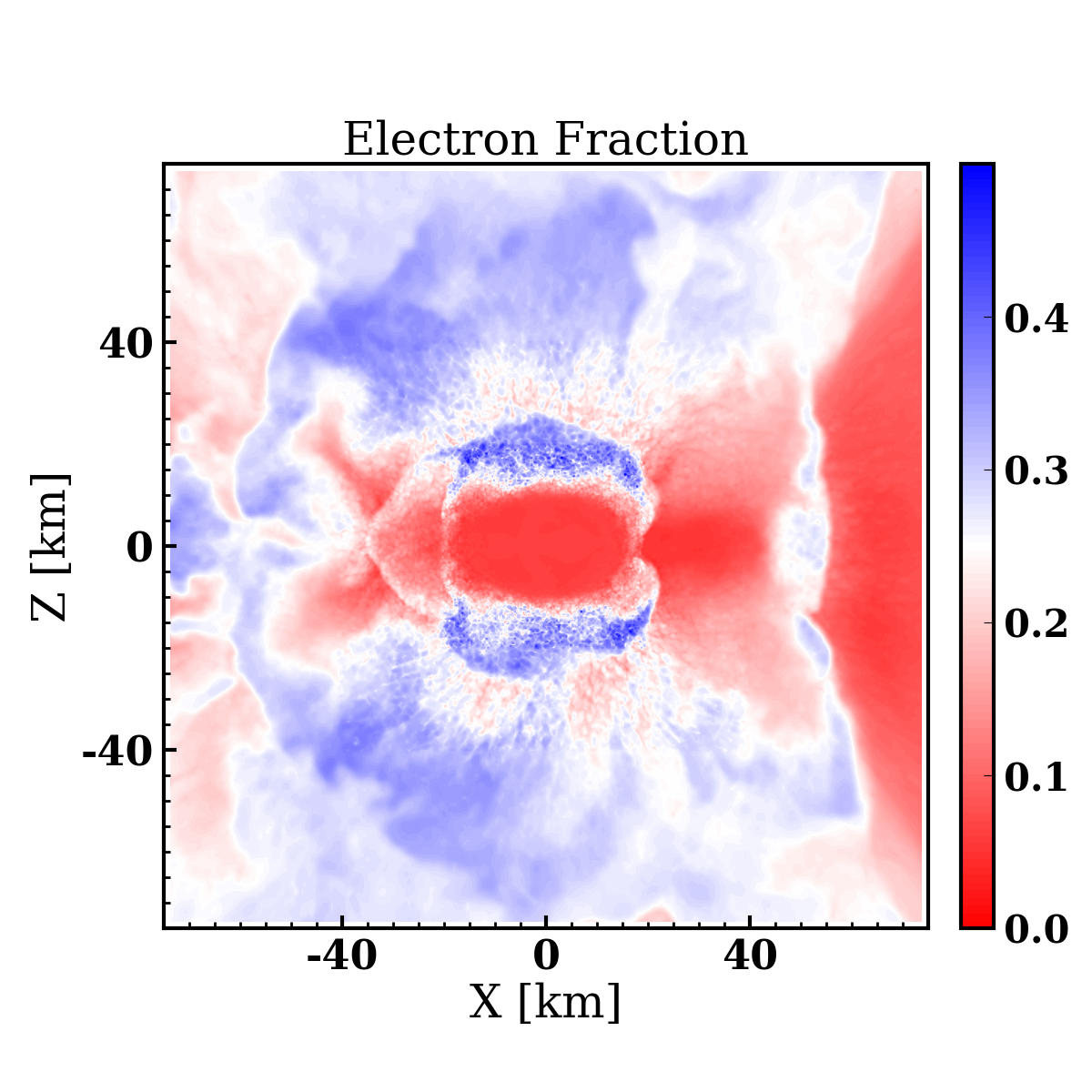}
 \caption{Same as Fig.~\ref{fig:vis_ver} but for simulations M1-Radice (top), MC-HR (middle) and MC-LR (bottom).}
\label{fig:vis_ver_mc}
\end{figure*}

The overall properties of the remnant between merger and collapse (specifically, $2.5\,{\rm ms}$ post-merger) are summarized on Table~\ref{tab:rem}, as well as through visualizations of horizontal and vertical slices through the remnant in Figs~\ref{fig:vis_hor}-\ref{fig:vis_ver_mc}. The bulk of the remnant has an average temperature of $(24-29)\,{\rm MeV}$ and an average electron fraction $(0.053-0.068)$. The main differences in temperature are between the rapidly collapsing MC simulations and the other configurations, which makes sense if the temperature of the high-density core of the remnant increases as it contracts. For the electron fraction, the main outliers are the M1-SpEC simulation, and the Hydro simulation, which obviously ignores weak interactions. The M1-SpEC simulation has clear issues with the evolution of $Y_e$ in the densest region, as the star neutronizes despite the neutrino luminosity being dominated by $\bar\nu_e$. This issue is not shared by any other simulation, including the M1-NuLib simulation which uses the same numerical methods in high-density regions, or previous simulations using the M1-SpEC methods. We will return to this issue in our discussion of the neutrinos within the remnant, but note here that this due to a failure of the M1-SpEC algorithm to satisfy lepton number conservation for a short period of time around merger. The M1-Radice and MC simulations, which should be considered as the `best' models for radiation transport in this work, are in much better agreement with $\langle Y_e \rangle = (0.062-0.068)$. We note that both the transport algorithms (M1-SpEC, M1-Radice, MC) and the chosen interaction rates (M1-NuLib vs M1-SpEC) appear to have an impact on our results here. Disentangling the impact of any specific interaction rate on these global quantities is however difficult in this nonlinear system in which only the sum of many energy-integrated interaction rates are used in the simulations. Here, we limit ourselves to assess the level at which these changes impact the results; a more detailed study of the impact of each assumption in the reaction rates would require more systematically modifying individual contributions to the interaction rates.

We can also look more specifically at the low-density matter that may eventually form a disk. Here, we use the arbitrary cutoff $\rho \leq 10^{13}\,{\rm g/cc}$ to define the ``disk'' material, noting that the ``disk'' material remains difficult at this point to distinguish from the surface of the remnant. We see in Table~\ref{tab:rem} that the amount of matter in the low-density regions is $(0.045-0.056)M_\odot$, with excellent agreement between the M1-Radice and MC methods which find $(0.054-0.056)M_\odot$. The same is true for the temperature and electron fraction of the low-density matter, with all simulations showing a range of temperature $(10.1-11.9)\,{\rm MeV}$ but the M1-Radice and MC simulations a smaller range $(10.8-11.0)\,{\rm MeV}$; and all simulations with neutrino transport showing a range of electron fraction $(0.118-0.145)$ but the M1-Radice and MC simulations showing only a range $(0.126-0.133)$. Unsurprisingly, the Hydro simulation remains at a much lower electron fraction $Y_e=0.046$ in the low-density regions. It is thus clear that all transport schemes used here capture the dominant effect of the neutrinos on the low-density matter, but with much smaller variations between our best radiation schemes than when we compared those best schemes to the more approximate M1 schemes. The chosen reaction rates additionally have a clear impact on both the mass and composition of the matter below $10^{13}\,{\rm g/cc}$ (comparing M1-SpEC and M1-NuLib in the table), with a stronger protonization of the low-density matter in the M1-NuLib simulation.

At even lower densities ($\rho \sim 10^{10-11}\,{\rm g/cc}$), most of the matter is in shocked tidal arms (see on Figs.~\ref{fig:vis_hor}-\ref{fig:vis_hor_mc}). The tidal arms will persist at least until collapse to a black hole. Fig.~\ref{fig:vis_hor} shows a horizontal slice through the remnant $2.5\,{\rm ms}$ post-merger for the M1 and Hydro simulations, while Fig.~~\ref{fig:vis_hor_mc} compares our best models M1-Radice and MC-HR, as well as the MC-LR simulation. We note that slight differences in the exact time at which snapshots were available leads to differences in the phase of the rotation of the remnant, which should be ignored in comparisons. At a qualitative level, we see broad agreements in the density and temperature of the remnants. The electron fraction is first impacted by the inclusion or not of neutrinos, with noticeable differences in the electron fraction of lower-density regions between different M1 simulations. These differences are particularly notable in the hot tidal arms, where they are $O(0.1)$. They are present in the temperature as well, but harder to see given the range of temperature that these plots must cover. We can for example look at the dominant tidal arm (above and right of the remnant on the figures) as it crosses the positive x-axis. There, it has temperature $(3-4)\,{\rm MeV}$ in the M1-Radice simulation, $(4-5)\,{\rm MeV}$ in the M1-SpEC and M1-NuLib simulation, $(6-7)\,{\rm MeV}$ in the MC simulations, and $(6-8)\,{\rm MeV}$ in the Hydro simulation -- much more significant variations than in higher-density regions. 

We can also take advantage of Fig.~\ref{fig:vis_hor_mc} to note that while noise due to the use of an MC algorithm is clearly visible in the electron fraction, the average properties of the matter in the MC-HR and MC-LR simulations are in much better agreement than between the M1-Radice and MC-HR simulations, indicating that MC noise does not have a significant impact on our final results. 

Taking both figures together, and focusing on the hot tidal arm region where differences are most notable, we see that in that region there is good agreement between the M1-NuLib and MC simulations, as well as between the M1-Radice and M1-SpEC simulations. This hints that the included reaction rates may have a more significant impact on the evolution of the electron fraction in those regions than the exact transport algorithm used. For simulations with neutrinos, there is a correlation between higher temperature and higher electron fraction. This is expected, as the fluid is driven towards a higher equilibrium $Y_e$ at higher temperatures. What drives differences in temperature is harder to ascertain, though it would be reasonable to guess that both the oscillations of the central compact object, which drive the production of tidal arms, and neutrino emission, which cools the tidal arms, play a role in setting the temperature of these tidal arms.

Figs.~\ref{fig:vis_ver}-\ref{fig:vis_ver_mc}, which show vertical slices through the post-merger remant for the same snapshots, show similar features. The phase of the evolution of the post-merger remnant is a confusing factor here when comparing density and temperature, yet we see broadly similar features in all simulations. The electron fraction, once again, shows the most significant differences -- especially in the tidal arms and polar regions. As for the horizontal slices, the two MC simulations are in good agreement with each other outside of the visible sampling noise, supporting the idea that the number of MC packets is not a critical source of error here. The MC simulations are once more closest to the M1-NuLib simulation, indicating again that in low-density regions the chosen interaction rates have a more significant impact on the electron fraction than the transport algorithm. 

Overall, we thus find that very early in the evolution of the post-merger remnant, the chosen transport algorithm (MC vs M1-NuLib; M1-Radice vs M1-SpEC) causes $\sim (20-30)\%$ relative changes in the average properties of the compact remnant (temperature, electron fraction). MC and M1 simulations in addition differ in the evolution of the densest region towards collapse. In the lower density regions close to the remnant, which are probed by our measurements of average temperature and electron fraction in the ``disk'', there is great agreement between the two best transport algorithms (M1-Radice, MC) -- as might be expected from regions where the decoupling of the neutrinos from the fluid is the driving factor for the evolution of the matter. In the tidal arms and polar regions, where neutrinos are farther from equilibrium, the interaction rates used in the simulations are the most important driver of differences in composition between simulations, and significant differences (factor of 2) are observed in the temperature of the tidal arms. The number of packets in the MC simulations does not seem to be a significant source of uncertainty, though we caution that this does not guarantee that MC errors are negligible in the hottest, densest region, where we rely on a low-accuracy implicit MC algorithm. The M1-SpEC simulations, for its part, has significant issues with the evolution of the electron fraction in dense regions, which erroneously decreases over the period considered here.

\subsection{Mass ejection}

\begin{table}
\begin{tabular}{c|c|c|c}
{\bf Sim} & $M_{\rm ej}(rp)$ & $\langle v_{\rm ej}(rp) \rangle$ & $\langle Y_e\rangle$\\
\hline
 & $M_\odot$ & $c$ & \\\hline
MC-HR & 0.0041 & 0.235 & 0.215\\
MC-LR & 0.0047 & 0.239 & 0.212\\
M1-Radice & 0.0037 & 0.204 & 0.209\\
M1-SpEC & 0.0036 & 0.219 & 0.187\\
M1-NuLib & 0.0031 & 0.214 & 0.210\\ 
Hydro & 0.0044 & 0.243 & 0.036 
\end{tabular}
\caption{Unbound mass $M_{\rm ej}$, as measured on the computational grid $2.5\,{\rm ms}$ after merger, as well as average velocity and electron fraction of the ejecta.}
\label{tab:ej}
\end{table}

The mass, composition, and geometry of the outflows is naturally one of the most important prediction of merger simulations. Over the first $\sim 3\,{\rm ms}$ of evolution, the system considered here first ejects a small amount of cold, neutron-rich tidal ejecta, then slighly larger amounts of hotter outflows whose ejection is tied to the core-bounce oscillations of the forming remnant. The former is largely confined to the equatorial plane, while the latter covers a broader range of polar angles. 

We can first compare the global properties of the outflows, summarized in Table~\ref{tab:ej}. The table considers as outflows all matter marked as unbound $2.5\,{\rm ms}$ post-merger, according to the criteria of~\cite{Foucart:2021ikp} which accounts for r-process heating, neutrino cooling, and the finite time available for heating before particles reach apoastron. We note that material will continue to be ejected over longer timescales, both due to continued core-bounces and the evolution of the remnant on longer timescales. This early time data however allows us to compare the different transport methods without the complications of different collapse times or the large errors associated on longer timescales with the absence of magnetic fields in our simulations. At $2.5\,{\rm ms}$ post-merger, the simulations flag $(0.0031-0.0047)M_\odot$ of material as unbound -- a larger relative uncertainty than for many of the remnant properties considered in the previous section, but not far outside of expected numerical error. Indeed, the finite resolution of the grid leads to $\sim 10\%$ relative uncertainty in the ejected mass for the M1 and Hydro simulations~\cite{Foucart:2024kci}, while the finite number of packets in the MC simulation leads to $\sim 25\%$ relative uncertainty in the ejected mass of the MC-HR simulation (assuming errors scaling as $\sqrt{N}$, with $N$ the number of packets, so that we estimate the total error in MC-HR as $1.7$ times the difference between MC-HR and MC-LR). The composition and velocity of the outflows are in better agreement ($[10-20]\%$ relative differences), with the easily explained exception of the composition of the outflows in the neutrinoless simulation. Otherwise, the higher velocity of the outflows in the MC simulations and the lower electron fraction of the M1-SpEC simulation are the main observable differences outside of expected numerical errors.

\begin{figure*}
\includegraphics[width=0.48\textwidth]{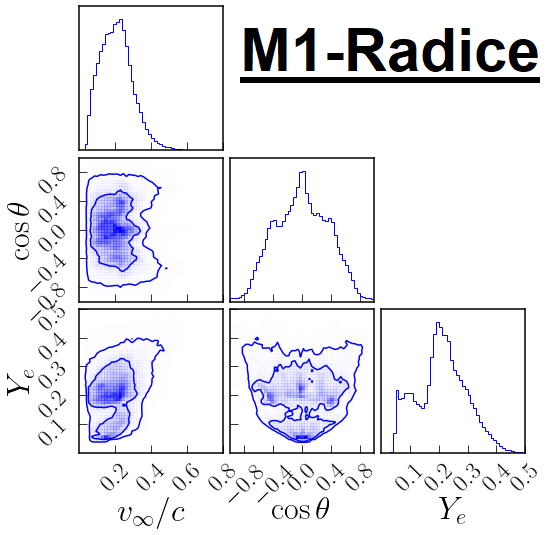}
\includegraphics[width=0.48\textwidth]{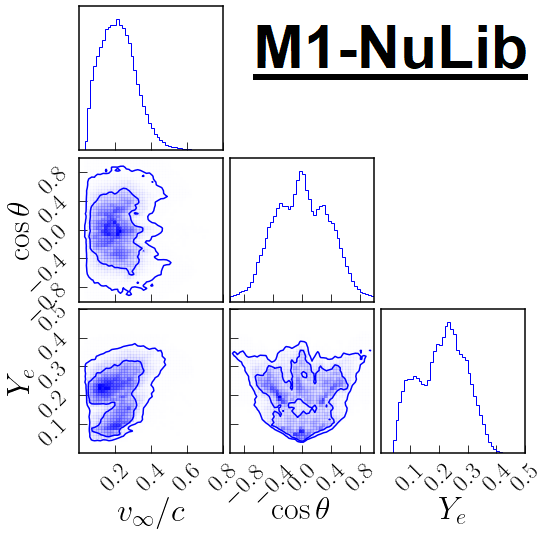}\\
\includegraphics[width=0.48\textwidth]{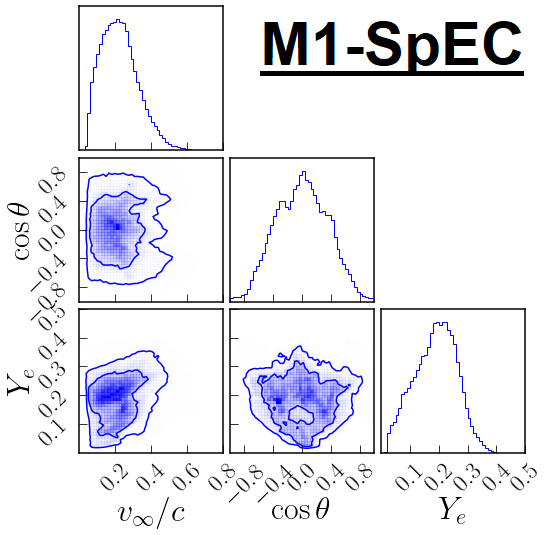}
\includegraphics[width=0.48\textwidth]{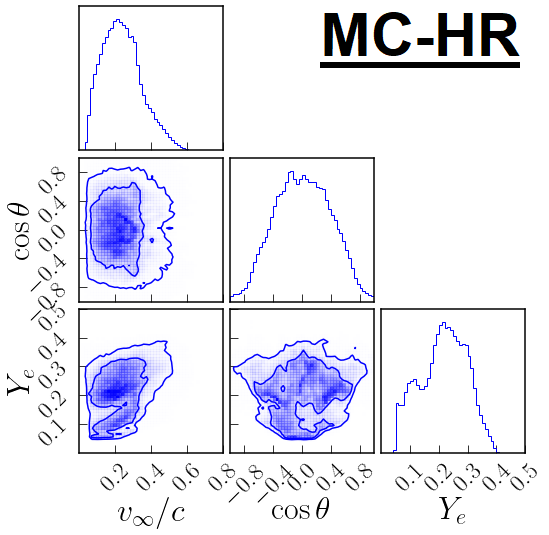}
 \caption{Distribution of the asymptotic velocity $v_\infty$, azimuthal angle $\theta$, and electron fraction $Y_e$ of the ejected matter $2.5\,{\rm ms}$ post-merger. We show simulations M1-Radice (top left), M1-NuLib (top right), M1-SpEC (bottom left), and MC-HR (bottom right). The MC-LR is, for these distributions, very similar to MC-HR while the neutrinoless simulation has obviously very different composition.}
\label{fig:corner_ej}
\end{figure*}

Fig.~\ref{fig:corner_ej} provides more fine-grained information about the outflows, for simulations with neutrinos. The corner plots show the distribution of the outflows in the three-dimensional parameter space of polar angle $\theta$, electron fraction $Y_e$, and predicted asymptotic velocity $v_\infty$. There is in general a fairly striking agreement between the four different neutrino transport schemes. The most notable differences are the underprodution of low-$Y_e$ tidal ejecta and high-$Y_e$ core-bounce ejecta in the M1-SpEC simulation, and a smoother angular distribution of the outflows in the MC simulation than in all M1 simulations. At the qualitative level however, all simulations predict that most of the outflows are within $\sim 60^\circ$ of the equator, with electron fraction peaking just above $Y_e\sim 0.2$ and extending to $Y_e\sim 0.4$, and velocity distribution peaking at $v_\infty\sim 0.2c$ and extending to $v_\infty \sim 0.6c$.

\subsection{Neutrinos in the remnant}

\begin{figure*}
\includegraphics[width=0.32\textwidth]{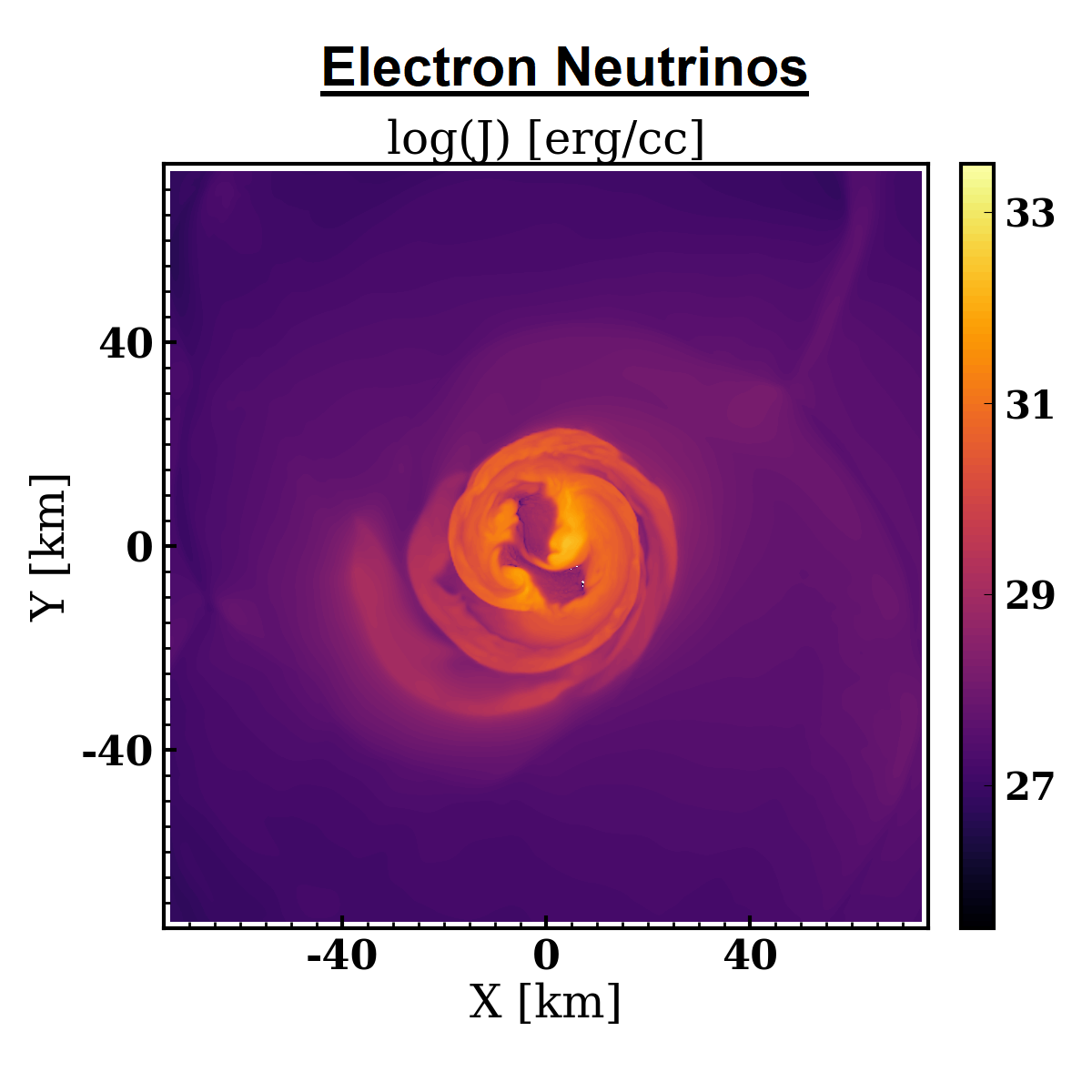}
\includegraphics[width=0.32\textwidth]{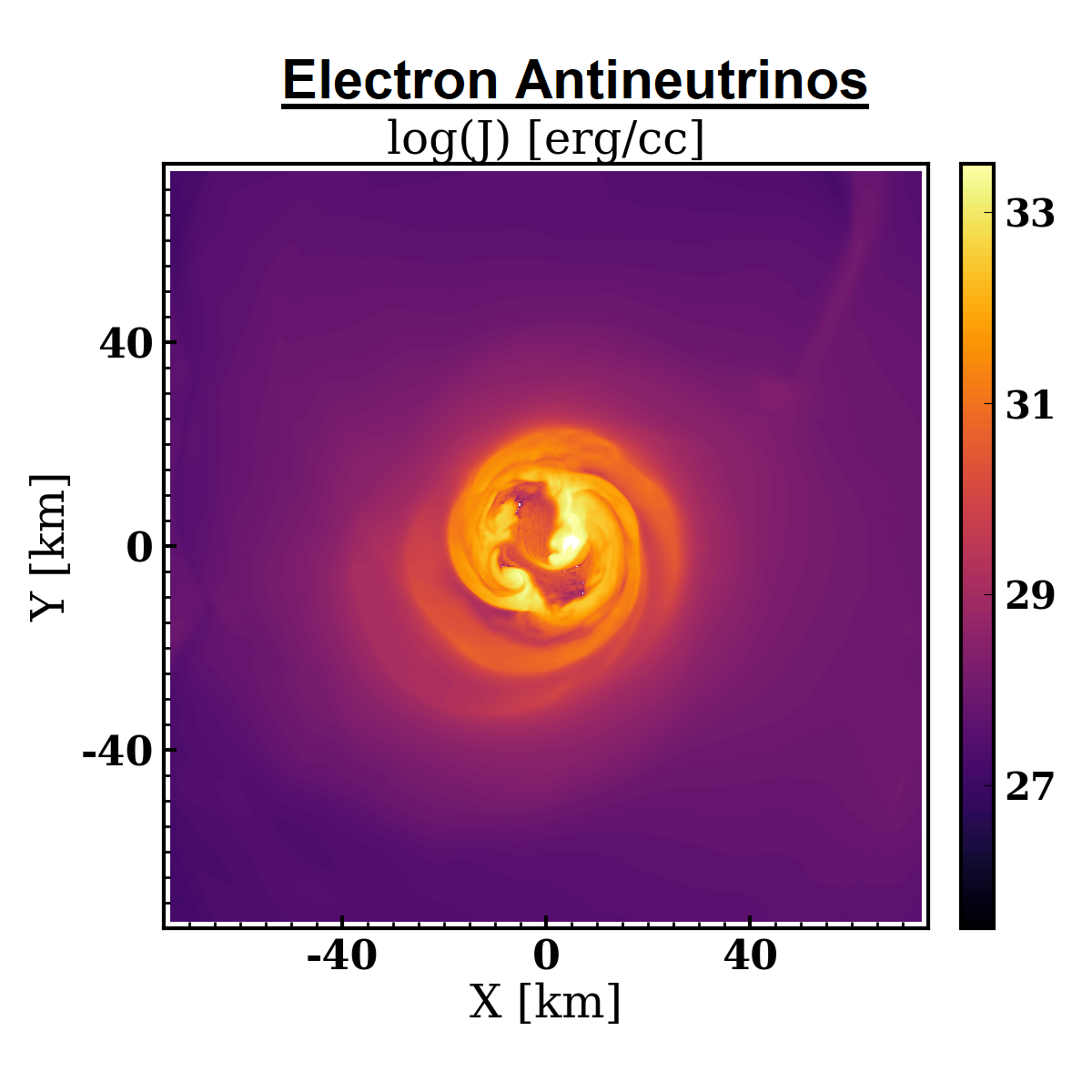}
\includegraphics[width=0.32\textwidth]{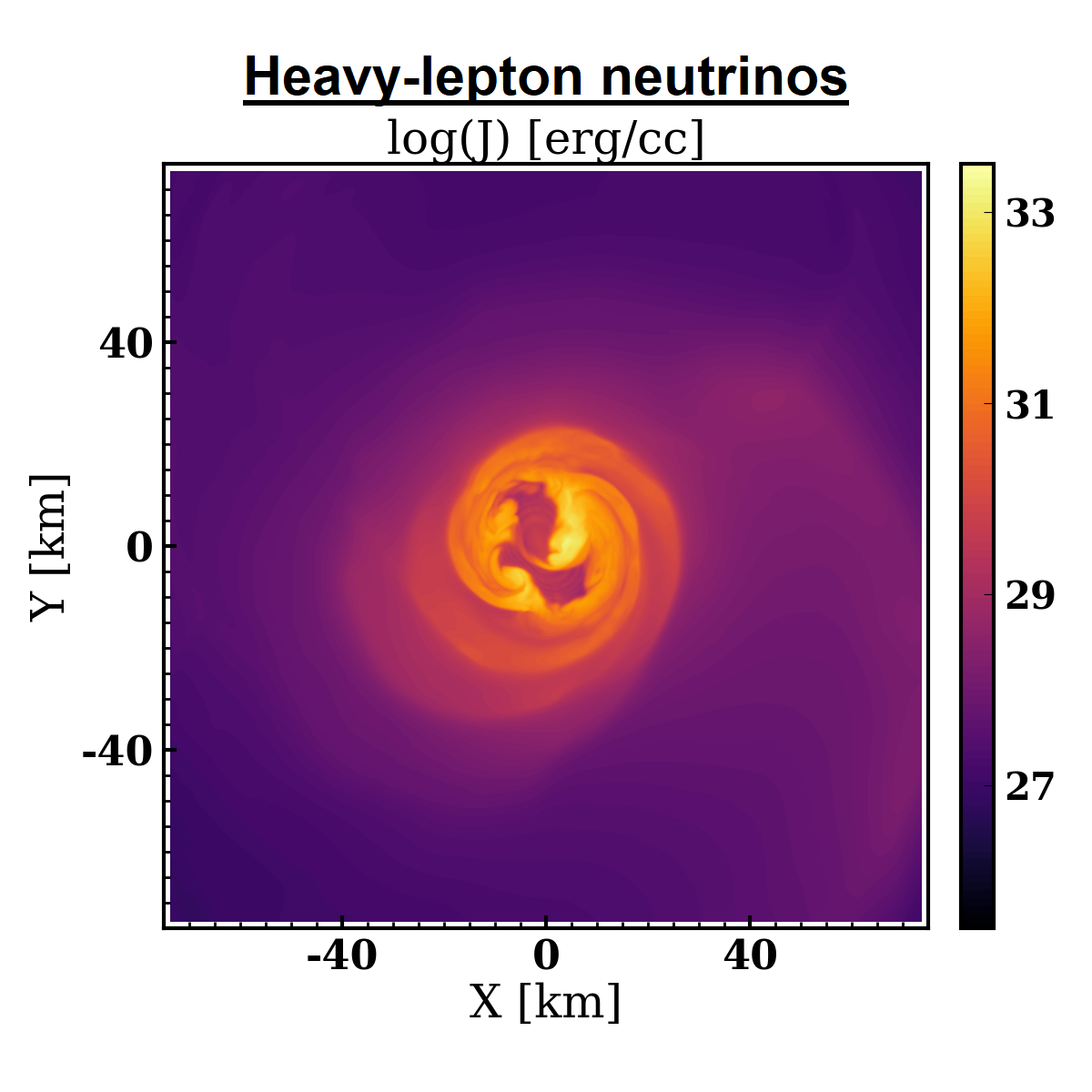}\\
\includegraphics[width=0.32\textwidth]{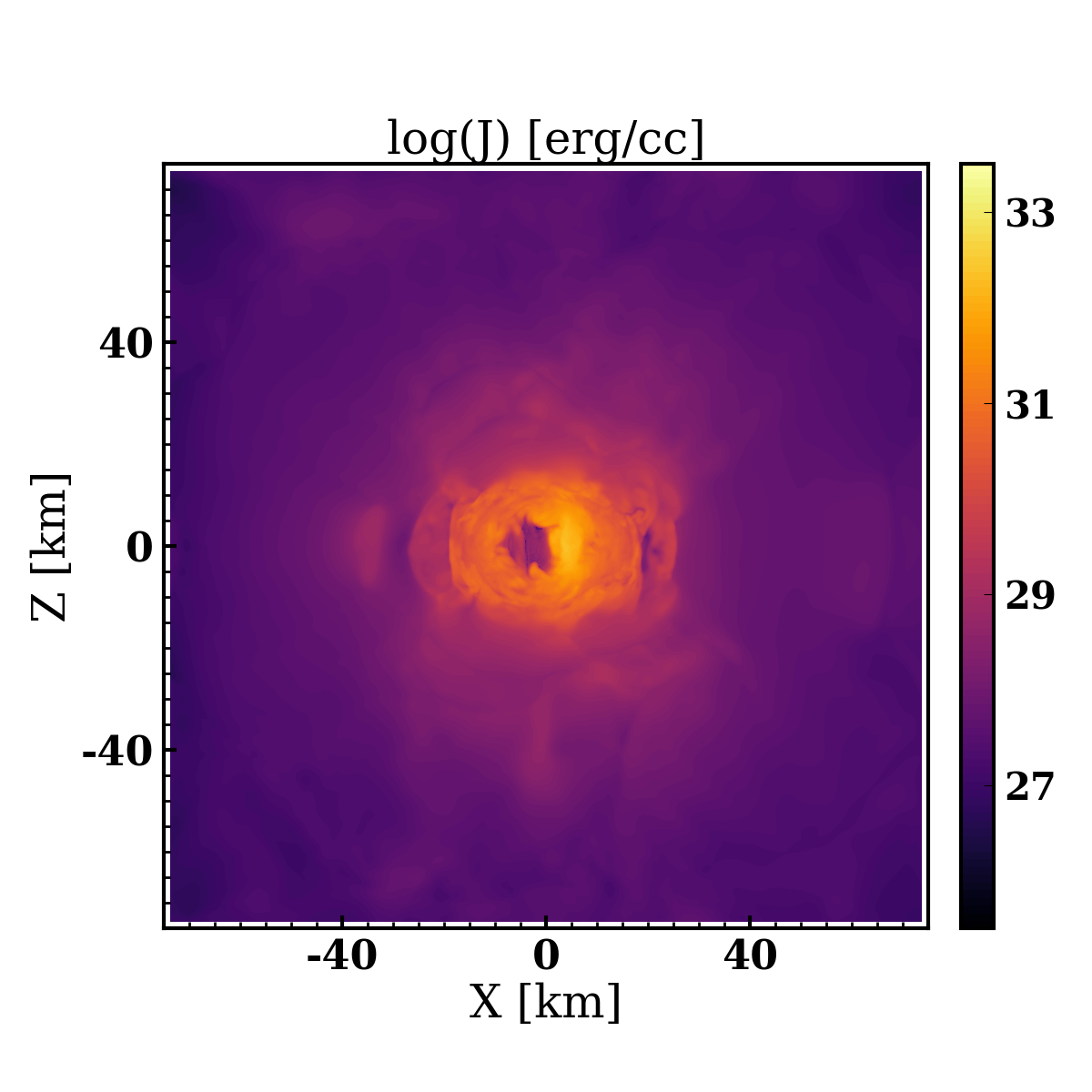}
\includegraphics[width=0.32\textwidth]{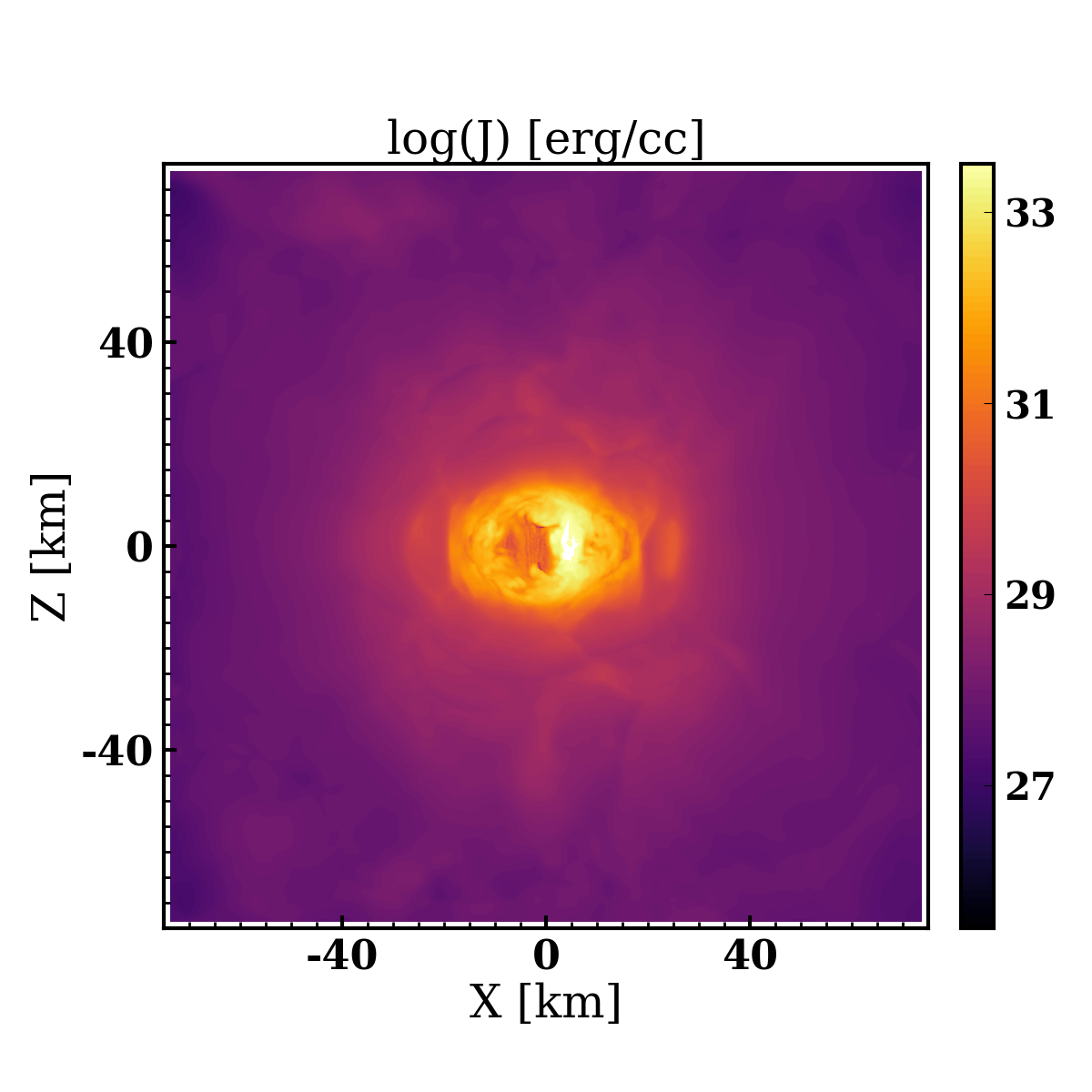}
\includegraphics[width=0.32\textwidth]{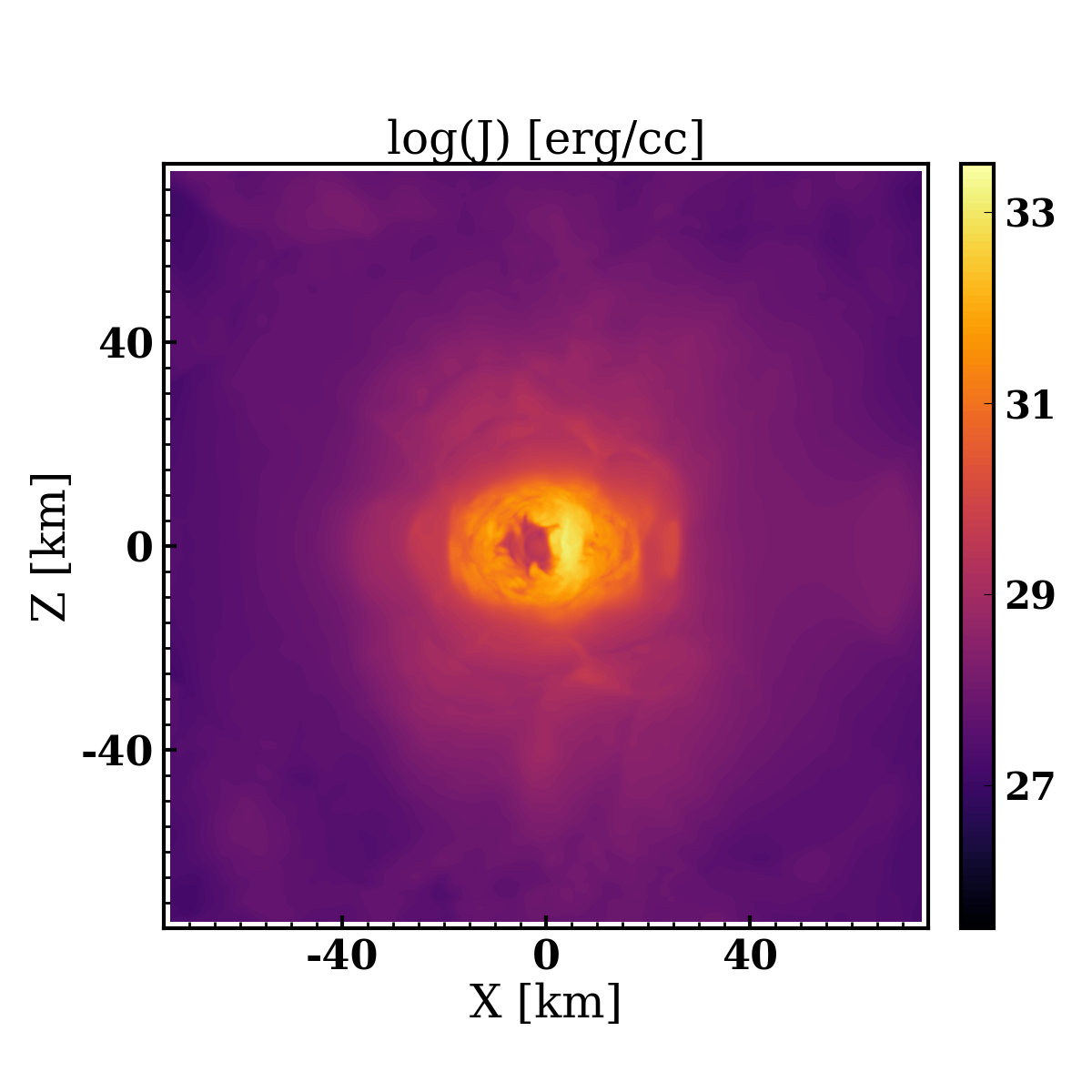}
 \caption{Energy density of neutrinos in the fluid frame for $\nu_e$ (left), $\bar\nu_e$ (middle), and each individual species of heavy-lepton neutrinos (right). We show results for the M1-Radice simulation $2.5\,{\rm ms}$ after merger and on a horizontal (top) and vertical (bottom) slice.}
\label{fig:jnu}
\end{figure*}

Some of the most important differences between the transport schemes used in this manuscript are mostly relevant in high-density, optically thick regions. In particular, one region of parameter space where the M1-Radice method is expected to perform better than the M1-SpEC is regions of high scattering opacity and low absorption opacity~\cite{Radice:2021jtw}. The implicit MC scheme used in our simulations is for its part active in any region where the absorption timescale is shorter than the simulation time step. Accordingly, we will now investigate in more details the distribution of neutrinos throughout the dense regions of the remnant.

To begin, we consider the fluid-frame energy density of neutrinos on horizontal and vertical slices on Fig.~\ref{fig:jnu}. Here, we use the M1-Radice simulation as reference considering that, a priori, it is the most accurate of the M1 methods (we will see below that getting clean information from the MC simulations is more difficult). For reference, the maximum energy density of the fluid at the beginning of the simulation (including rest mass energy density), at the center of the most massive neutron star, is $J_{\rm fl}\sim 8\times 10^{36}\,{\rm erg/cc}$. Neutrinos are mainly created in the dense, hot regions of the remnant. The $\bar \nu_e$ species dominates, with energy density rising to $O(10^{-3})$ of the fluid energy density. Heavy-lepton neutrinos have slightly lower energy density, while $\nu_e$ production is relatively suppressed. The same hierarchy is visible in shocked regions of the tidal tail, and in the free-streaming regions. A comparison with Figs.~\ref{fig:vis_hor}-\ref{fig:vis_ver_mc} shows that the neutrino energy follows the fluid temperature, as expected. In particular, at this early time in the post-merger evolution, cold dense regions remain within the remnant where very few neutrinos are created or able to diffuse into.

\begin{figure*}
\includegraphics[width=0.31\textwidth]{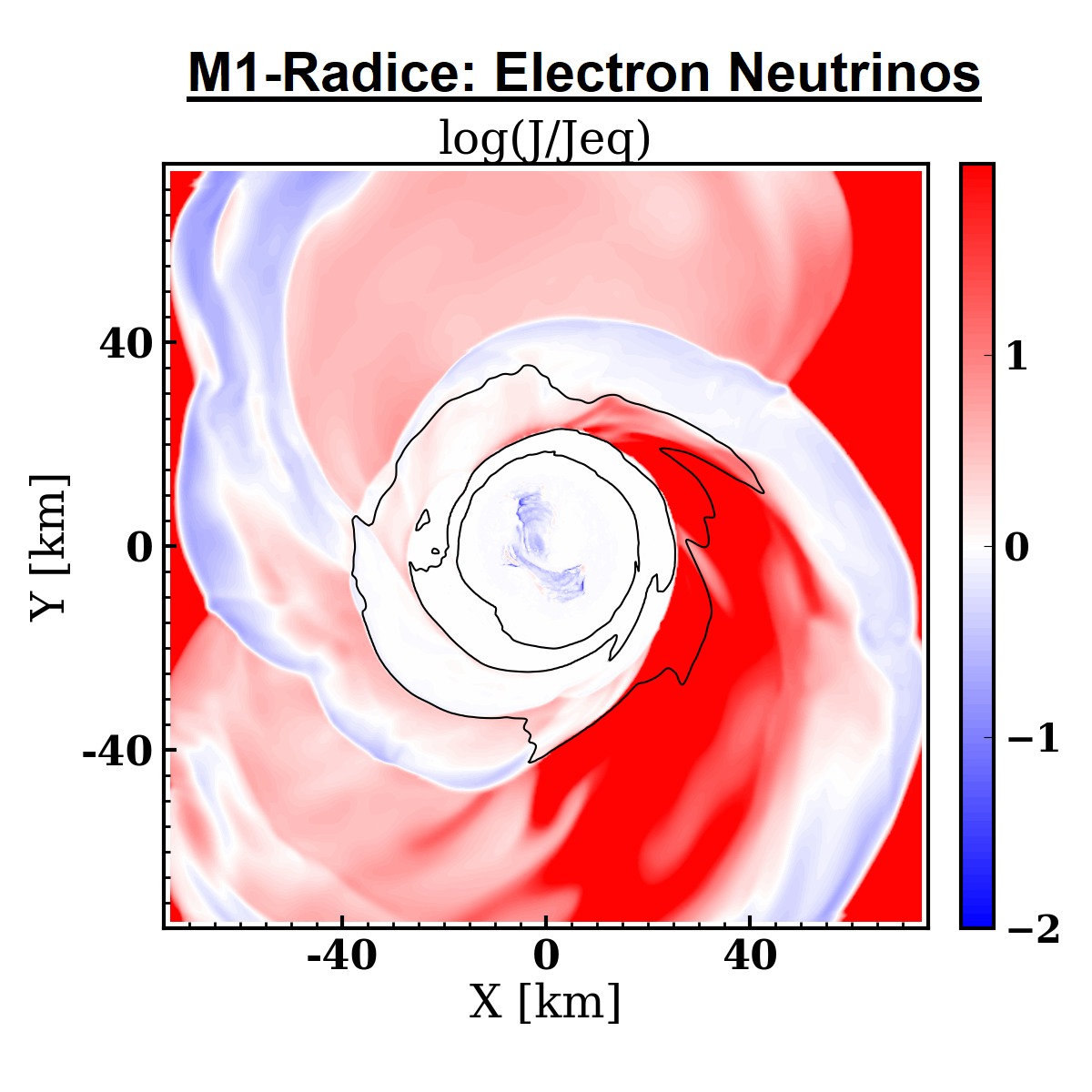}
\includegraphics[width=0.31\textwidth]{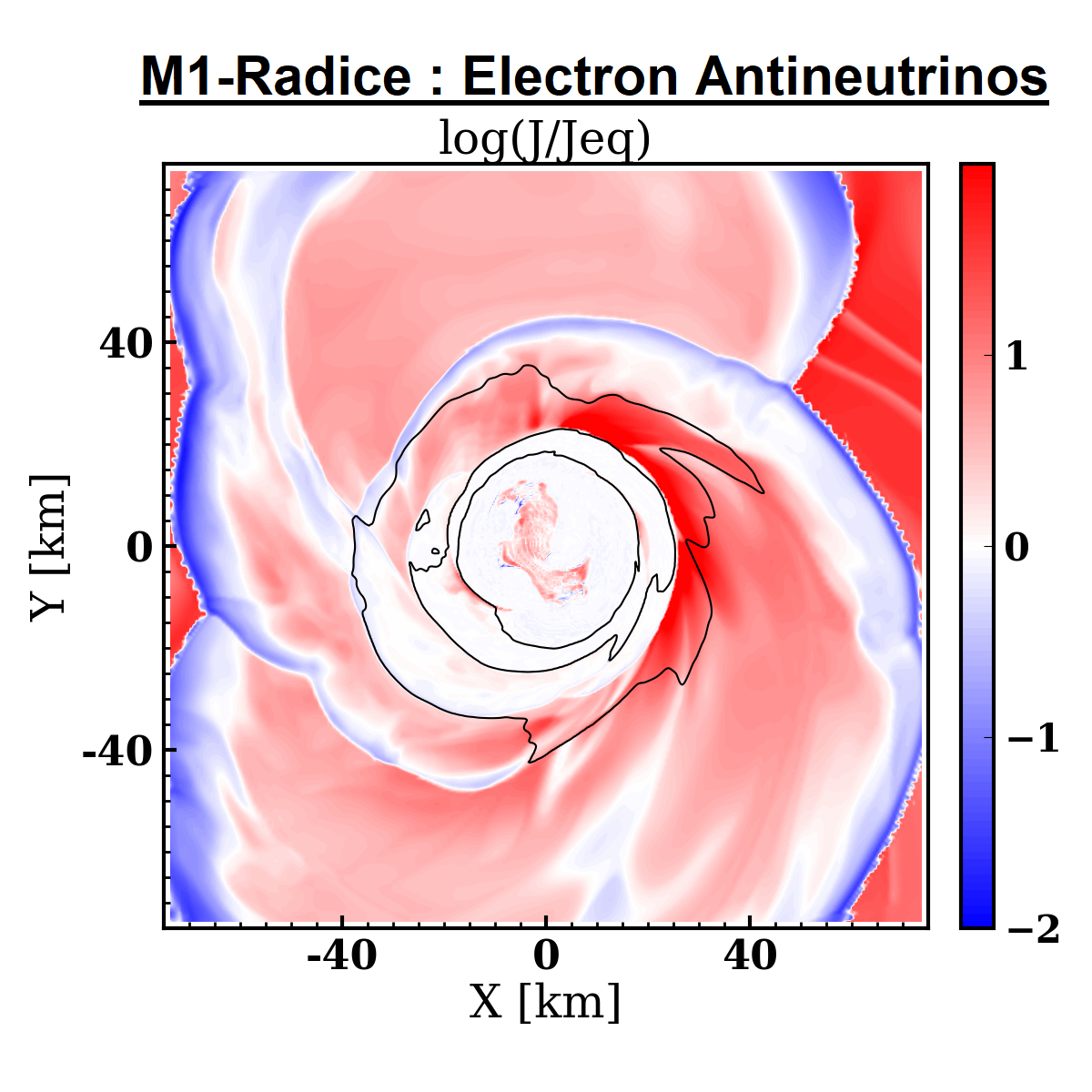}
\includegraphics[width=0.31\textwidth]{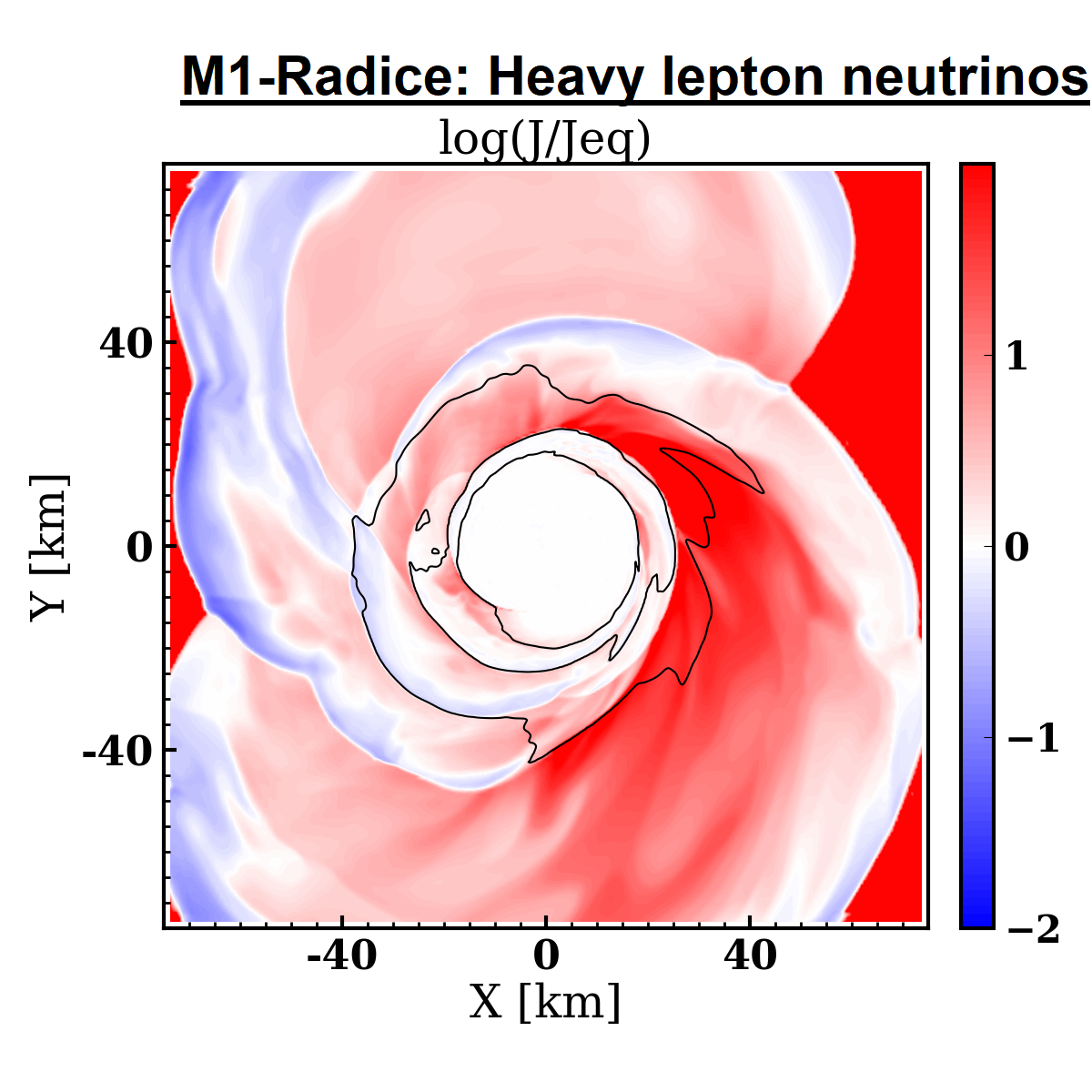}\\
\includegraphics[width=0.31\textwidth]{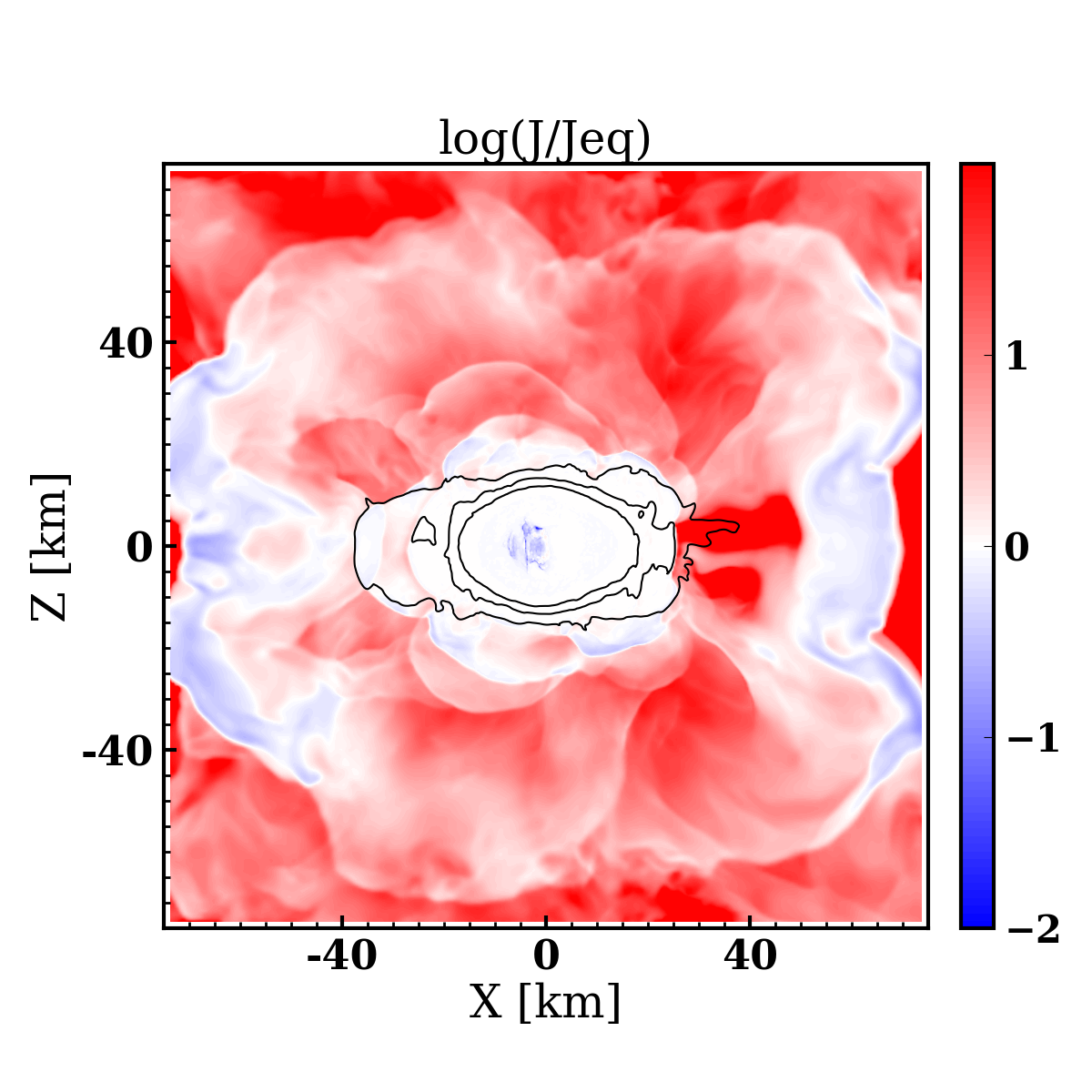}
\includegraphics[width=0.31\textwidth]{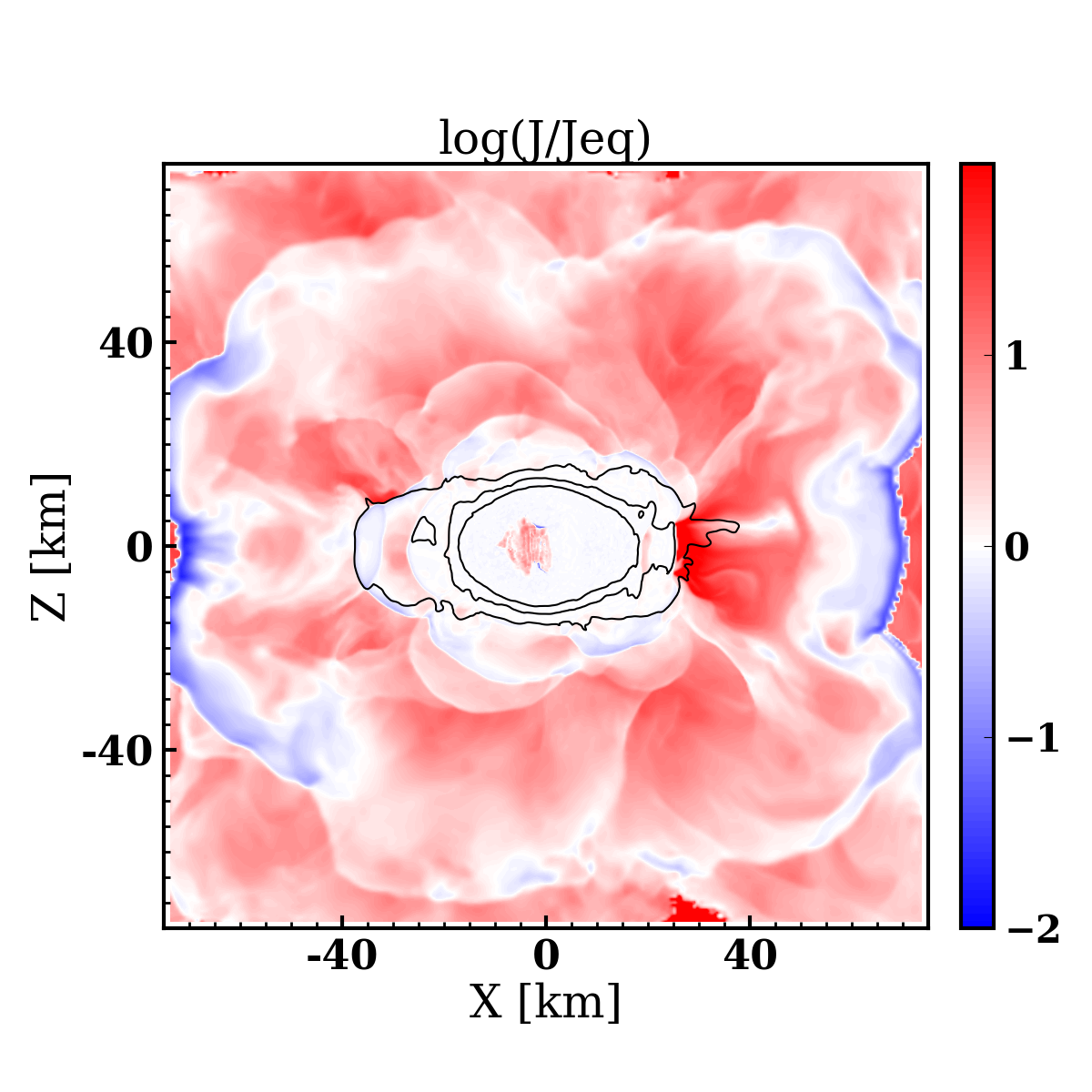}
\includegraphics[width=0.31\textwidth]{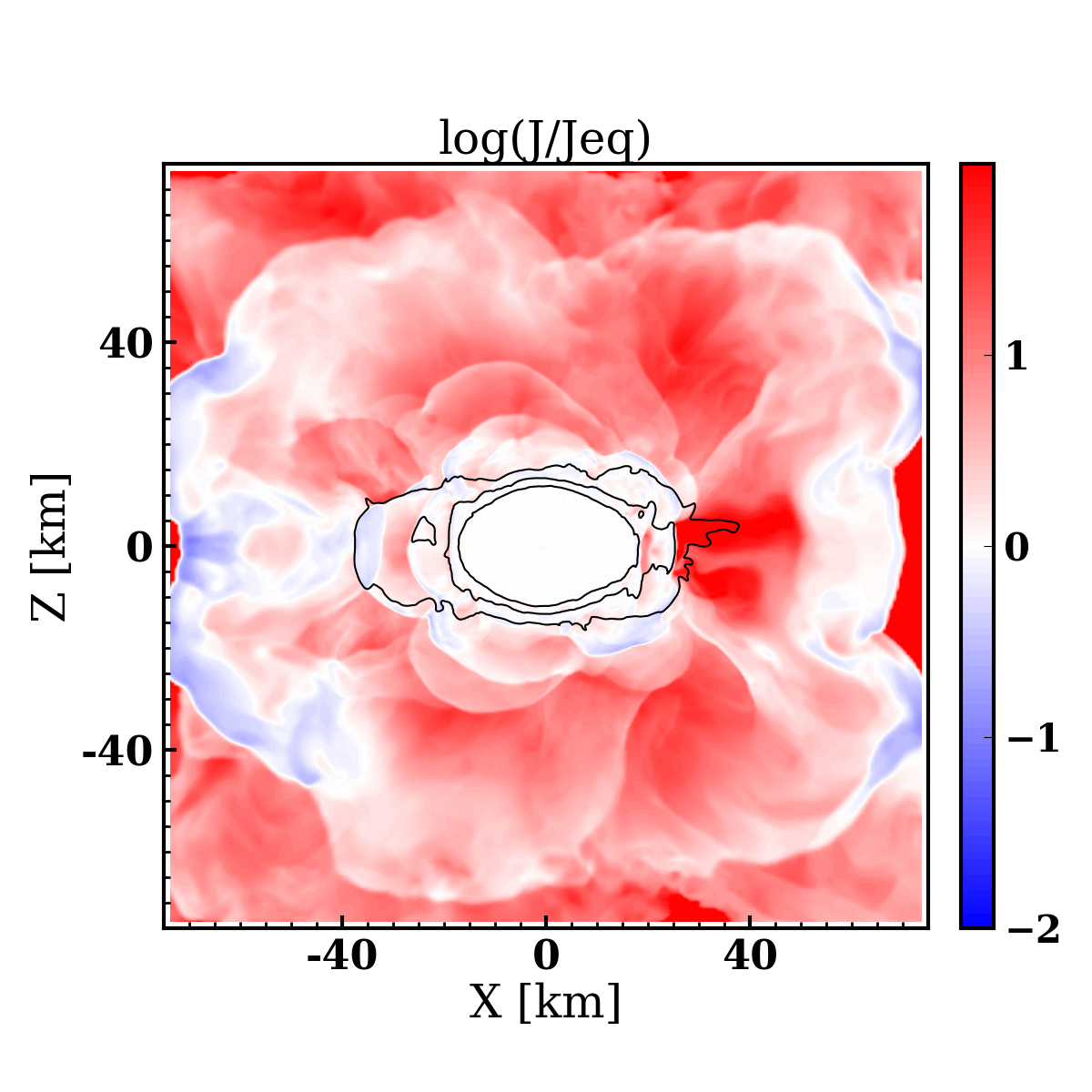}\\
\includegraphics[width=0.31\textwidth]{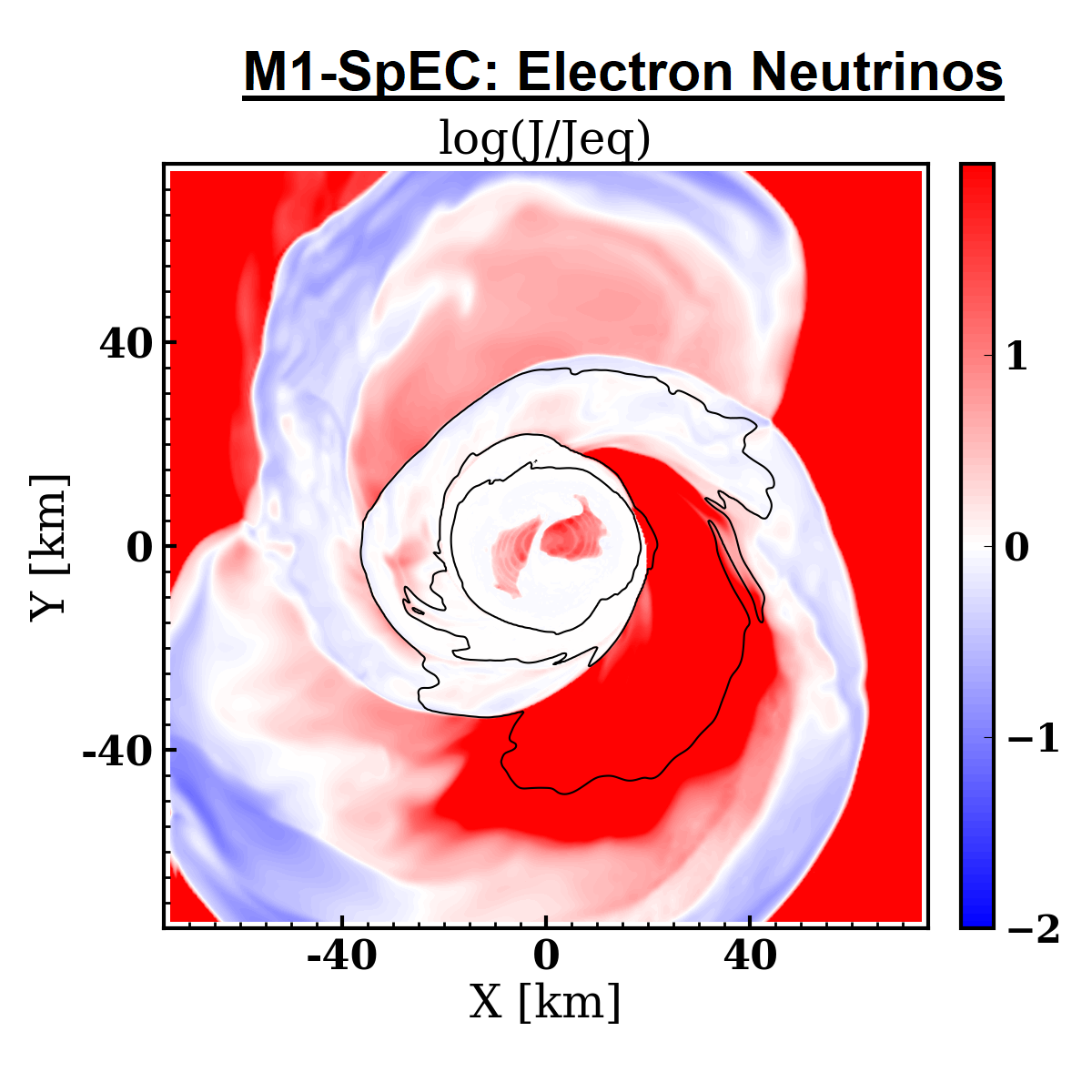}
\includegraphics[width=0.31\textwidth]{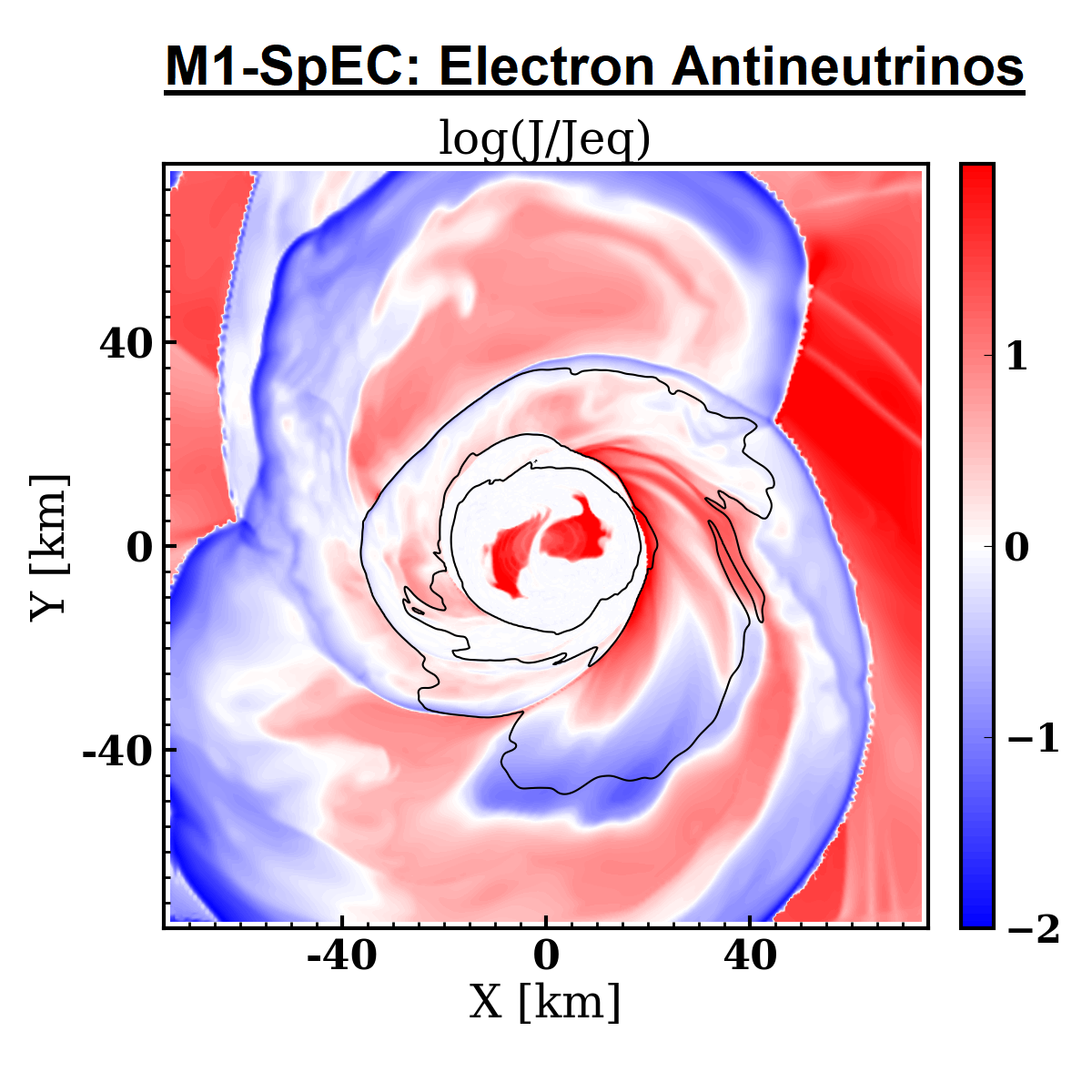}
\includegraphics[width=0.31\textwidth]{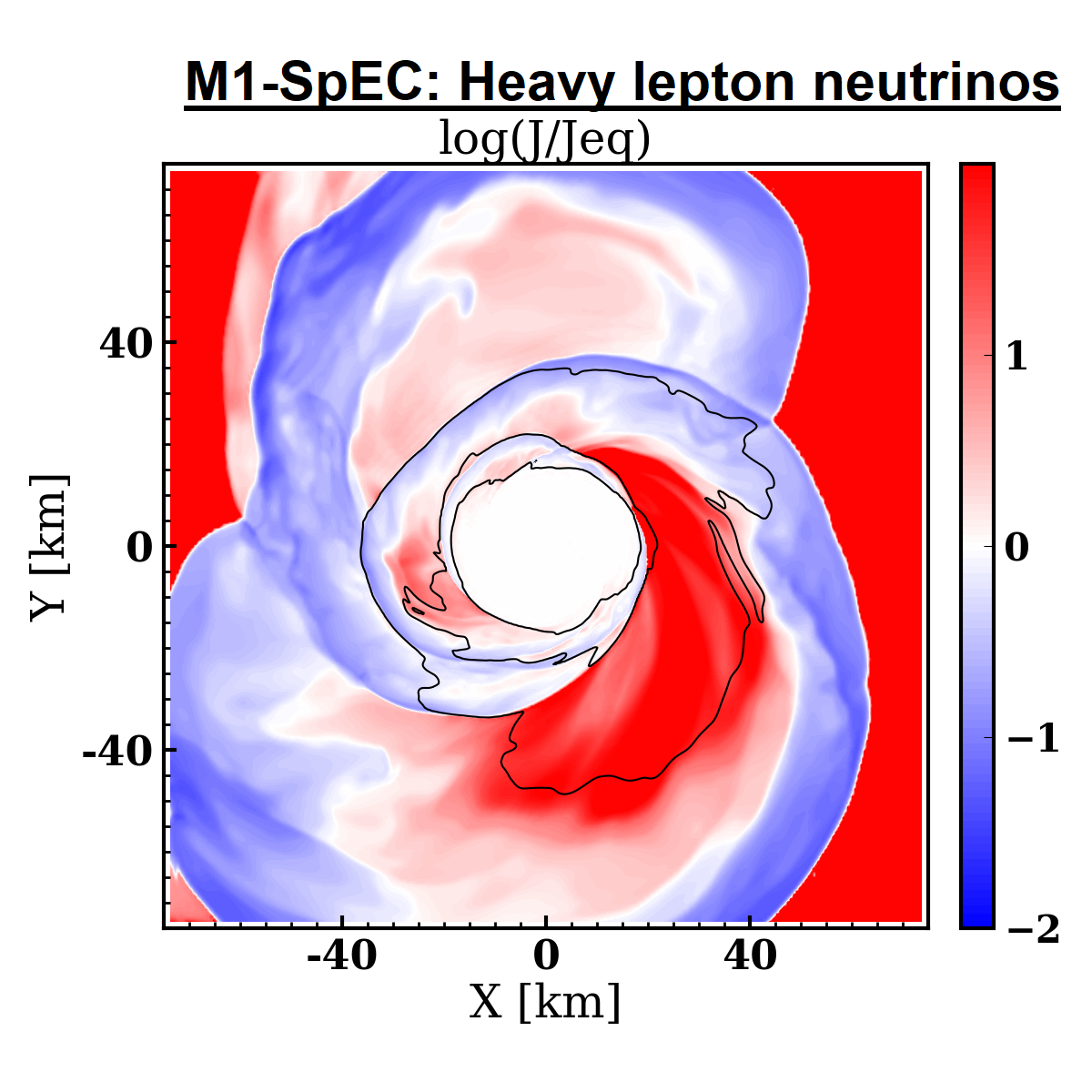}\\
\includegraphics[width=0.31\textwidth]{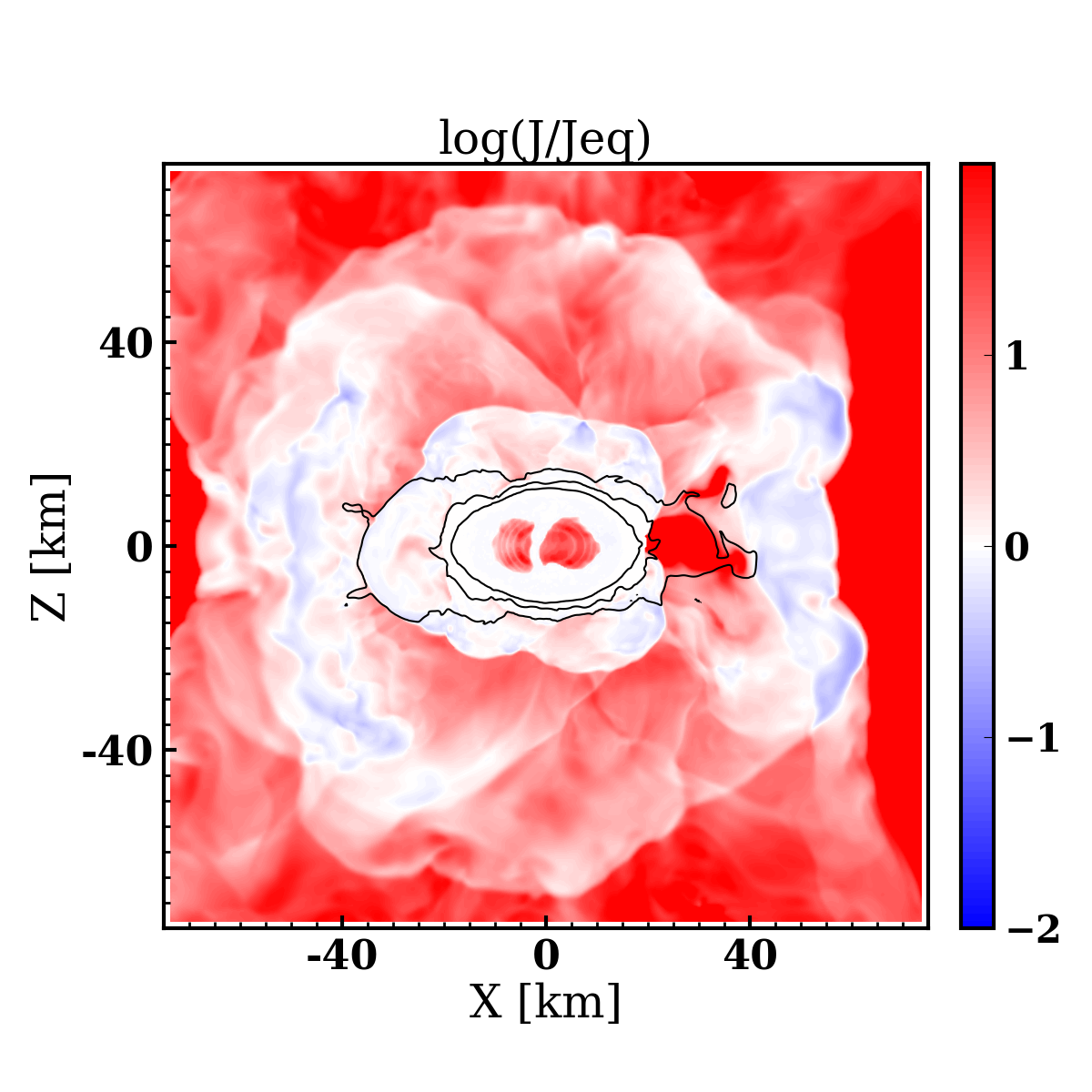}
\includegraphics[width=0.31\textwidth]{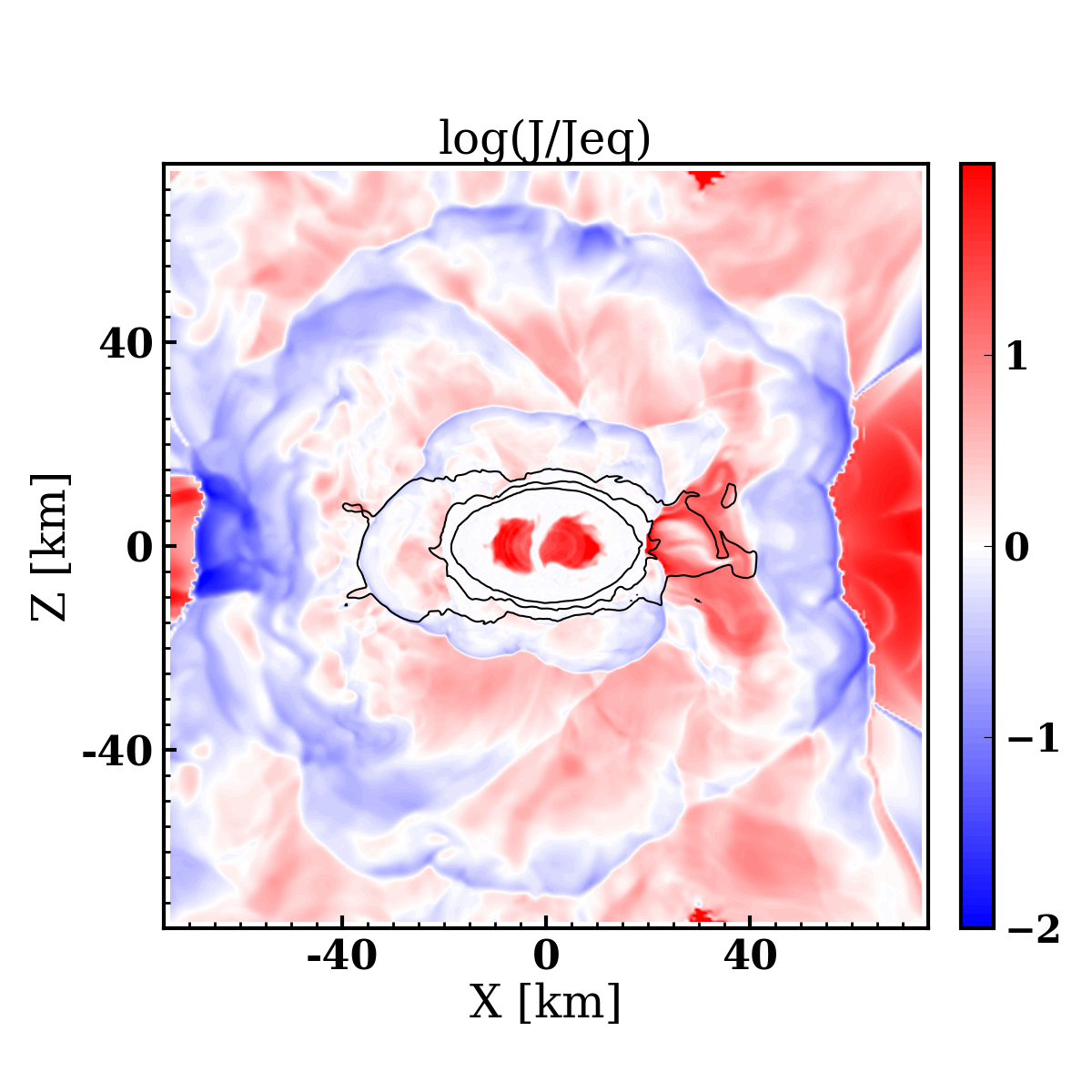}
\includegraphics[width=0.31\textwidth]{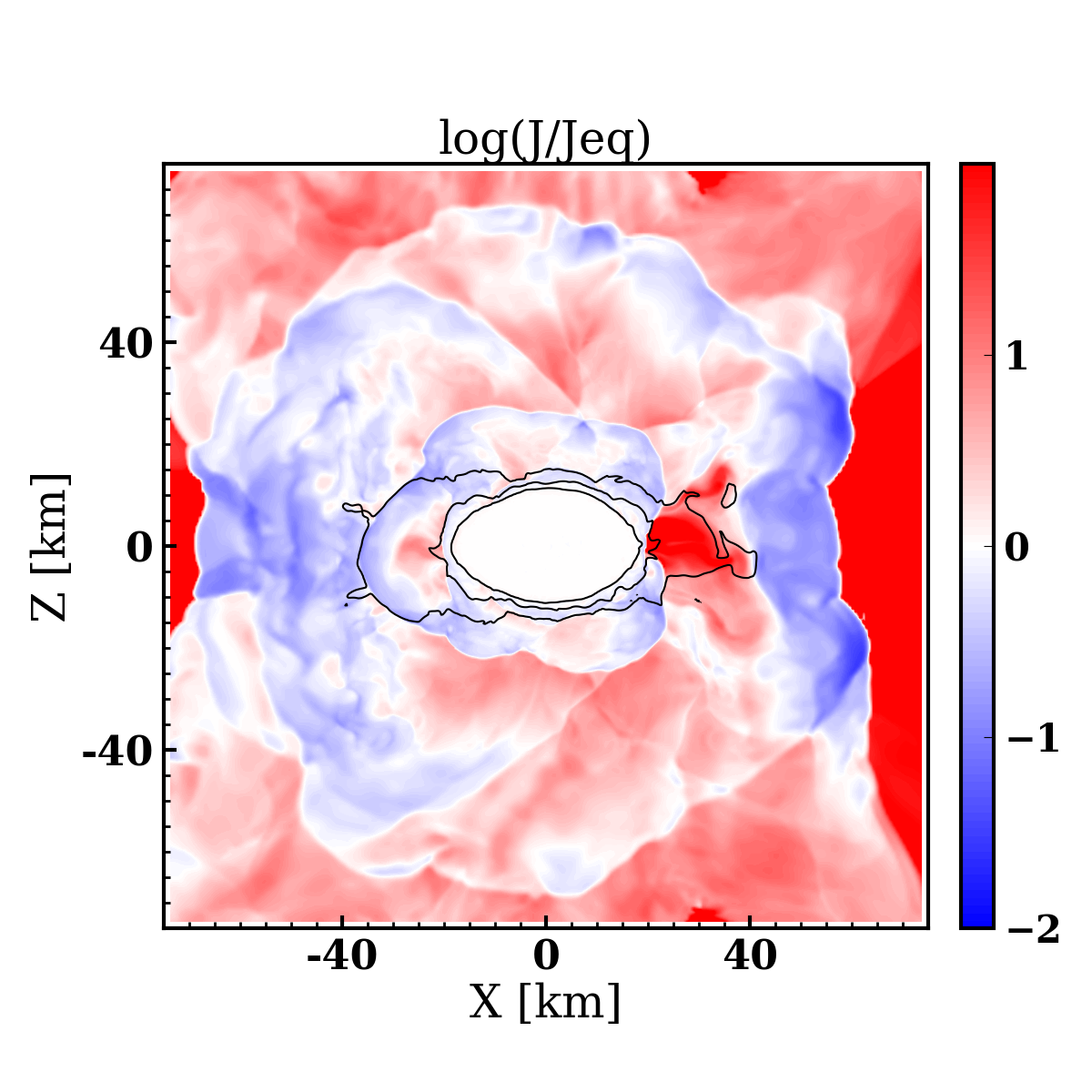}
 \caption{Logarithm of the ratio of the energy density of neutrinos in the fluid frame to the energy density that neutrinos would have if they were in equilibrium with the fluid. We show results for $\nu_e$ (left), $\bar\nu_e$ (center) and $\nu_x$ (right), in a horizontal (top) and vertical (bottom) slice of the remnant and for the M1-Radice simulation (first two rows) and M1-SpEC simulation (last two rows). Black contours correspond to densities of $10^{11,12,13}\,{\rm g/cc}$. In white regions, neutrinos are close to being in equilibrium with the fluid. Blue regions are semi-transparent emission region, while red regions are irradiated by other regions of the simulations.}
\label{fig:jratio}
\end{figure*}

\begin{figure*}
\includegraphics[width=0.31\textwidth]{Jratio_Nue_Hor_Radice}
\includegraphics[width=0.31\textwidth]{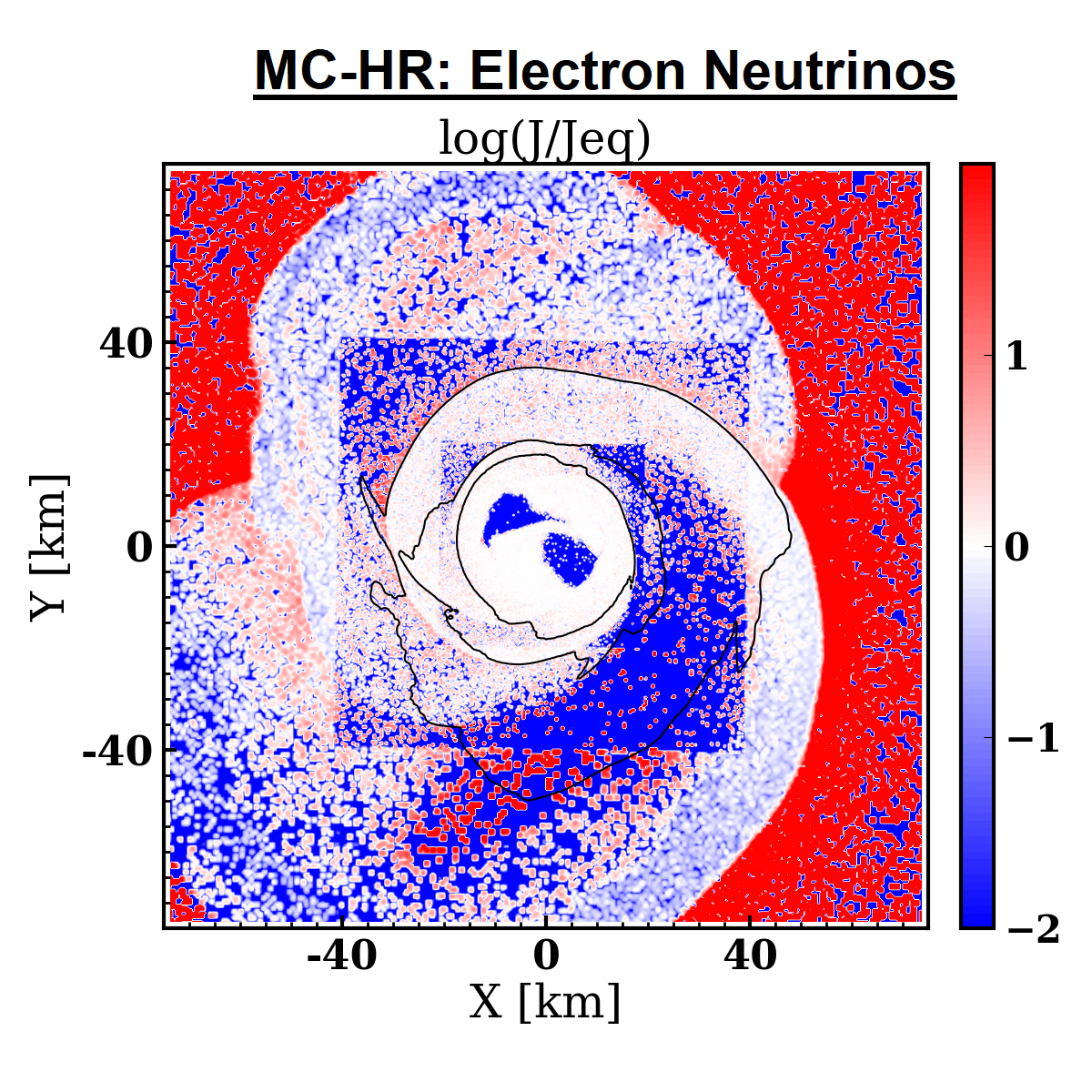}
\includegraphics[width=0.31\textwidth]{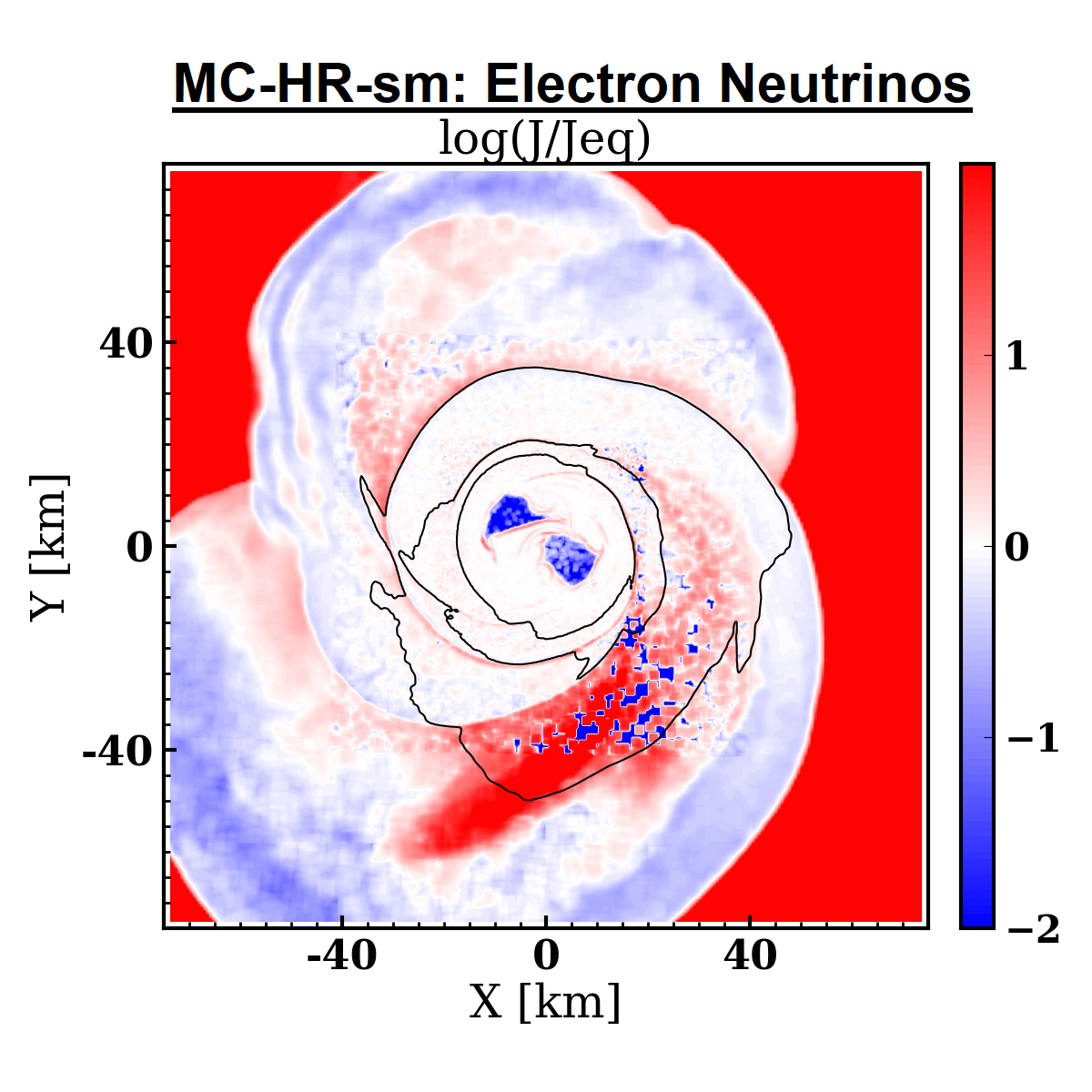}\\
\includegraphics[width=0.31\textwidth]{Jratio_Nux_Hor_Radice}
\includegraphics[width=0.31\textwidth]{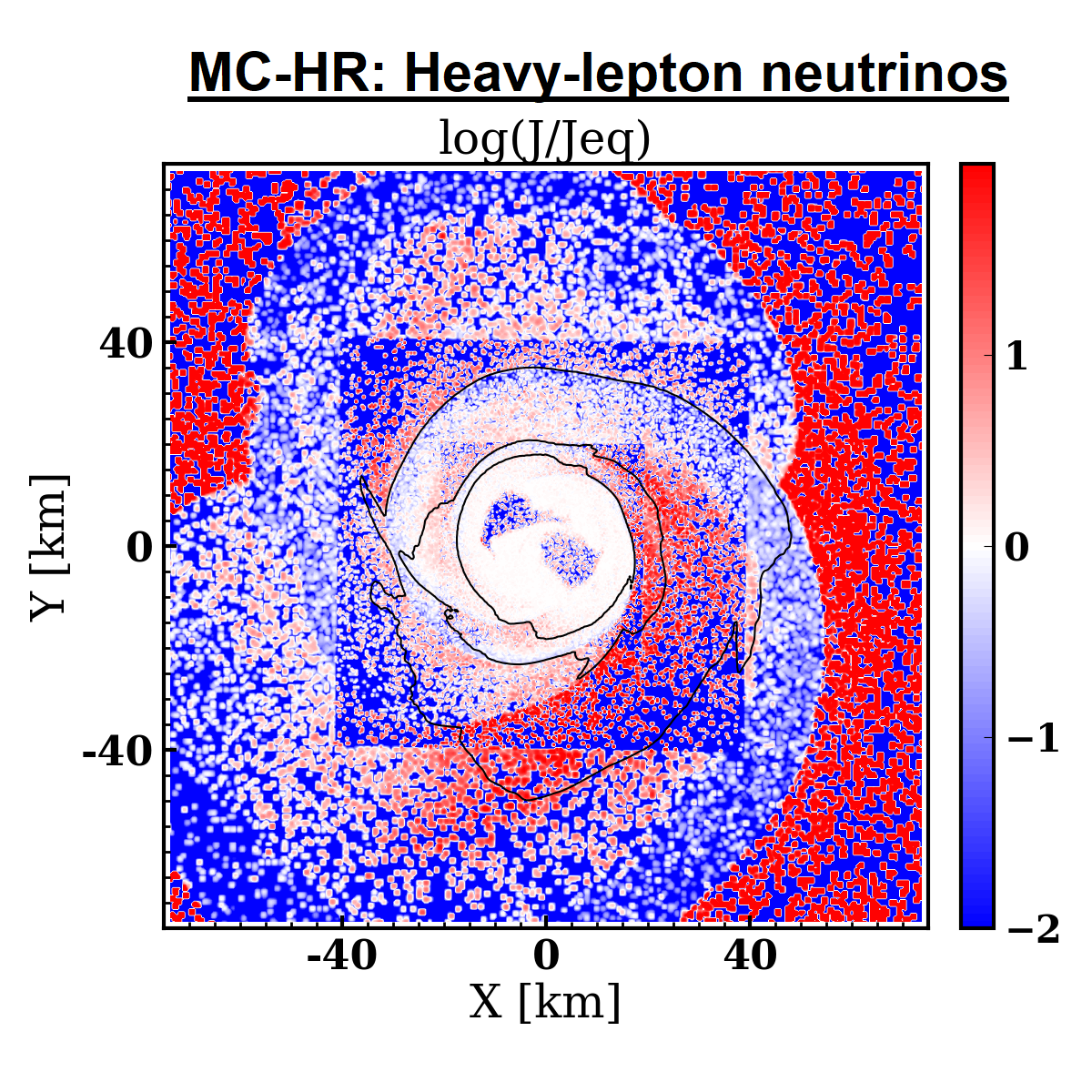}
\includegraphics[width=0.31\textwidth]{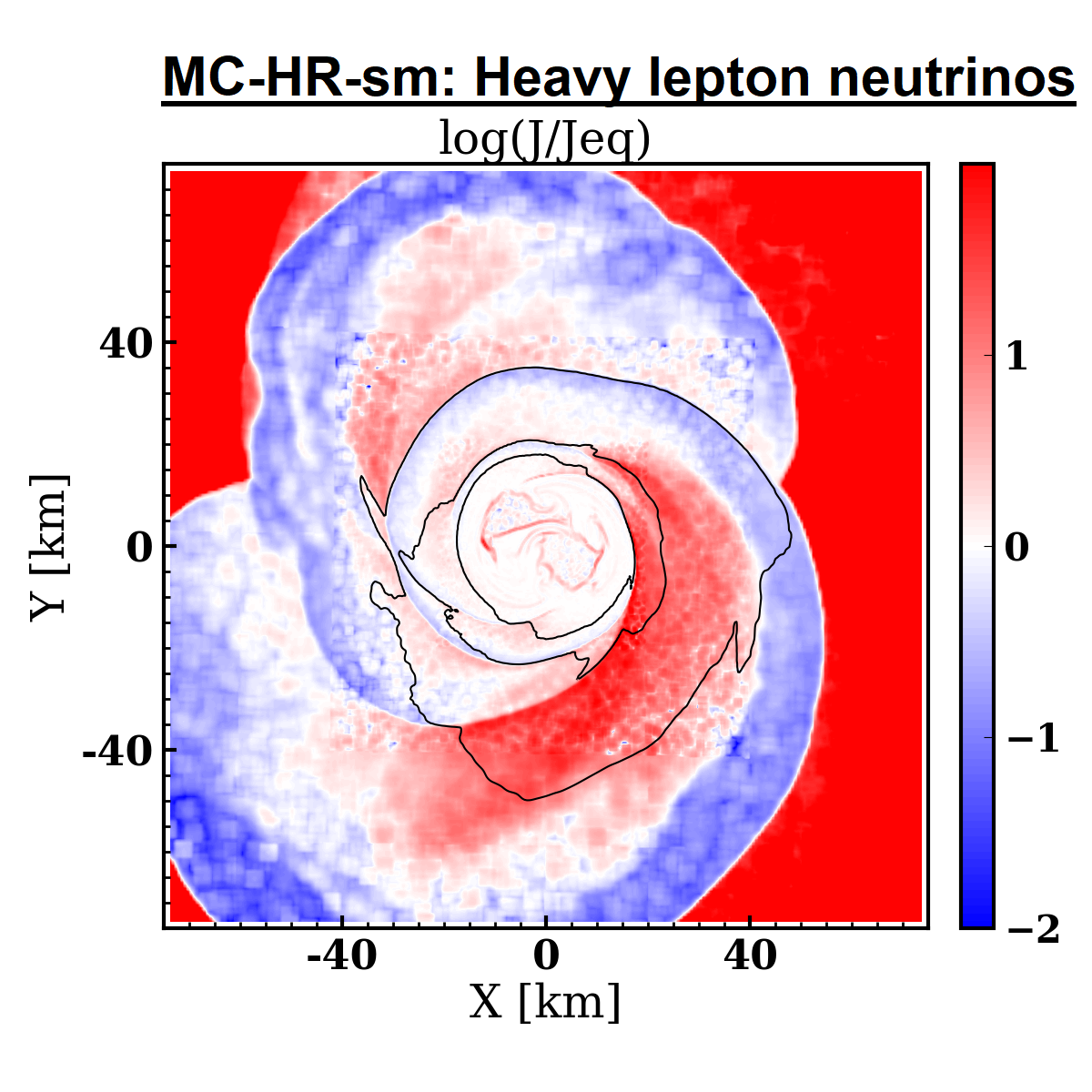}
 \caption{Same as Fig.~\ref{fig:jratio} but for simulation M1-Radice (left) and MC-HR (center). Note that for the MC simulation, we calculate the energy density from the packets existing during a single step $dt\sim 25\,{\rm ns}$ in a single grid cell, which introduces large sampling noise. In particular, saturated colors (dark red, dark blue) are typically associated with cells with $0-1$ packets, making them entirely unreliable. The right panel shows the data from the central panel smoothed over $5^3$ cells, to limit the impact of that noise. The top row shows results for $\nu_x$, and the bottom row for $\nu_x$. The boundary between refinement level is clearly visible on the MC plots because coarser cells have lower sampling noise than immediately neighboring finer cells.}
\label{fig:jratio_mc}
\end{figure*}

To compare different radiation transport methods, we find it easier to consider the ratio of the neutrino energy density in the fluid frame to the expected energy density in equilibrium. Indeed, this tends to show more clearly regions where neutrinos are in equilibrium ($J\approx J_{\rm eq}$), regions irradiated by hotter parts of the remnants ($J>J_{\rm eq}$), and (semi-)transparent regions from which neutrinos are partially free-streaming away ($J<J_{\rm eq}$). These appear, respectively, as white, red, and blue regions on Figs~\ref{fig:jratio}-\ref{fig:jratio_mc}. We note however that $J\approx J_{\rm eq}$ is not a sufficient condition to guarantee that the neutrinos are trapped and in thermal equilibrium with the fluid; it is only a necessary condition.

In Fig.~\ref{fig:jratio}, we compare the M1-Radice and M1-SpEC simulations. The M1-NuLib simulation is quite similar to M1-SpEC when considering the neutrino energy density, despite the use of different reaction rates. We will see below however that the M1-SpEC simulation has problematic behavior in the evolution of the number density at the time of merger. We note a few important regions where the M1-Radice and M1-SpEC schemes disagree. First, the neutrinos are generally closer to equilibrium in the hot tidal tails in the M1-Radice simulation than in the M1-SpEC simulation. This is visible as blue regions in the tidal tail for M1-SpEC, especially for the heavy-lepton neutrinos and electron antineutrinos. Second, dense regions remain in equilibrium down to lower density in the M1-Radice simulations. Third, cold regions inside the neutron star are more strongly irradiated in the M1-SpEC simulation. The last two could naturally be understood as the inability of the M1-SpEC simulation to properly advect neutrinos along with the fluid in dense regions. The first difference is more puzzling. We can estimate the length scale over which neutrinos are expected to equilibrate in the tidal tail as $L_{\rm eq}\approx \sqrt{\kappa_a (\kappa_a + \kappa_s)}$ with $\kappa_{a,s}$ averaged over the equilibrium energy spectrum of neutrinos. For the conditions in the tidal tail of the M1-Radice simulations ($\rho\approx 4\times 10^{10}\,{\rm g/cc}$, $T\approx 3\,{\rm MeV}$, $Y_e\approx 0.15$ at the points where the tidal tail crosses the positive x-axis), we have $L_{\rm eq}=(20,120,2700)\,{\rm km}$ for $(\nu_e,\bar\nu_e,\nu_x)$ respectively. Considering that the tidal tail has width $\sim 20\,{\rm km}$ and height $\sim 50\,{\rm km}$, we would expect that in the tidal tail the $\nu_e$ would be partially trapped, the $\bar \nu_e$ would be in the semi-transparent regime, and the $\nu_x$ would be nearly free-streaming. The numerical results are reasonably consistent with these predictions for $\nu_e$ and $\bar\nu_e$, but clearly not for $\nu_x$. In the M1-SpEC simulations, and for different conditions in the tidal tail 
($\rho\approx 5.5\times 10^{10}\,{\rm g/cc}$, $T\approx 6\,{\rm MeV}$, $Y_e\approx 0.21$), we estimate $L_{\rm eq}=(4,17,180)\,{\rm km}$ instead. We observe trapped $\nu_e$, partially trapped $\bar\nu_e$, and free streaming $\nu_x$, in better agreement with expectations. One possible explanation, considering that this issue is only observed in $\nu_x$ in the tidal tail, is that the equilibrium energy density in the M1-Radice tail is a factor of $\sim 15$ lower than in the M1-SpEC tail, and thus that energy density may be more easily reached through irradiation from other regions of the simulation -- i.e. this might be an example of the fact that $J\sim J_{\rm eq}$ can be true even if neutrinos are not trapped in a given region. 

Finally, we turn to Fig.~\ref{fig:jratio_mc}, which performs the same comparison for the M1-Radice and MC-HR simulations. We only show $\nu_e$ and $\nu_x$ here (the $\bar\nu_e$ show a level of agreement between simulations comparable to the $\nu_e$). We note that getting an estimate of the energy density of neutrinos in the MC simulations is more difficult than for the M1 simulations. When computing moments in MC simulations, we currently integrate the contribution of all packets over a 4-volume corresponding to a grid cell of proper spatial volume $\Delta V$ evolved over a time $\Delta \tau$, then divide the result by $\Delta \tau$ to estimate the energy density (using the proper time of an inertial observer, $\Delta \tau = \alpha \Delta t$, with $\Delta t$ the elapsed simulation time). We keep track of moments of the distribution function of neutrinos integrated over the latest simulation time step, which at $2.5\,{\rm ms}$ post-merger leads to $\Delta t\sim 25\,{\rm ns}$. Without additional averaging, this leads to significant shot noise everywhere but in the densest and hottest regions of the remnant (middle panels of Fig.~\ref{fig:jratio_mc}; we note in particular that the grid structure is clearly visible on these plots, because on coarser grids we average over a larger volume, thus lowering the MC noise). To improve on this issue, we average the MC results over a $5\times 5 \times 5$ group of cells (right panels of Fig.~\ref{fig:jratio_mc}). This improves the accuracy of our prediction in lower-density regions, at the cost of inaccuracies in regions where the energy density varies rapidly; e.g. at the boundary between hot and cold regions inside the remnant. We see that the MC simulation agrees quite well with the M1-Radice simulation in dense regions. In particular, the region where neutrinos are in equilibrium with the fluid extends to similar densities, and the cold regions of the remnant are devoid of $\nu_e$ (i.e. blue on the figure). This tends to confirm that the differences observed in these regions between M1-Radice and M1-SpEC are indeed due to the limitations of the M1-SpEC simulations in dense regions. In the hot tidal tail, the MC simulations behave closer to the M1-SpEC simulation. As discussed in the previous paragraph, the behavior of the MC simulations is also what we might intuitively expect from the calculation of the equilibration length $L_{\rm eq}$ of neutrinos. As the MC simulation also has a hotter tidal tail than the M1-Radice simulation, this does not however provide new information on the origin of the apparently-trapped $\nu_x$ neutrinos in M1-Radice.

Fig.~\ref{fig:jratio_mc} also provides a good illustration of some of the strength and weaknesses of the MC code. The fact that, in the absence of smoothing, the shot noise in the calculation of the energy density is so large in many regions of the simulation is a good reminder that directly estimating the distribution function of neutrinos (or its moments) from an MC simulation can be very inaccurate. This can be problematic for example when attempting to calculate $\nu\bar\nu$ annihilation rate in polar regions, where very few packets are located, or in any region where the neutrino pressure may become dynamically important. The impact of these errors when using MC as a closure for M1 schemes~\cite{Izquierdo:2023fub} at realistic number of packets remains an interesting open question for hybrid methods as well. On the other hand, the very simple smoothing performed on this figure illustrates why MC methods with relatively low packet numbers work quite well in binary merger simulations. Indeed, for the output of simulations to be minimally impacted by shot noise we only need the distribution function (or its relevant moments) to be well-modeled by the neutrino packets averaged over a timescale $\tau_{\rm coupling}$ over which the fluid variables are modified by neutrino-matter interactions. Going from the middle to the right panels of Fig.~\ref{fig:jratio_mc} is similar to estimating the neutrino energy density by averaging over a time $125\Delta t \approx 3\,{\rm \mu s}$, which is still much shorter than $\tau_{\rm coupling}$ ($\sim$ ms or more) except in regions where neutrinos are trapped, where the MC sampling noise is generally lower even without smoothing. Averaging the neutrino energy density over $1\,{\rm ms}$ would in theory reduce shot noise by an additional factor of $\sim 20$, assuming that the noise scales as $N^{-1/2}$ when using $N$ packets. The same scaling also explains why getting ten percent-level errors from MC simulations is cost-effective, but reducing that error by multiple orders of magnitude would be prohibitively expensive.

\begin{figure}
\includegraphics[width=0.95\columnwidth]{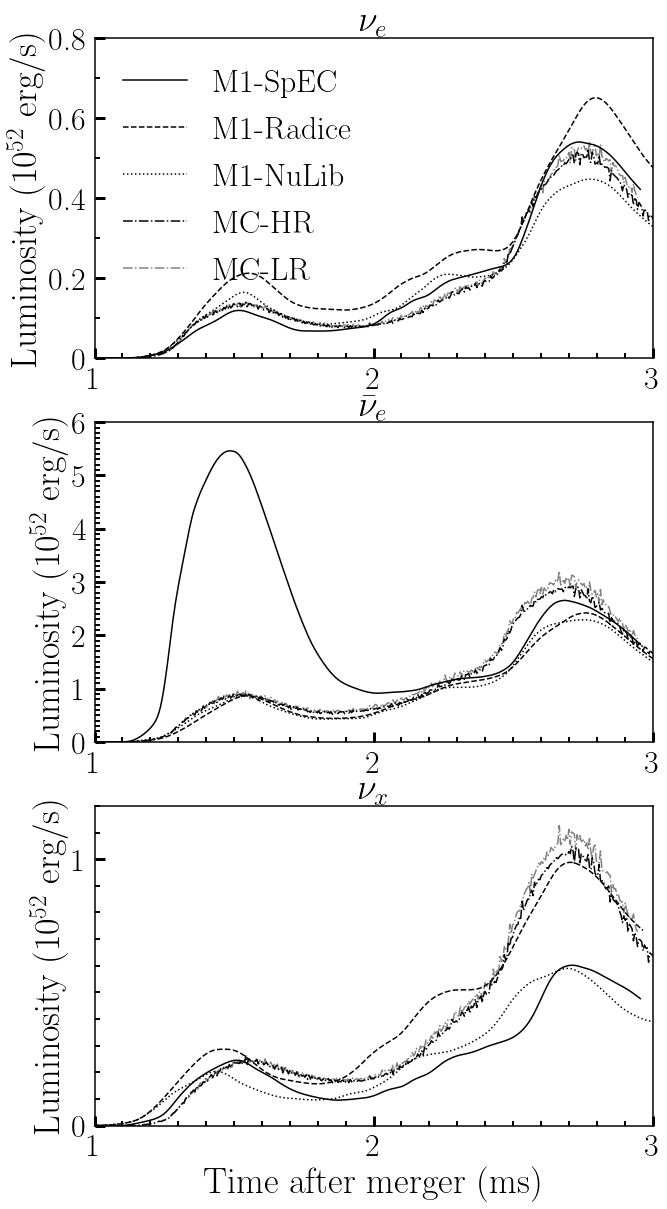}
 \caption{Luminosity of neutrinos for all simulations, for $\nu_e$ (top panel), $\bar \nu_e$ (middle panel), and each species of $\nu_x$ individually (bottom panel), measured at the outer boundary of our computational domain (a box of side $\sim 320\,{\rm km}$). Note the different scales for each species.}
\label{fig:lum}
\end{figure}

On Fig.~\ref{fig:lum} we report the luminosity of neutrinos in the $(1-3)\,{\rm ms}$ post-merger for which data is available for all simulations. The most noticeable feature here is obviously the large burst of $\bar\nu_e$ emission from simulation M1-SpEC over a $0.5\,{\rm ms}$ period following the merger. We already noted that the high-density regions of the remnant erroneously neutronize in this simulation; this is another indication that the early post-merger composition evolution of the high-density regions is not well captured in M1-SpEC. A more detailed analysis of the neutrino distribution in post-processing shows that during a $\sim 0.5\,{\rm ms}$ period around merger, the M1-SpEC simulation created large numbers of low-energy $\bar \nu_e$ in the dense, cold regions of the merging neutron star cores, and that lepton number conservation was not satisfied during that time period. These low-energy $\bar\nu_e$ are the source of the extra emission of $\bar \nu_e$ at early times, but also of the decrease in the electron fraction observed in dense regions, as the conservation of lepton number was satisfied during their partial reabsorption into the remnant. The exact source of this violation of lepton number conservation is unclear, but likely due to a combination of the inaccuracies of the M1-SpEC scheme in dense cold regions and of corrections applied to the evolution of the number density to guarantee that the average energy of neutrinos remain between $0.01\,{\rm MeV}$ and $1000\,{\rm MeV}$. This is of course a clear indication that the results of the other simulations should be trusted over the M1-SpEC simulation in any region impacted by this problem.

The second main differences between simulations is that both M1-NuLib and M1-SpEC underproduces $\nu_x$ by the end of the evolution. Considering that these two simulations use the same inaccurate methods for the implicit solve in the two-moment equation, that these inaccuracies are expected to be particularly important for regions of high scattering opacities but low absorption opacities, and that $\nu_x$ neutrinos specifically go through such regions as they exit the remnant, these differences can be reasonably attributed to the limitations of that approximate implicit solve. We note that a similar difference in the luminosity of $\nu_x$ had also been noted as the main difference between M1 and MC in our earlier comparison of transport schemes in a low mass binary neutron star system in~\cite{Foucart:2020qjb}. Our results here indicate that this was likely an effect of the approximations made in the implicit solver within the M1 algorithm, rather than a limitation of the M1 formalism itself.

The three simulations that we would consider most reliable in the neutrino sector (MC-HR, MC-LR, M1-Radice) are in fairly good agreement. The M1-Radice simulation produces slightly less $\bar \nu_e$ and slightly more $\nu_e$ than the MC simulations, with $\sim 20\%$ relative differences, while the two MC simulations have nearly identical neutrino luminosities. This is in this case consistent with the M1-Radice protonizing less in the dense regions (less $\bar\nu_e$ emission) and protonizing more in the outflows (more $\nu_e$ absorption).

\begin{figure}
\includegraphics[width=0.95\columnwidth]{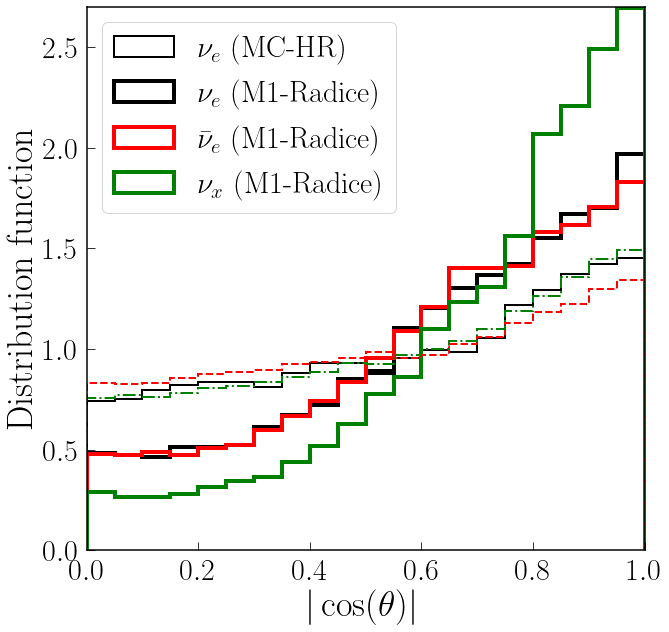}
 \caption{Distribution of neutrinos with respect to polar angle of propagation $\theta$, for each species of neutrinos. We show results for the M1-Radice (thick lines) and MC-HR (thin lines) simulations, for neutrinos leaving the computational domain during the $\sim 3\,{\rm ms}$ preceding the collapse of the neutron star in MC-HR. We measure the angle between the neutrino momentum and the z-axis for MC, and between the flux vector and the z-axis for M1. The M1 simulation has significantly larger neutrino density in the polar direction, and significantly smaller neutrino density in the equatorial direction.}
\label{fig:theta}
\end{figure}

This agreement between M1 and MC results on the total luminosity does not imply that the distributions of neutrinos in low-density regions are similar. In Fig.~\ref{fig:theta}, we show the distribution of neutrinos with respect to the angle between the neutrino momentum (or, for M1, flux) and the polar axis. We see that the M1 simulation has significantly more neutrinos moving in the polar direction, and significantly less in the equatorial direction, especially for $\nu_x$. This is a known consequence of the use of the approximate Minerbo closure, which results in neutrinos coming from different regions of the remnant colliding and forming a beam of neutrinos in the polar direction. Such an effect has also been seen for different M1 implementations in e.g.~\cite{Foucart:2020qjb,Gizzi:2021ssk}.

\begin{table}
\begin{tabular}{c|c|c|c|c}
{\bf Sim} & $N_l/N_l^0$ & $N_{l,fluid}/N_l^0$ & $N_{l,\nu}/N_l^0$ & $N_{l,outflow}/N_l^0$ \\
\hline
MC-HR & 1.003 & 1.149 & -0.146 & -0.009\\
M1-Radice & 0.985 & 1.089 & -0.104 & -0.015\\
M1-NuLib & 1.010 & 1.120 & -0.110 & -0.015\\
M1-SpEC & 0.677 & 0.932 & -0.255 & -0.289\\
\end{tabular}
\caption{Net lepton number on the grid $N_l=N_{e^-}-N_{e^+}+N_{\nu_e}-N_{\bar\nu_e}$ $2.5\,{\rm ms}$ post-merger, normalized by its initial value $N_l^0$. We also show the net lepton number of the fluid (electrons and positrons) and neutrinos ($\nu_e$ and $\bar\nu_e$) on the grid, and of the neutrinos that left the grid. For exact conservation of lepton number, the second column plus the last column should be exactly 1.}
\label{tab:Yl}
\end{table}

Finally, we can look more carefully at lepton number evolution and conservation in the simulations, using information from the neutrino sector. In Table~\ref{tab:Yl}, we show the total lepton number $N_l=N_{e^-}-N_{e^+}+N_{\nu_e}-N_{\bar\nu_e}$ in all simulations $2.5\,{\rm ms}$ post-merger, normalized by its initial value. We also show how that lepton number is divided between the neutrino and fluid sector, as well as the net lepton number that exited the grid through neutrino emission. After $2.5\,{\rm ms}$, the lepton number lost to matter outflows is negligible. We see that the MC-HR simulation has relatively small violations of lepton number conservation, at the level of $0.5\%$ of the initial number of leptons. The M1-Nulib simulation performs similarly in that respect, and the M1-Radice simulation slightly worse ($3\%$ lepton number violation). This is better than for schemes that do not evolve the number density of neutrinos~\cite{Foucart:2016rxm}, though still a measurable error over the course of the entire simulation and likely indicative of the fact that inaccuracies created by the joint evolution of the number and energy density of neutrinos occasionally lead to unphysical neutrino energies and corrections applied to the number density (as the M1 schemes are exactly lepton-number conserving up to such corrections). It is maybe surprising that the MC simulation shows better lepton number conservation than most M1 simulations, despite using a scheme that only conserve lepton number up to statistical noise in neutrino-matter interactions. For comparision, errors in the conservation of the total baryon number are $O(10^{-4})$ or less. As already mentioned, the M1-SpEC simulation has large violations of the lepton number conservation, to an unacceptable level in dense regions. Another notable difference between the MC-HR simulation and the M1-NuLib and M1-Radice simulation is that the former has $\sim 50\%$ more $\bar\nu_e$ neutrinos within the remnant than the latter. All simulations on the other hand have similar numbers of $\nu_e$ (and an order of magnitude less $\nu_e$ than $\bar\nu_e$). It is unclear whether such differences can explain the more rapid collapse of the neutron star in MC simulations, or if instead they are a consequence of the more rapid evolution of these simulations towards collapse... or if the larger number of neutrinos in MC is simply due to the fact that the MC code is energy-dependent and thus captures more accurately the evolution of low-energy neutrinos within the remnant.

\subsection{Cost of simulations}

In order to determine how convenient any of these methods is to perform surveys of the binary parameter space, it is also useful to compare computational costs. These simulations were performed on the Plasma cluster at UNH and on the Wheeler cluster at Caltech, which have similar performance for our simulations (at the $\sim 20\%$ level). We report costs from the beginning of the simulation to $3\,{\rm ms}$ post-merger. The cheapest simulations are the pure hydrodynamics simulation and the MC-LR simulation, with costs of $194k$ and $211k$ CPU-hrs respectively. The simulations using the cheaper but less accurate implicit time stepping for the two moment scheme (M1-SpEC, M1-NuLib) as well as the high packet count MC simulation MC-HR are $\sim 50\%$ more costly ($295k$, $309k$, and $288k$ respectively). This is consistent with our previous findings that our old two-momenst scheme and our MC scheme with $\sim 10^8$ packets per species have comparable computational costs~\cite{Foucart:2020qjb}. Finally, the two-moments simulation using the more accurate but more expensive implicit time stepping of~\cite{Radice:2021jtw} (M1-Radice) costs about twice as much as the pure hydrodynamics simulation ($412k$). We note of course that the slow convergence of Monte-Carlo methods means that significantly improving the accuracy of the Monte-Carlo simulations would inevitably require simulations much more expensive than the two-moments simulations, and that the cost of the M1-Radice might be impacted by details of the implementation of the implicit solver that have not been fully optimized in SpEC so far. On the other hand, while the two-moment scheme converges faster than the Monte-Carlo scheme, it does not converge to the correct solution to the transport equations.

\section{Discussion}

In this manuscript, we performed simulations of a single binary neutron star merger system that forms a compact remnant with expected lifetime of $O(10\,{\rm ms})$, using a range of state-of-the art neutrino transport schemes as well as two sets of neutrino-matter interactions. We find that for this metastable system, small differences in the evolution of dense regions right after merger lead to significant difference in the time required for the system to collapse to a black hole. Simulations using an energy-dependent, low-resolution Monte-Carlo transport collapse after $\sim 4\,{\rm ms}$, while the simulation using a gray approximate two-moment scheme with similar reactions rates collapses after $\sim 8.5\,{\rm ms}$ -- in reasonable agreement with preexisting simulations of a similar system with a two-moment scheme by Sekiguchi {\it et al}~\cite{Sekiguchi:2016}. These differences can be contrasted with the good agreement of the evolution of dense regions between MC and two-moment algorithms for lower mass remnants found in~\cite{Foucart:2020qjb}. We note however that the collapse time to a black hole for these metastable systems is also known to be quite sensitive to grid resolution and magnetic field evolution (see e.g.~\cite{Mosta:2020hlh} for similar effects in a system with a $\sim (20-30)\,{\rm ms}$ collapse time), and may be seen mainly as a confirmation of the high sensitivity of such systems to small changes in the evolution of dense regions.

Besides this very visible difference between MC and two-moment simulations, we find that all neutrino transport used here capture the qualitative effects of neutrino-matter interactions on important observables, with quantitative differences between simulations at the level of $(10-30)\%$ relative uncertainties for the global properties of the remnant and outflows and significant difference in the temperature and composition of lower-density regions, especially the polar regions and hot tidal arms. In the tidal arms, the temperature varies by about a factor of $2$ between simulations. These differences are seen when varying the transport method (MC vs M1), the detailed implementation of the moment scheme (M1-Radice vs M1-SpEC), and the interaction rates used in the simulation (M1-SpEC vs M1-NuLib). These results provide us with a useful starting point to determine whether the details of neutrino transport in simulations are important for any given use of simulation results; e.g. predicting kilonova lightcurves or r-process yields at a desired accuracy level.

Our simulations also allow us to assess more directly the impact of improvements to the M1 scheme recently proposed by Radice {\it et al}~\cite{Radice:2021jtw}. We find indications that these improvements do indeed modify the behavior of neutrinos in regions with high scattering optical depths, as predicted in~\cite{Radice:2021jtw}, and generally lead to differences with our preexisting M1 code that are often comparable to or larger than other sources of uncertainties in the neutrino transport algorithm. We note in addition that the M1-SpEC simulation has a peculiar behavior leading to the emission of low-energy $\bar\nu_e$ during merger, violating conservation of electron lepton number. Considering that it occurs in the same high density, cold regions where the M1-SpEC is expected to be least accurate, it is likely a result of the inaccuracies of that M1 scheme. It is also interesting to note that, somewhat surprisingly, even the better behaved M1 simulations (M1-NuLib, M1-Radice) at best conserve lepton number as well as the MC scheme, even though the MC scheme is only constructed to conserve lepton number up to statistical noise in neutrino-matter interactions. The $O(1\%)$ violations of total lepton number conservation observed here are not expected to be significant for the short simulations presented in this manuscript, but may be a worry when applying these methods to much longer numerical evolution without additional care to ensure lepton number conservation.

The M1-Radice and MC simulations show $\sim 20\%$ level agreements in early-time neutrino luminosities, while more significant differences are observed between those three simulations and those using the approximate implicit solve previously implemented in SpEC. Interestingly, this difference between M1-Radice and M1-SpEC also appears sufficient to explain the largest difference between M1 and MC results observed in our previous comparison of low-mass binary neutron star systems, i.e. the fact that the M1 scheme was underproducing $\nu_x$ neutrinos by up to $\sim 50\%$. Other differences between M1 and MC remain however notable: the M1 simulations result in overdensities of neutrinos in the polar regions, as was already known, and the MC simulation show higher densities of $\bar\nu_e$ within the remnant itself. The latter effect is however currently impossible to disentangle from the differing evolution of the MC and M1 simulations towards collapse (and seems equally likely to be a cause or an effect of that rapid collapse, or entirely unrelated to that question).

Finally, we perform simulations with Monte-Carlo transport for two different number of packets. We find that at the number of packets currently in use in our simulations, Monte-Carlo sampling noise is an important contribution to the uncertainty in the total amount of matter ejected by the merger ($\sim 25\%$ relative error here). For all other observables, packet count is a negligible contribution to the error budget. However, we caution that a more detailed testing of the impact of the use of implicit Monte-Carlo in the dense regions is likely warranted before concluding that these simulations are converged in other observables. In particular, it is at this point unclear whether differences between M1 and Monte-Carlo in dense regions are due to Monte-Carlo shot noise and the use of implicit Monte-Carlo in dense regions, or to the energy and pressure closure used by the M1 code in dense regions where neutrinos are not in equilibrium with the fluid.

\begin{acknowledgments}
F.F. thanks Alexander Haber, Peter Hammond, Philipp Moesta, David Radice and Federico Schianchi for useful discussions over the duration of this project. This research was supported in part by grant NSF PHY-2309135 to the Kavli Institute for Theoretical Physics (KITP).
F.F. gratefully acknowledges support from the Department of Energy, Office of Science, Office of Nuclear Physics, under contract number
DE-AC02-05CH11231 and from the NSF through grant AST-2107932. M.D. gratefully acknowledges support from the NSF through grant PHY-2110287.  M.D. and F.F. gratefully acknowledge support from NASA through grant 80NSSC22K0719. M.S. acknowledges funding from the Sherman Fairchild Foundation
and by NSF Grants No. PHY-1708212, No. PHY-1708213, and No. OAC-1931266
at Caltech.  L.K. acknowledges funding from the Sherman Fairchild Foundation
and by NSF Grants No. PHY-1912081, No. PHY-2207342, and No. OAC-1931280
at Cornell.
P.C. gratefully acknowledges support from NSF Grant PHY-2020275 (Network for Neutrinos, Nuclear Astrophysics, and Symmetries (N3AS)).
Computations for this manuscript were performed on the Plasma cluster, a Cray CS500 supercomputer at UNH supported by the NSF MRI program under grant AGS-1919310, and on the Wheeler cluster at Caltech, supported by the Sherman Fairchild Foundation. 
\end{acknowledgments}

\bibliography{References/References.bib}

\end{document}